\documentclass[12pt]{article}
\usepackage{amsfonts}
\usepackage{amsmath,amsthm,amssymb}
\usepackage{mathptmx}
\usepackage{geometry}
\usepackage{graphicx}
\usepackage{subfigure}
\usepackage{hyperref}
\usepackage{cancel}
\usepackage{textcomp}
\usepackage{tikz}
\usepackage{bm}
\usepackage{mathrsfs}
\usepackage{filecontents}
\usepackage{times}
\usepackage{epsfig}
\usepackage{xcolor}
\usepackage{slashed}
\usepackage{booktabs}
\usepackage{latexsym}
\usepackage{verbatim}
\usepackage{extarrows}
\usepackage{multirow}
\usepackage{rotating}
\usepackage{colortbl}
\usepackage{indentfirst}
\usepackage[numbers,sort&compress]{natbib}
\usepackage{soul}
\definecolor{mygray}{gray}{.9}
\definecolor{intnull}{RGB}{213,229,255}
\usepackage{arydshln}
\usepackage{diagbox}
\usepackage{appendix}

\newcommand{\dif}{\mathrm{d}}

\usepackage{caption}
\usepackage{tikz}
\usetikzlibrary{arrows,shapes,chains}
\begin{document}
\renewcommand{\thefootnote}{\fnsymbol{footnote}}
\baselineskip=16pt
\pagenumbering{arabic}
\vspace{1.0cm}
\begin{center}
{\Large\sf Superradiance of massive scalar particles around rotating regular black holes }
\\[10pt]
\vspace{.5 cm}

{Hao Yang\footnote{E-mail address: hyang@mail.nankai.edu.cn} and Yan-Gang Miao\footnote{Corresponding author. E-mail address: miaoyg@nankai.edu.cn}}
\vspace{3mm}

{\em School of Physics, Nankai University, Tianjin 300071, China}

\vspace{4.0ex}
\end{center}
\begin{center}
{\bf Abstract}
\end{center}

Regular black holes, as an important attempt to eliminate the singularities in general relativity, have been widely concerned. 
Due to the fact that the superradiance associated with rotating regular black holes plays an indispensable role in black hole physics, 
we calculate the superradiance related effects, {\em i.e.}, the superradiance instability and the energy extraction efficiency for a scalar particle with a small mass around a rotating regular black hole, where the rotating regular black hole is constructed by the modified Newman-Janis algorithm. We analytically give the eigenfrequency associated with instability and the  amplification factor associated with energy extraction.
For two specific models, the rotating Hayward and Bardeen black holes, we investigate how their regularization parameters affect the growth of instability and the efficiency of energy extraction from the two rotating regular black holes.
We find that the regularization parameters give rise to different modes on the superradiance instability and the energy extraction when the rotation parameters are varying. There are two modes for the growth of superradiance  instability, and four modes for the energy extraction.
Our results show the diversity of superradiance in the competition between the regularization parameter and the rotation parameter for rotating regular black holes.

\renewcommand{\thefootnote}{\arabic{footnote}}


\section{Introduction}
\label{sec:intr}
Due to the progress made by the gravitational wave detectors~\cite{A1} and Event Horizon Telescope~\cite{A2,A3}, we have more and more profound understanding of black holes.
Nowadays the general relativity remains the most widely accepted theory of gravity.
However, the singularities are unavoidable in terms of the law of singularity~\cite{A4,A5},which can be proved if the spacetime described by the general relativity has a certain overall structure and meets the strong energy condition.
The existence of singularities not only destroys the completeness of spacetime, but also makes all physical laws including the general  relativity itself fail at the singularities. Therefore, it is highly expected to eliminate the singularities.
A black hole that has no singularities in spacetime, {\em i.e.}, its center is not a singularity, is called a ``regular black hole"~\cite{A6}.
The first regular black hole was constructed by Bardeen~\cite{A7}, which was later interpreted~\cite{A8} as the solution of the gravitational field coupled with the magnetic monopole field of nonlinear electrodynamics.
Subsequently, more regular black holes were constructed under a variety of physical motives. 
Among them, some representative regular black holes are the Hayward black hole~\cite{A9}, the noncommutative spacetime inspired black hole~\cite{A10}, and the regular black hole from the loop quantum gravity~\cite{A11}, {\em etc}.
We note that these regular black holes are static and spherically symmetric, while the regular black holes that would exist in nature are most probably rotating and axially symmetric. For a static and spherically symmetric black hole with singularities,
the Newman-Janis algorithm (NJA) can be used~\cite{A12} to transform it into a rotating and axially symmetric black hole.
However, for a static and spherically symmetric regular black hole without singularities, the NJA does not give a unique rotating and axially symmetric regular black hole because the complex parameterization of radial coordinates has flexibility to a certain extent~\cite{A13,A14,A15}.
To this end, the modified NJA was proposed~\cite{A16} for a static and spherically symmetric black hole, which constrains the complex parameterization of radial coordinates through the global nature of coordinate transformations and thus leads to a unique rotating black hole solution.
In this paper, we analyze rotating regular black holes in terms of the modified NJA.

As is already known, singular black holes (the black holes with singularities) have different spacetime structures from those of regular black holes~\cite{A17,A18,A19,A20}.
Since the event horizon of a black hole shields the region inside a black hole~\cite{A21}, the internal structure of a black hole cannot be detected directly. 
Therefore, the description of internal structures needs to rely on the interaction between a black hole and its external matter or the gravitational field.
Here we focus on an interesting phenomenon, the ``superradiance"~\cite{A22,A23,A24,A25,A26}, which refers to the phenomenon that the energy of a particle can be amplified and reflected back by the black hole when the particle incident on a black hole meets certain conditions.
Based on the superradiance, if a particle is reflected by a ``mirror" outside a black hole, the  particle's energy will increase exponentially near the event horizon of the black hole, which is also known as a ``black hole bomb" or ``superradiance instability"~\cite{A27,A28,A29}.
When a massive scalar particle is at its bound state, the mass term can act as a mirror~\cite{A30,A31,A32,A33,A34}, making the superradiance instability possible.
The superradiance related effects, {\em i.e.}, the superradiance instability and the energy extraction efficiency are mainly affected by the properties of black hole spacetime, so the behavior of scalar particles satisfying the conditions of superradiance will reveal the properties of black hole spacetime.
In the past, the  superradiance was studied~\cite{A35,A36,A37,A38,A39,A40} mainly for singular black holes, but rarely for rotating regular black holes. Considering the fact that the superradiance associated with rotating regular black holes plays an indispensable role in black hole physics, we cover the shortage.
Moreover, we shall better understand the properties of regular black holes when we study the superradiance phenomena affected by regularization parameters. This is because the spacetime singularities can be removed by introducing regularization parameters into the metrics of singular black holes.
Therefore, starting from the bound state of a massive scalar particle, we deal with analytically the superradiance instability under the background of a rotating regular black hole constructed by the modified NJA, and then from the free state of the scalar particle we calculate the energy amplification factor.
For two specific models, the rotating Hayward and Bardeen black holes, we make the numerical calculations and image analyses in order to show the influence of regularization parameters on the superradiance effects of the two models. By comparing the results of the two specific models with that of the Kerr black hole, we can show the differences in superradiance between rotating regular black holes and rotating  singular black holes.


This paper is organized as follows. In Sec.~\ref{sec: Rotating_metric}, we briefly introduce rotating regular black holes and massive scalar field equations. In Sec.~\ref{sec: QBS}, we determine the conditions of superradiance instability by using the asymptotic matching method when a massive scalar particle is at its bound state.
For the scattering process of a free state particle, we calculate the superradiance amplification factor in Sec.~\ref{sec: Free-State}.
For the two specific  models, we discuss in Sec.~\ref{sec: model_analysis} how the superradiance behaviors of the rotating Hayward and Bardeen black holes vary with respect to their respective regularization parameters and rotation parameters, and compare them with the superradiance behavior of Kerr black holes (a kind of rotating singular black holes).
Finally, we give our conclusions in Sec.~\ref{sec:conclusion}, where some comments and further extensions are included. 
Throughout the paper we adopt the natural units, $\hbar=c=G=1$.

\section{Master equations of massive scalar fields in rotating regular black holes}
\label{sec: Rotating_metric}
It is considerably difficult to obtain the solutions of rotating black holes from the Einstein field equations because the complexity of Einstein's field equations in the rotating case is much greater than that in the static case.
Therefore, the widely-used method for constructing rotating black holes is the NJA~\cite{A12}.
This algorithm originated from the connection between a static black hole and a rotating one in general relativity.
It is well known that the Schwarzschild, Reissner-Nordstr\"om (RN), Kerr, and Kerr-Newman (KN) black holes were obtained by solving Einstein's field equations in vacuum. 
By comparing the metrics of these black holes, Newman and Janis proposed the transformation from the static and spherically symmetric Schwarzschild black hole to the rotating and axially symmetric Kerr black hole. Moreover,
the algorithm can also realize the transformation from the RN black hole to the KN black hole.

However, for most static regular black holes with the spherical symmetry, the corresponding rotating  regular black holes with the axial symmetry cannot be uniquely determined due to the uncertainty of the complex parametrization of radial coordinates. 
Based on the constraint of the global coordinate transformation, the modified NJA without complexification was proposed~\cite{A16}. 
Here we briefly introduce the rotating regular black holes constructed by this algorithm. 

As far as we know, the regular black holes with the spherical symmetry in general relativity are described by the following line element,
\begin{equation}
{\rm d}s^2=-F(r){\rm d}t^2+\frac{{\rm d}r^2}{F(r)}+r^2({\rm d}\theta^2+\sin^2\theta {\rm d}\phi^2),\label{sphesymmmetr}
\end{equation}
where $F(r)$ contains the black hole mass $M$.
In the advanced null coordinates $(u,r,\theta,\phi)$ defined by
\begin{equation}
\dif u=\dif t-\frac{\dif r}{F(r)},
\end{equation}
one expresses the contravariant form of the metric in terms of a null tetrad,
\begin{equation}
g^{\mu\nu}=-l^\mu n^\nu-l^\nu n^\mu+m^\mu m^{*\nu}+m^\nu m^{*\mu},
\end{equation}
where 
\begin{subequations}\label{eq:tetrad}
\begin{equation}
l^\mu=\delta^\mu_r,
\end{equation}
\begin{equation}
n^\mu=\delta^\mu_u-\frac{F}{2}\delta^\mu_r,
\end{equation}
\begin{equation}
m^\mu=\frac{1}{\sqrt{2r^2}}\left(\delta^\mu_\theta+\frac{i}{\sin\theta}\delta^\mu_\phi\right),
\end{equation}
\begin{equation}
l_\mu l^\mu=m_\mu m^\mu=n^\nu n_\nu=l_\mu m^\mu=n_\mu m^\mu=0,
\end{equation}
\begin{equation}
l_\mu n^\mu=-m_\mu m^{*\mu}=1,
\end{equation}
\end{subequations}
and ``$*$" means complex conjugate.
Then, one introduces the rotation via the complex transformation,
\begin{equation}
r\rightarrow r+ia\cos\theta,\qquad  u\rightarrow u-i a\cos\theta,
\end{equation}
where $a$ is rotation parameter, and requires that $\delta^\mu_\nu$ transform as a vector under this complex transformation,
\begin{equation}\label{eq:transform-delta}
\delta^\mu_r\rightarrow \delta^\mu_r,
\qquad \delta^\mu_u\rightarrow\delta^\mu_u,
\qquad \delta^\mu_\theta\rightarrow\delta^\mu_\theta+ia\sin\theta(\delta^\mu_u-\delta^\mu_r),
\qquad \delta^\mu_\phi\rightarrow\delta^\mu_\phi.
\end{equation}

For a singular black hole, the metric function of its rotating counterpart can be determined under this complex transformation rule.
However, such a transformation rule does not work well for a regular black hole.
Thus, one assumes that $\{F,r^2\}$ transform to $\{B,\Psi\}$:
\begin{equation}\label{eq:transform-function}
\{F(r),r^2\}\rightarrow\{B(r,\theta,a),\Psi(r,\theta,a)\},
\end{equation}
where $\{B,\Psi\}$ are real functions to be determined and should recover their static counterparts in the limit of $a\rightarrow 0$, namely, 
\begin{equation}
\lim_{a\to 0}B(r,\theta,a)=F(r),\qquad \lim_{a\to 0}\Psi(r,\theta,a)=r^2.
\end{equation}
According to Eq.~\eqref{eq:transform-delta} and Eq.~\eqref{eq:transform-function}, the null tetrad becomes
\begin{subequations}
\begin{equation}
l^\mu=\delta^\mu_r,
\end{equation}
\begin{equation}
n^\mu=\delta^\mu_u-\frac{B}{2}\delta^\mu_r,
\end{equation}
\begin{equation}
m^\mu=\frac{1}{\sqrt{2\Psi}}\left[\delta^\mu_\theta+ia\sin\theta(\delta^\mu_u-\delta^\mu_r)+\frac{i}{\sin\theta}\delta^\mu_\phi\right],
\end{equation}
\end{subequations}
and the corresponding line element with rotation takes the form,
\begin{equation}
\begin{split}
\dif s^2=& -B\dif u^2-2\dif u\dif r-2a\sin^2\theta\left(1-B\right)\dif u\dif \phi+2a\sin^2\theta\dif r\dif \phi\\
& +\Psi\dif\theta^2+\sin^2\theta\left[\Psi+a^2\sin^2\theta\left(2-B\right)\right]\dif\phi^2.
\end{split}
\end{equation}

Next, one rewrites the above line element with the Boyer-Lindquist coordinates, and lets the metric have only one non-vanishing off-diagonal term $g_{t\varphi}$.
To reach the aim, one needs the following coordinate transformation,
\begin{equation}
\dif u=\dif t+\lambda(r)\dif r,\qquad \dif\phi=\dif\varphi+\chi(r)\dif r,
\end{equation}
where $\{\lambda(r),\chi(r)\}$ depend only on $r$ in order to ensure integrability.
If the transformation Eq.~\eqref{eq:transform-function} is a priori determined, $\{\lambda(r),\chi(r)\}$ may not exist.
Considering these constraints, one has the formulations of $\{B(r,\theta,a), \Psi(r,\theta), \lambda(r), \chi(r)\}$,
\begin{subequations}
\begin{equation}
B(r,\theta)=\frac{Fr^2+a^2\cos^2\theta}{\Psi},
\end{equation}
\begin{equation}
\Psi(r,\theta)=r^2+a^2\cos^2\theta,
\end{equation}
\begin{equation}
\lambda(r)=-\frac{r^2+a^2}{Fr^2+a^2},
\end{equation}
\begin{equation}
\chi(r)=-\frac{a}{Fr^2+a^2}.
\end{equation}
\end{subequations}
As a result, one obtains the line element for rotating regular black holes with the Kerr-like form,
\begin{equation}
\label{rotating_metric}
{\rm d}s^2=-\left(1-\frac{2f}{\rho^2}\right){\rm d}t^2+\frac{\rho^2}{\Delta}{\rm d}r^2-\frac{4af\sin^2\theta}{\rho^2}{\rm d}t{\rm d}\varphi+\rho^2{\rm d}\theta^2+\frac{\Sigma \sin^2\theta}{\rho^2}{\rm d}\varphi^2,
\end{equation}
where 
\begin{eqnarray}
\rho^2&=&r^2+a^2\cos^2\theta,\label{rho}\\
2f&=&r^2(1-F),\label{fandF} \\
\Delta&=&r^2F+a^2=r^2-2f+a^2,\\
\Sigma&=&(r^2+a^2)^2-\Delta\, a^2\sin^2\theta=(r^2+a^2)\rho^2+2fa^2\sin^2\theta.
\end{eqnarray}
The regularity of the above rotating regular black holes can be verified because the Ricci scalar $R$ and Kretschmann scalar $K$ are finite everywhere. The reason is that the static regular black hole as a seed has the de Sitter-like behavior in the limit of $r\rightarrow 0$~\cite{A16,A16A}:
\begin{equation}
F(r)\sim 1-Cr^2\qquad {\rm{with}}\qquad C>0.\label{asybehrgotozerorbh}
\end{equation}
On the other hand, one can determine the horizon $r_{\rm H}$ by solving the algebraic equation, $g_{tt}g_{\varphi\varphi}-g_{t\varphi}^2=-\Delta\sin^2\theta=0$, which can be reduced to $\theta=0$ or
\begin{equation}
 r^2-2f+a^2=0.\label{horizonequ}
\end{equation}

In the following of this section, we turn to  massive scalar field equations.
In a curved spacetime, a massive scalar field $\Phi(t, r, \theta, \varphi)$ with the mass $\mu$ is described by
\begin{equation}
\nabla^\nu\nabla_\nu\Phi=\mu^2\Phi.
\end{equation}
In order to separate variables in the rotating regular black hole spacetime described by Eq.~(\ref{rotating_metric}), 
we make the assumption,
\begin{equation}
\Phi(t, r, \theta, \varphi)={\rm e}^{-i\omega t+im\varphi}S(\theta)R(r),\label{sepvar}
\end{equation}
and then obtain the equations that govern $S(\theta)$ and $R(r)$, respectively,
\begin{equation}\label{S(theta)}
\frac{1}{\sin\theta}\frac{{\rm d}}{{\rm d}\theta}\left[\sin\theta\frac{{\rm d}}{{\rm d}\theta}S(\theta)\right]+\left[a^2(\omega^2-\mu^2)\cos^2\theta-\frac{m^2}{\sin^2\theta}+\lambda\right]S(\theta)=0,
\end{equation}
and
\begin{equation}\label{R(r)}
\Delta\frac{{\rm d}}{{\rm d}r}\left[\Delta \frac{{\rm d}}{{\rm d}r}R(r)\right]+\left[\omega^2(r^2+a^2)^2-4afm\omega+a^2m^2-\Delta(\mu^2r^2+\lambda+a^2\omega^2)\right]R(r)=0,
\end{equation}
where $\omega$ is the frequency of the massive scalar field, $m$ the azimuthal number with respect to the rotation axis, and $\lambda$ the separation parameter which will be fixed approximately as an eigenvalue of Eq.~(\ref{S(theta)}).

\section{Superradiance instability of massive scalar fields}
\label{sec: QBS}
We discuss the eigenvalue of Eq.~(\ref{R(r)}) under the boundary conditions of outgoing waves at infinity and ingoing waves at an event horizon. If the mass $\mu$ is small (see the following details), we shall show that the eigenvalue of Eq.~(\ref{R(r)}) is exactly $\omega^2$, where $\omega$ is also called eigenfrequency and it is complex,
\begin{equation}
\omega=\omega_{\rm R}+i\omega_{\rm I}.
\end{equation}
If $\omega_{\rm I}>0$, $\phi(t, r, \theta, \varphi)$ increases exponentially with time, see Eq.~(\ref{sepvar}), which leads to instability, namely, the superradiance instability. In order to analyze the condition that the superradiance instability happens, we deal with $\omega_{\rm I}$ as a positive quantity in the following discussions. In general, we have the condition, $\omega_{\rm R}\gg \omega_{\rm I}$, which means that a black hole is perturbed by a scalar field, leading to a slow change of $\phi(t, r, \theta, \varphi)$ with respect to time, that is, $\omega_{\rm I}$ is small when compared with $\omega_{\rm R}$.

\begin{figure}[htbp]
	\centering
	\begin{minipage}[t]{0.8\linewidth}
		\centering
		\includegraphics[width=1\linewidth]{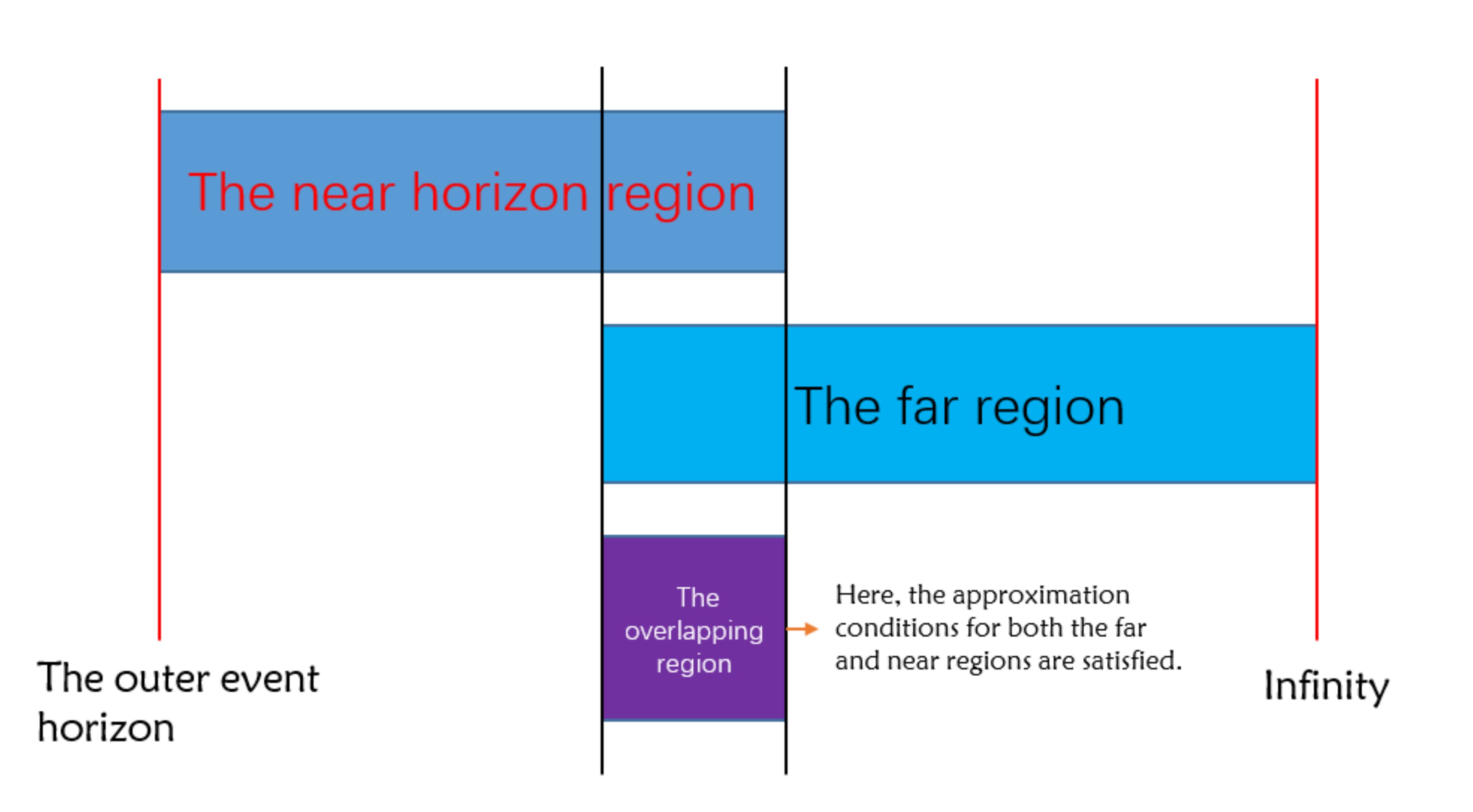}
	\end{minipage}
	\caption{The schematic diagram for showing the asymptotic matching method. In the near horizon region and the far region, the wave function behaves as two different solutions, respectively. The asymptotic behavior of the two solutions should be consistent in the overlapping region. Through this consistency, we can determine the coefficients of the wave function.}
	\label{fig:Schematic diagram}
\end{figure}

In order to solve Eq.~(\ref{R(r)}) analytically in terms of the asymptotic matching  method~\cite{A32}, see Fig. \ref{fig:Schematic diagram} for the illustration of this method, we have to make the approximations,
$\mu M\ll 1$ and $|\omega| M\ll 1$, {\em i.e.}, $\mu$ is small and takes the same order of magnitude as $|\omega|$. 
Using these approximations and considering that $a$ has the same {\color{red} or lower} order of magnitude {\color{red} as} $M$, {\color{red}which means $\mu a\ll1$ and $|\omega|a\ll 1$}, we find that the solution of Eq.~(\ref{S(theta)}),  $S(\theta)$, is just the spherical harmonics and simultaneously we fix the separation parameter, $\lambda\approx l(l+1)$, where $l=0, 1, 2,...$ is called multipole number.
Next we focus on Eq.~(\ref{R(r)}). The key idea of the asymptotic matching  method is to solve Eq.~(\ref{R(r)}) approximately in the near horizon region and in the far region, respectively, and then match the coefficients of the two solutions in the overlapping region of the two regions, so as to determine the general solution.

In the first step of the asymptotic matching  method, we search for the asymptotic form of the far region solution in the near horizon region, which includes two sub-steps: The first is to compute the far region solution and the second sub-step is to give its asymptotic form in the near horizon region. In the far region, $r\gg M$, also for the quantities of the same {\color{red} or lower} order of magnitude {\color{red} as} $M$, such as $a$, {\em i.e.}, $r\gg a$, Eq.~(\ref{R(r)}) can be approximately simplified to be
\begin{equation}\label{approx_1}
r^2\frac{{\rm d}}{{\rm d}r}\left[r^2\frac{{\rm d}}{{\rm d}r}R(r)\right]+\left[\omega^2r^4-\mu^2r^4+2\mu^2r^2f(r)-\lambda r^2\right]R(r)=0,
\end{equation}
where $\omega^2$ acts as the eigenvalue. In the above equation, the only uncertain factor is $f(r)$ which satisfies $2f(r)=r^2[1-F(r)]$, see Eq.~(\ref{fandF}). 
Notice that $F(r)\sim 1$ in the far region because the spacetime is asymptotically flat, we expand $F(r)$ as the series of $1/r$,
\begin{equation}
F(r)=1+A\cdot\frac{1}{r}+B\cdot\frac{1}{r^2}+O\left(\frac{1}{r^3}\right),\label{Fexpand}
\end{equation}
where $A=\left.-r^2\frac{{\rm d}F(r)}{{\rm d}r}\right|_{r\rightarrow\infty}$ and $B=\left.\frac{r^2}{2}\frac{{\rm d}}{{\rm d}r}\left[r^2\frac{{\rm d}F(r)}{{\rm d}r}\right]\right|_{r\rightarrow\infty}$.
Here we consider the situation where both $A$ and $B$ converge.
Note that $A$ has the same order of magnitude as $M$, and $B$ as $M^2$ because the far region can also be written as $M/r\ll 1$.
After Eq.~(\ref{fandF}) and Eq.~(\ref{Fexpand}) are considered together with $\lambda\approx l(l+1)$, Eq.~(\ref{approx_1}) can be further simplified to be
\begin{equation}\label{approx_2}
\frac{{\rm d^2}}{{\rm d}r^2}\left[rR(r)\right]+\left[\omega^2-\mu^2-\frac{A\mu^2}{r}-\frac{l(l+1)}{r^2}\right]rR(r)=0.
\end{equation}
When we use the following definitions and notations,
\begin{eqnarray}
& & k^2\equiv \mu^2-\omega^2,\label{Def_1}\\
& & \nu\equiv -\frac{A\mu^2}{2k},\label{Def_2}\\
& & x\equiv 2kr,\label{Def_x}
\end{eqnarray}
Eq.~(\ref{approx_2}) becomes
\begin{equation}\label{approx_3}
\frac{{\rm d^2}}{{\rm d}x^2}\left[xR(x)\right]+\left[-\frac{1}{4}+\frac{\nu}{x}-\frac{l(l+1)}{x^2}\right]xR(x)=0.
\end{equation}
And its solution with the boundary condition of an outgoing wave in the far region takes the form, 
\begin{equation}
R(x)= C_1 x^l e^{-x/2}U(l+1-\nu,2l+2,x),\label{farsolu}
\end{equation}
where $U(a,b,x)$ is the second Kummer function~\cite{A41} and $C_1$ integration constant.
At the moment, we finish the first sub-step mentioned above. We make some comments to Eq.~(\ref{farsolu}). It turns exactly to the solution of hydrogen atoms when $\nu$ takes an integer.
When we solve Eq.~(\ref{approx_3}) for bound states, the biggest difference between a hydrogen atom  and the present case lies in the asymptotic behavior for a small $|x|$. 
For a hydrogen atom, where $x$ and $\nu$ are real, the wave function is required to be finite at $x=0$; while for the present case, the scalar field interacts with the rotating regular black hole, leading to a slow change of the wave function, {\em i.e.}, to the appearance of a small $\omega_{\rm I}$.
Thus, $\nu$, like $\omega$, is also complex because of the relationship between $\omega$ and $\nu$, see Eq.~(\ref{Def_1}) and Eq.~(\ref{Def_2}). If we define $\delta\nu$ as
\begin{equation}
	n+\delta \nu\equiv \nu-l-1,\label{Def_n}
\end{equation}
where $n$ is a non-negative integer, $\delta\nu$ is a complex number with a small modulus. And we can rewrite Eq.~(\ref{farsolu}) to be
\begin{equation}
	R(x)= C_1 x^l e^{-x/2}U(-n-\delta \nu,2l+2,x).\label{farsolu2}
\end{equation}

Now we turn to the second sub-step mentioned above, that is, to give the asymptotic form of the far region solution (Eq.~(\ref{farsolu2})) in the near horizon region.
That is, considering the condition,\footnote{Note that the condition of the near horizon region, $r\ll {\rm Max}(l/\mu,l/|\omega|)$, is compatible with the condition of the far region, $r\gg M$. That is, there exists the overlapping region of the near horizon region and the far region, in which the inequality, $M\ll r\ll {\rm Max}(l/\mu,l/|\omega|)$, holds. Here we give a brief proof. Considering the definition of $k$, see Eq.~(\ref{Def_1}), and the above-mentioned assumption that $\mu$ takes the same order of magnitude as $|\omega|$, we know that the order of magnitude of $|k|$ is smaller than $\mu$ and $|\omega|$, so the near horizon region means $ |k|r\ll {\rm Max}(l|k|/\mu,l|k|/|\omega|)<l$. Again considering $l\ge 1$ in the near horizon region, we deduce that the near horizon region can equivalently be expressed by $|k|r\ll 1$. In addition, due to our approximations,
$\mu M\ll 1$ and $|\omega| M\ll 1$, i.e., $ M\ll 1/\mu\le l/\mu$ and $ M\ll 1/|\omega|\le l/|\omega|$, which gives rise to $M\ll {\rm Max}(l/\mu,l/|\omega|)$. As a result, the overlapping region exists and stays in $M\ll r\ll {\rm Max}(l/\mu,l/|\omega|)$.} $|x|\ll 1$ or $|k|r\ll 1$, and expanding the second Kummer function in Eq.~(\ref{farsolu2}), we obtain the asymptotic  behavior in the near horizon region,
\begin{eqnarray}\label{far_small}
R(x)&\approx& C_1\left[ (-1)^n\frac{(2l+1+n)!}{(2l+1)!}x^l+(-1)^{n+1}(2l)!n!(\delta\nu) x^{-l-1}\right]\nonumber \\
&=&C_1(-1)^n\frac{(2l+1+n)!}{(2l+1)!}(2k)^l\cdot r^l+C_1(-1)^{n+1}(2l)!n!\delta\nu (2k)^{-l-1}\cdot r^{-l-1}.
\end{eqnarray}
In order for the two terms in the above expansion to have the same order of magnitude, we deduce $\delta\nu\sim (kr)^{2l+1}$, which coincides with the above assumption that $|\delta\nu|$ is much smaller than one.

In the second step of the asymptotic matching  method, which also contains two sub-steps, we give the asymptotic equation of Eq.~(\ref{R(r)}) in the near horizon region and solve it, and then match the asymptotic solution with Eq.~(\ref{far_small}) in the overlapping region, so as to determine $\delta\nu$. 
In the first sub-step, we expand $f(r)$ near the outer event horizon $r_{\rm H}^+$,
\begin{equation}
f(r)\approx f(r_{\rm H}^+)+f'(r_{\rm H}^+)(r-r_{\rm H}^+)+O(r^2),
\end{equation}
where the prime means the derivative with respect to $r$, and compute $\Delta(r)$ to the first order of  
$(r-r_{\rm H}^+)$,
\begin{eqnarray}
\Delta(r)&=&r^2-2f+a^2\nonumber \\
&\approx& r^2-2rf'(r_{\rm H}^+)+a^2-2f(r_{\rm H}^+)+2r_{\rm H}f'(r_{\rm H}^+)\nonumber \\
&=&(r-r_{\rm H}^+)[r-2f'(r_{\rm H}^+)+r_{\rm H}^+].
\end{eqnarray}
Using the definitions,
\begin{eqnarray}
& &P\equiv\frac{ma-2\omega f(r_{\rm H}^+)}{2r_{\rm H}^+-2f'(r_{\rm H}^+)},\label{Def_P}\\
& & z\equiv\frac{r-r_{\rm H}^+}{2r_{\rm H}^+-2f'(r_{\rm H}^+)},\label{Def_Z}
\end{eqnarray}
we rewrite Eq.~(\ref{R(r)}) to be
\begin{equation}
z(z+1)\frac{{\rm d}}{{\rm d}z}\left[z(z+1)\frac{{\rm d}R(z)}{{\rm d}z}\right]+\left[P^2-\lambda z(z+1)\right]R(z)=0.\label{equwPz}
\end{equation}
By introducing a new function $G(z)$ that is related to $R(z)$ as follows,
\begin{equation}
R(z)=\left(\frac{z}{z+1}\right)^{iP}G(z),
\end{equation}
we further simplify Eq.~(\ref{equwPz}) to be
\begin{equation}
z(z+1)G''(z)+(1+2iP+2z)G'(z)-\lambda G(z)=0.
\end{equation}
It is exactly the hypergeometric differential equation with the general solution,
\begin{equation}
G(z)=A_1\cdot {_2{F}_1}(-l,l+1,1+2iP;-z)+A_2(-z)^{-2iP} {_2{F}_1}(-l-2iP,l+1-2iP,1-2iP;-z),
\end{equation} 
where $A_1$ and $A_2$ are integration constants.
Considering the boundary condition of ingoing waves at the event horizon, i.e., $R(z)\sim z^{iP}$ at $z\sim 0$, we obtain the corresponding solution,
\begin{equation}
R(z)=A_1\left(\frac{z}{z+1}\right)^{iP}{_2F_1}(-l,l+1,1+2iP;-z),\label{nhrsolu}
\end{equation}
which ends the first sub-step.
In the second sub-step, we figure out the asymptotic form of Eq.~(\ref{nhrsolu}) in the far region from, {\em i.e.}, from the solution of the near horizon region. That is, Eq.~(\ref{nhrsolu}) takes the following form for a large $z$,
\begin{eqnarray}\label{near_large}
R(z)&\approx& A_1\left[\frac{\Gamma(1+2iP)\Gamma(2l+1)}{\Gamma(l+1)\Gamma(l+1+2iP)}z^l+\frac{\Gamma(1+2iP)\Gamma(-2l-1)}{\Gamma(-l)\Gamma(2iP-l)}z^{-l-1}\right]\nonumber \\
&=&  A_1\left\{\frac{\Gamma(1+2iP)\Gamma(2l+1)}{\Gamma(l+1)\Gamma(l+1+2iP)}2^{-l}\left[r_{\rm H}^+-f'(r_{\rm H}^+)\right]^{-l}\cdot r^l\right.\nonumber \\
& &\left.+\frac{\Gamma(1+2iP)\Gamma(-2l-1)}{\Gamma(-l)\Gamma(2iP-l)}2^{l+1}\left[r_{\rm H}^+-f'(r_{\rm H}^+)\right]^{l+1}\cdot r^{-l-1}\right\}.
\end{eqnarray}
At the moment we finish the second sub-step.

Up to now we are ready to achieve the expected asymptotic matching, {\em i.e.}, to determine $\delta\nu$ by comparing  Eq.~(\ref{far_small}) with Eq.~(\ref{near_large}), where both the equations are valid in the overlapping region.
We note that Eq.~(\ref{far_small}) is the solution for a small $r$ approximation in the far region, $r\gg M$, and Eq.~(\ref{near_large}) is the solution for a large $r$ approximation in the near horizon region, $r\ll {\rm Max}(l/|\omega|,l/\mu)$, and that both the solutions are worked out under the approximations,
$\mu M\ll 1$ and $|\omega| M\ll 1$.
Thus, there exists the common solution in the overlapping region of the near horizon region and the far region. In this overlapping region, the inequality, $M\ll r\ll {\rm Max}(l/|\omega|,l/\mu)$, holds and both Eq.~(\ref{far_small}) and Eq.~(\ref{near_large}) are applicable.  
After matching the coefficients in Eq.~(\ref{far_small}) and Eq.~(\ref{near_large}), we obtain 
\begin{equation}
\delta\nu=2iP\left\{ 4k\left[r_{\rm H}^+-f'(r_{\rm H}^+)\right]\right\}^{2l+1}\cdot\frac{(2l+1+n)!}{n!}\cdot\left[\frac{l!}{(2l)!(2l+1)!}\right]^2\prod_{j=1}^{l}(j^2+4P^2).\label{imanu}
\end{equation}

With the performance above, we are able to derive\footnote{See Appendix~\ref{appendix:A} for the analysis in detail.} the condition the superradiance instability occurs. By using Eq.~(\ref{Def_1}), Eq.~(\ref{Def_2}), and Eq.~(\ref{Def_n}), 
we obtain
\begin{equation}
	\mu^2-\omega^2=\frac{A^2\mu^4}{4(l+1+n+\delta \nu)^2}.\label{muomere}
\end{equation}
Considering $\omega_{\rm I}\ll\omega_{\rm R}$ and  $|\delta\nu|\ll 1$, we deduce the following two relations from Eq.~(\ref{muomere}),
\begin{equation}
\omega_{\rm R}\approx\mu ,
\end{equation}
and
\begin{equation}
\omega_{\rm I}=\frac{A^2\mu^3}{4(l+1+n)^3}\delta \nu_{\rm I}.\label{imaomega}
\end{equation}
Further considering $k_{\rm R}\gg |k_{\rm I}|$ and $P_{\rm R}\gg |P_{\rm I}|$ and combining Eq.~(\ref{imanu}) and Eq.~(\ref{imaomega}), we finally derive the imaginary part of $\omega$,
\begin{equation}\label{eq:Omega_I}
\begin{split}
\omega_{{\rm I}(l,m)}=&|A|\left[ma-2\mu f(r_{\rm H}^+)\right]\cdot 2^{2l-1}A^{2l+2}\mu^{4l+5}\frac{(2l+1+n)!}{(l+1+n)^{2l+4}n!}\left[\frac{l!}{(2l)!(2l+1)!}\right]^2\\
&\times\prod_{j=1}^{l}\left\{j^2\left[r_{\rm H}^+-f'(r_{\rm H}^+)\right]^2+\left[ma-2\mu f(r_{\rm H}^+)\right]^2\right\}.\\
\end{split}
\end{equation}
Note that the superradiance instability will occur when $\omega_{\rm I}>0$.
From the above equation, we can conclude the condition,
\begin{equation}
 \frac{ma}{2f(r_{\rm H}^+)}=\frac{ma}{{r_{\rm H}^+}^2+a^2}>\mu,
\end{equation}
or we can rewrite it as
\begin{equation}
m\Omega>\mu,\label{superradinscon}
\end{equation}
where
\begin{equation}
\Omega\equiv \frac{a}{{r_{\rm H}^+}^2+a^2}
\end{equation}
is the angular velocity of the outer horizon of a black hole.
We emphasize that the superradiance instability will only occur in the quasi-bound states of the scalar particles satisfying $\mu M\ll 1$ and $|\omega| M\ll 1$, where a quasi-bound state requires $\omega_{\rm R}<\mu$.

\section{Superradiance amplification factors}
\label{sec: Free-State}
We discuss the scattering process of massive scalar particles and calculate the amplification factors.

In the scattering process, the scalar particles are at free states, which means that the frequency is greater than the mass, $\omega>\mu$.\footnote{Note that the frequency of particles, $\omega$, is real for the scattering process.}
Therefore, Eq.~\eqref{approx_2} becomes 
\begin{equation}\label{5}
\frac{\rm{d}^2}{{\rm{d}}y^2}\left[yR(y)\right]+\left[-\frac{1}{4}+\frac{i\Lambda}{y}-\frac{l(l+1)}{y^2}\right]yR(y)=0,
\end{equation}
where we have defined new notations,
\begin{eqnarray}
K^2&\equiv & \omega^2-\mu^2,\label{Def_3}\\
\Lambda &\equiv& \frac{A\mu^2}{2K},\label{Def_4}\\
 y&\equiv &2iKr.
\end{eqnarray}
We only need to consider the ingoing waves at the event horizon, but not the outgoing waves at infinity in the scattering process.  
As a result, we obtain the general solution of Eq.~(\ref{5}),
\begin{equation}\label{R(y)}
R\left(y\right)=B_1 e^{-y/2}y^lM(l+1-i\Lambda,2l+2,y)+B_2e^{-y/2}y^{-l-1}M(-l-i\Lambda,-2l,y),
\end{equation}
where $M(a,b,z)$ is the first Kummer function. 
Next, we determine the superradiance amplification factors still by using the asymptotic matching  method.

On the one hand, we deduce the asymptotic solution in the near horizon region. Using the property of $M(a,b,z)$, we give the asymptotic behavior of the general solution at a small $y$,
\begin{equation}
\lim_{y\rightarrow 0} R(y)=B_1y^l+B_2y^{-l-1}
\label{far_small_2}.
\end{equation}
Then matching Eq.~\eqref{far_small_2} with Eq.~\eqref{near_large} in the overlapping region, we derive the coefficients, 
\begin{equation}
B_1=A_1\frac{\Gamma(1+2iP)\Gamma(2l+1)}{\Gamma(l+1)\Gamma(l+1+2iP)}(4iK)^{-l}\left[r_{\rm H}^+-f'(r_{\rm H}^+)\right]^{-l},\label{coefb1}
\end{equation}
\begin{equation}
B_2=A_1\frac{\Gamma(1+2iP)\Gamma(-2l-1)}{\Gamma(-l)\Gamma(2iP-l)}(4iK)^{l+1}\left[r_{\rm H}^+-f'(r_{\rm H}^+)\right]^{l+1}.\label{coefb2}
\end{equation}

On the other hand, we consider the asymptotic solution of Eq.~\eqref{R(r)} in the large $r$ limit together with the low frequency limit,
\begin{equation}
R(r)\approx \mathcal{I}e^{-iKr}r^{i\Lambda-1}+\mathcal{R}e^{iKr}r^{-i\Lambda-1}=R_{\rm in}(r)+R_{\rm re}(r),\label{farregionsol}
\end{equation}
where $R_{\rm in}(r)$ and $R_{\rm re}(r)$ denote the incident wave and  reflected wave with their amplitudes $\mathcal{I}$ and $\mathcal{R}$, respectively. We then compute their fluxes as follows,
\begin{equation}
\Phi_{\rm in}(r)=\frac{1}{2i\mu}\left[R_{\rm in}^*(r)\frac{\partial}{\partial r}R_{\rm in}(r)-R_{\rm in}(r)\frac{\partial}{\partial r}R_{\rm in}^*(r)\right]
=\frac{\Lambda-Kr}{\mu r^3}|\mathcal{I}|^2,
\end{equation}
\begin{equation}
\Phi_{\rm re}(r)=\frac{1}{2i\mu}\left[R_{\rm re}^*(r)\frac{\partial}{\partial r}R_{\rm re}(r)-R_{\rm re}(r)\frac{\partial}{\partial r}R_{\rm re}^*(r)\right]
=\frac{Kr-\Lambda}{\mu r^3}|\mathcal{R}|^2,
\end{equation}
and determine the superradiance amplification factor in terms of the amplitudes,
\begin{equation}
Z_{lm}=\left|\frac{\Phi_{\rm re}(r)}{\Phi_{\rm in}(r)}\right|-1=\frac{|\mathcal{R}|^2}{|\mathcal{I}|^2}-1.\label{amplifact}
\end{equation}
Next we turn to the calculation of the amplitudes. Expanding Eq.~\eqref{R(y)} at infinity to obtain its  asymptotic form and matching the asymptotic solution with  Eq.~(\ref{farregionsol}) in the overlapping region, we obtain the amplitudes,
\begin{equation}\label{I}
\mathscr{I}=(-1)^l\frac{i}{2K}e^{i\Lambda\ln{(2K)}}e^{\frac{\Lambda\pi}{2}}\left[\frac{\Gamma(2l+2)}{\Gamma(l+1+i\Lambda)}B_1-\frac{\Gamma(-2l)}{\Gamma(-l+i\Lambda)}B_2\right],
\end{equation}
\begin{equation}\label{R}
\mathscr{R}=-\frac{i}{2K}e^{-i\Lambda\ln{(2K)}}e^{\frac{\Lambda\pi}{2}}\left[\frac{\Gamma(2l+2)}{\Gamma(l+1-i\Lambda)}B_1+\frac{\Gamma(-2l)}{\Gamma(-l-i\Lambda)}B_2\right],
\end{equation}
where $B_1$ and $B_2$ have been given in Eqs.~(\ref{coefb1}) and (\ref{coefb2}). In the following, we analyze the amplification factor in two different cases.

The first case is that the frequency is very close to the mass, namely, $\omega^2-\mu^2\ll\mu^2$, which leads to the result that $\Lambda$ cannot be ignored, see Eqs.~(\ref{Def_3}) and (\ref{Def_4}). We thus turn to the second terms of Eq.~\eqref{I} and Eq.~\eqref{R} in the brackets. As $l$ is a  positive integer,  
$\Gamma(-2l)$ tends to infinity, {\em i.e.}, the second terms dominate the amplitudes. Therefore,
we find that the amplification factor is vanishing in this case when we substitute Eq.~\eqref{I} and Eq.~\eqref{R} into Eq.~(\ref{amplifact}).
The vanishing amplification factor means that a rotating regular black hole will reflect all the incoming particles back, which is usually called ``the total reflection process".

The second case is that the frequency is not very close to the mass, which means $\Lambda\ll 1$, or $\Lambda$ can be ignored in Eq.~\eqref{I} and Eq.~\eqref{R}.
Substituting Eqs.~(\ref{coefb1}), (\ref{coefb2}), \eqref{I}, and \eqref{R} into Eq.~(\ref{amplifact}), we give the amplification factor,
\begin{eqnarray}
\label{Zlm}
Z_{lm}&=&8f(r_{\rm H}^+)\left[\frac{ma}{2f(r_{\rm H}^+)}-\omega\right]\cdot\left[\frac{(l!)^2}{(2l)!(2l+1)!!}\right]^2\cdot\left\{2\left[r_{\rm H}^+-f'(r_{\rm H}^+)\right]\right\}^{2l}\nonumber \\
& & \times \left(\sqrt{\omega^2-\mu^2}\right)^{2l+1}\prod^l_{j=1}\left\{1+\frac{\left[ma-2f(r_{\rm H}^+)\omega\right]^2}{j^2\left[r_{\rm H}^+-f'(r_{\rm H}^+)\right]^2}\right\}.
\end{eqnarray}
It is obvious that the necessary condition for $Z_{lm}$ to be an amplification factor is  $Z_{lm}>0$, which means 
\begin{equation}\label{Super_Gene_con}
m\Omega>\omega>\mu.
\end{equation}
Comparing Eq.~(\ref{Super_Gene_con}) with the superradiance instability condition given by Eq.~(\ref{superradinscon}), we can see that the superradiance instability of bound states is closely related to the scattering progress of free states.
In conclusion, for a scalar particle with a small mass, the superradiance instability of bound states will occur if the scattering process of free states happens, 
and vice versa.

\section{Two specific models}
\label{sec: model_analysis}
\subsection{Background}\label{{sec: model_analysis_Back}}
We give the background of two regular black hole models, the rotating Hayward and Bardeen black holes, by  clarifying the regularity in spacetime, which is helpful for us to present the significance of the  results obtained in Sec.~\ref{sec: model_analysis_Hay} and Sec.~\ref{sec: model_analysis_Bar}.

The static and spherically symmetric Hayward black hole is constructed with a source consisting of high-density matter, which makes its spacetime flat at the center, that is, the asymptotic behavior of the shape function in Eq.~(\ref{sphesymmmetr}) takes the form,
\begin{equation}
F_{\rm Hay}(r)\sim 1-\frac{r^2}{L^2} \quad {\rm as} \quad r\rightarrow 0,\label{asybehhayrgotozero}
\end{equation}
where the regularization parameter $L$ is a convenient encoding of the central energy density $3/(8\pi L^2)$. 
Thus, the shape function satisfying the above asymptotic condition reads~\cite{A9},
\begin{equation}
F_{\rm Hay}(r)=1-\frac{2Mr^2}{r^3+2ML^2}.\label{stsphsymmhay}
\end{equation}

By using the modified NJA, we can derive the line element of the rotating and axially symmetric Hayward black hole, see Eq.~\eqref{rotating_metric} together with Eq.~\eqref{stsphsymmhay}.  We can verify that the Ricci scalar $R$ and Kretschmann scalar $K$ of the rotating Hayward black hole are finite everywhere since this black hole has de Sitter-like behavior near $r=0$, see Eq.~\eqref{asybehrgotozerorbh} and Eq.~\eqref{asybehhayrgotozero}.
Therefore, the rotating Hayward black hole is still regular black hole.
Because the source of the  static Hayward black hole consists of high-density matter, the rotating counterpart corresponds to a fluid rotating about $z$ axis.
Let us give a brief explanation. At first, we construct the dual basis of orthonormal tetrad,
\begin{equation}
e^\mu_t=\frac{(r^2+a^2,0,0,a)}{\sqrt{\rho^2\Delta}},\quad e^\mu_r=\frac{\sqrt\Delta(0,1,0,0)}{\sqrt{\rho^2}},\quad
e^\mu_\theta=\frac{(0,0,1,0)}{\sqrt{\rho^2}},\quad e^\mu_\phi=-\frac{(\sin^2\theta,0,0,1)}{\sqrt{\rho^2}\sin\theta}.
\end{equation}
Then we obtain the expression of $T^{\mu\nu}$ by using the Einstein field equations, $G_{\mu\nu}=T_{\mu\nu}$,
\begin{equation}
T^{\mu\nu}=\epsilon e^\mu_t e^\nu_t+p_r e^\mu_r e^\nu_r+p_\theta e^\mu_\theta e^\nu_\theta+p_\phi e^\mu_\phi e^\nu_\phi,
\end{equation}
where $\epsilon$ is energy density and $(p_r,p_\theta,p_\phi)$ are components of pressure.
We further give the relations between these quantities and $f(r)$,
\begin{equation}
\epsilon=-p_r=\frac{2(rf'-f)}{\rho^4},\qquad p_\theta=p_\phi=-p_r-\frac{f''}{\rho^2},
\end{equation}
where $f$ and $\rho$ are shown in Eq.~\eqref{rho} and Eq.~\eqref{fandF}.
Due to the above relations, 
we deduce that the source associated with the rotating Hayward black hole is an imperfect fluid.

Different from the static and spherically symmetric Hayward black hole, the static and spherically symmetric Bardeen black hole was considered~\cite{A8} as a magnetic solution to the Einstein field equations coupled to nonlinear electrodynamics, where the action reads
\begin{equation}
I=\int{\rm d}x^4\left[\frac{1}{16\pi}R-\frac{1}{4\pi}\mathscr{L}(\mathscr{F})\right],
\end{equation}
$\mathscr{L}$ is the Lagrangian density of electromagnetic fields, $\mathscr{F}\equiv\frac{1}{4}F_{\mu\nu}F^{\mu\nu}$, and $F_{\mu\nu}=2\nabla_{[\mu}A_{\nu]}$ is the electromagnetic strength.
More specifically, the source of the static Bardeen black hole is determined by the following $\mathscr{L}$:
\begin{equation}
	\mathscr{L}=\frac{3M}{|g|^3}\left(\frac{\sqrt{2g^2\mathscr{F}}}{1+\sqrt{2g^2\mathscr{F}}}\right)^{5/2},
\end{equation}
where $g$ is the monopole charge of a self-graving magnetic field described by nonlinear electrodynamics and the gauge field takes the form,
\begin{equation}
	A_\mu=g\cos\theta\delta^\varphi_\mu.
\end{equation}
As shown in Refs.~\cite{A7,A8}, the static and spherically symmetric Bardeen black hole solution takes the form of Eq.~(\ref{sphesymmmetr}) and the corresponding shape function reads 
\begin{equation}
F_{\rm B}(r)=1-\frac{2Mr^2}{(r^2+g^2)^{3/2}},\label{stasphsymmbard}
\end{equation}
where $g$ acts as the regularization parameter and the asymptotic behavior of $F_{\rm B}(r)$ has the de Sitter-like form,
\begin{equation}
	F_{\rm B}(r)\sim1-\frac{2M}{g^3}r^2 \quad {\rm as} \quad r\rightarrow 0.
\end{equation}   

In terms of the modified NJA, the gauge field changes~\cite{A19} as follows,
\begin{equation}
A_\mu=-\frac{ga\cos\theta}{\rho^2}\delta^t_\mu+\frac{g(r^2+a^2)\cos\theta}{\rho^2}\delta^\varphi_\mu,
\end{equation}
and correspondingly $\mathscr{L}$ becomes
\begin{eqnarray}
\mathscr{L}&=&\frac{r^2}{2\rho^8}\left\{\left[15a^4-8a^2r^2+8r^4+4a^2(5a^2-2r^2)\cos2\theta+5a^4\cos4\theta\right]\left(\frac{f}{r}\right)'\right.\nonumber \\
& &\left.+16a^2r\rho^2\cos^2\theta \left(\frac{f}{r}\right)''\right\},
\end{eqnarray}
where the prime represents the derivative with respect to $r$. Correspondingly, the line element of the rotating and axially symmetric Bardeen black hole can be obtained, see Eq.~\eqref{rotating_metric} together with Eq.~\eqref{stasphsymmbard}.
It is easy to verify that the Ricci scalar $R$ and Kretschmann scalar $K$ of the rotating Bardeen black hole are finite everywhere.

Although the physical mechanisms of the rotating Hayward and Bardeen black holes are not the same, their eliminations of singularity are the same from the perspective of metrics, i.e., the introduction of an extra regularization parameter into their metrics. 
When the regularization parameters go to zero, the singularity will reappear, and the two black holes will return to the Kerr black hole.
Therefore, the regularization parameters are the key factors affecting the properties of spacetime. 
We note that the asymptotic behaviors of the rotating Hayward and Bardeen black holes at infinity are consistent with that of the Kerr black hole, and that the asymptotic behaviors of the two black holes at the center are completely different from that of the Kerr black hole, where 
the closer to the center the position is, the greater differences the metrics of spacetime present. 
The limitation of observations is the outer horizon of black holes, which indicates that the influence of regularization parameters can be best reflected at the outer horizon. 
However, we cannot directly observe metrics, but can understand the properties of spacetime from the interaction between spacetime and external particles. 
As we have shown in Eq.~\eqref{eq:Omega_I} and Eq.~\eqref{Zlm}, the superradiance instability and amplification factors of scalar fields are closely related to the properties of outer horizons. 
Therefore, we shall better understand the properties of regular black holes when we study the superradiance phenomena affected by regularization parameters.
Next, we analyze in detail how the regularization parameters affect the superradiance phenomenon of scalar particles in the rotating Hayward and Bardeen black holes.

\subsection{Hayward black holes}\label{sec: model_analysis_Hay}

We derive the equation that governs the event horizons of the rotating Hayward black hole,
\begin{equation}
\label{Hayward_Horizon}
r_{\rm H}^5-2Mr_{\rm H}^4+a^2r_{\rm H}^3+2ML^2r_{\rm H}^2+2ML^2a^2=0.
\end{equation}
This equation has two real and positive roots, $r_{\rm H}^-$ and $r_{\rm H}^+$, which correspond to the inner and outer horizons, respectively.
When the inner and outer horizons are equal, the black hole reaches its extreme configuration. The derivative of Eq.~\eqref{Hayward_Horizon} with respect to $r_{\rm H}$ gives rise to the equation that governs $r_{\rm H}^{\rm ext}$, the horizon of the extreme configuration,
\begin{equation}
5\left(r_{\rm H}^{\rm ext}\right)^3-8M\left(r_{\rm H}^{\rm ext}\right)^2+3a^2r_{\rm H}^{\rm ext}+4ML^2=0.\label{derhorequ}
\end{equation}

\begin{figure}[htbp]
	\centering
	\begin{minipage}[t]{0.5\linewidth}
		\centering
		\includegraphics[width=1\linewidth]{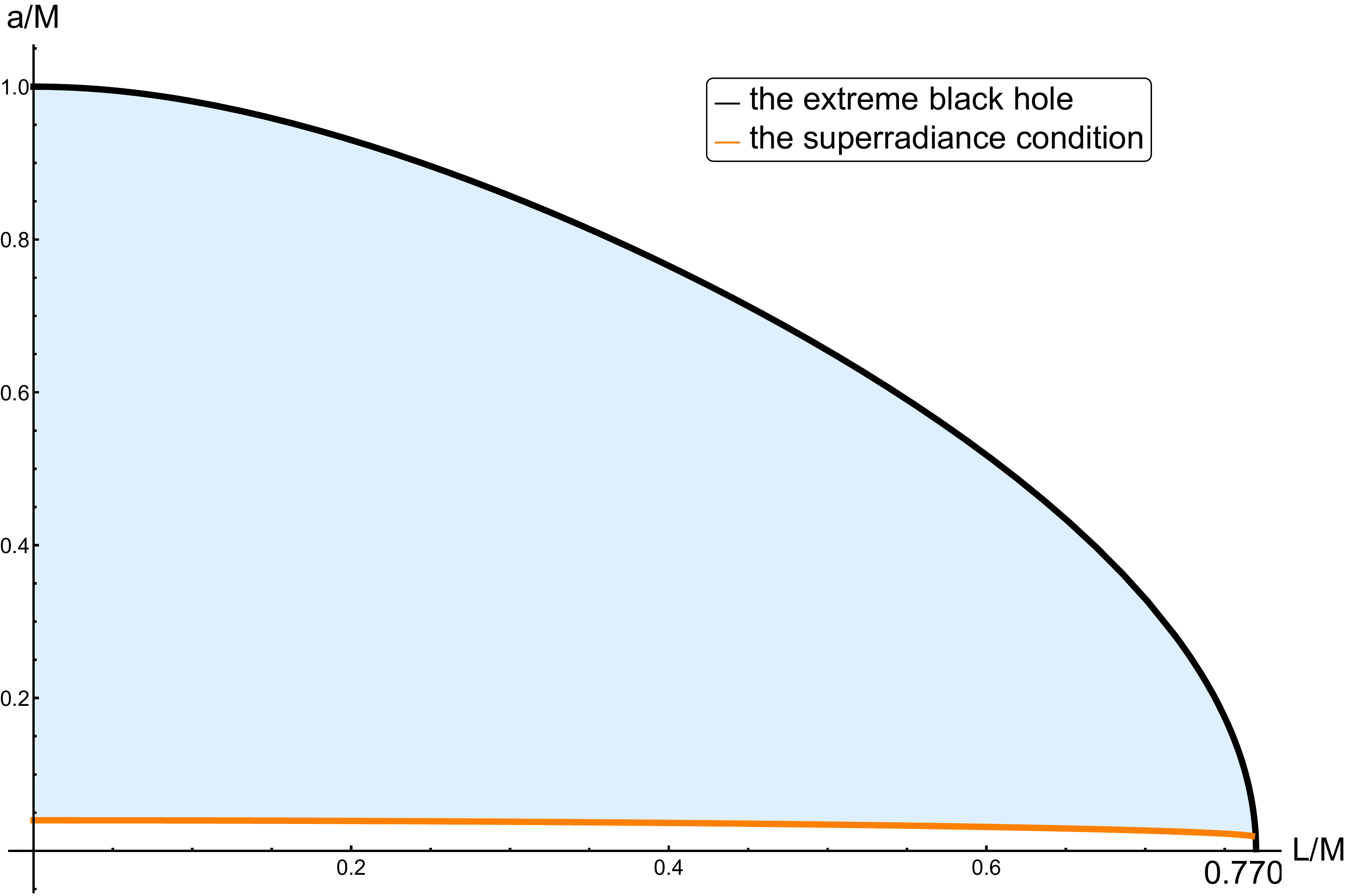}
	\end{minipage}
	\caption{The relationship between the regularization parameter $L$ and the rotation parameter $a$ in the rotating Hayward black holes, where the abscissa is $L/M$, the ordinate is $a/M$, and $M$ is the mass of the black hole. The black curve (depicted by Eqs.~(\ref{Hayward_Horizon}) and (\ref{derhorequ})) represents the extreme black hole, the orange curve (depicted by Eqs.~(\ref{Hayward_Horizon}) and (\ref{case1orglin})) is the lower bound for the occurrence of superradiance, and the blue area bounded by the two curves is the region of black hole parameters in which the superradiance can occur.}
	\label{fig:Hayward_H}
\end{figure}

In Fig.~\ref{fig:Hayward_H}, the black curve  is described by Eqs.~(\ref{Hayward_Horizon}) and (\ref{derhorequ}) and the orange one by Eq.~(\ref{superradinscon}) together with the threshold of the occurrence of superradiance, and therefore the blue area surrounded by the two curves represents the parameter region, $(L/M, a/M)$, in which the superradiance phenomenon can happen.
In order to simplify the following analysis, we make the parameters dimensionless by using the black hole mass $M$, such as $L/M$, $a/M$, etc., where
$M$ is fixed by default since we focus on the effects of the rotation parameter and the others.
From the figure we can see that the regularization parameter $L$ decreases as the rotation parameter $a$ increases, and $a$ also decreases as $L$ increases.
When $L$ approaches to zero, $a$ reaches its maximum, $M$;
while $a$ approaches to zero, $L$ reaches its maximum, $0.770 M$.

Now we discuss in detail how the regularization parameter $L$, together with the rotation parameter $a$, affects the evolution of  amplitude of massive scalar particles in the quasi-bound state.
Since we are concerned with the effects of the parameters $a$ and $L$, we fix the particle mass to be $M\mu=0.01$ in the following discussion.
Because $n$ affects only the constant term in Eq.~\eqref{eq:Omega_I}, we take the fundamental order $n=1$ without loss of generality.
According to Eq.~\eqref{Super_Gene_con}, the critical condition that the superradiance phenomenon occurs is
\begin{equation}
\frac{a}{\left(r_{\rm H}^{+}\right)^2+a^2}=\frac{0.01}{M},\label{case1orglin}
\end{equation}
in which $r_{\rm H}^{+}$ is given by Eq.~(\ref{Hayward_Horizon}). That is, Eq.~(\ref{Hayward_Horizon}) and Eq.~(\ref{case1orglin}) depict the orange curve in Fig.~\ref{fig:Hayward_H}. Therefore, the blue area surrounded by the black and orange curves is the parameter region in which the superradiance can occur.
Note that the intersection of the black and orange curves is $(0.769, 0.018)$, which indicates that there will be no superradiance effects when $L>0.769M$ or $a<0.018M$, that is, the superradiance instability of quasi-bound particles and the superradiance scattering of free particles will not exist.

\begin{figure}[htbp]
	\centering
	\subfigure[]{
		\begin{minipage}[t]{0.4\linewidth}
			\centering
			\includegraphics[width=1\linewidth]{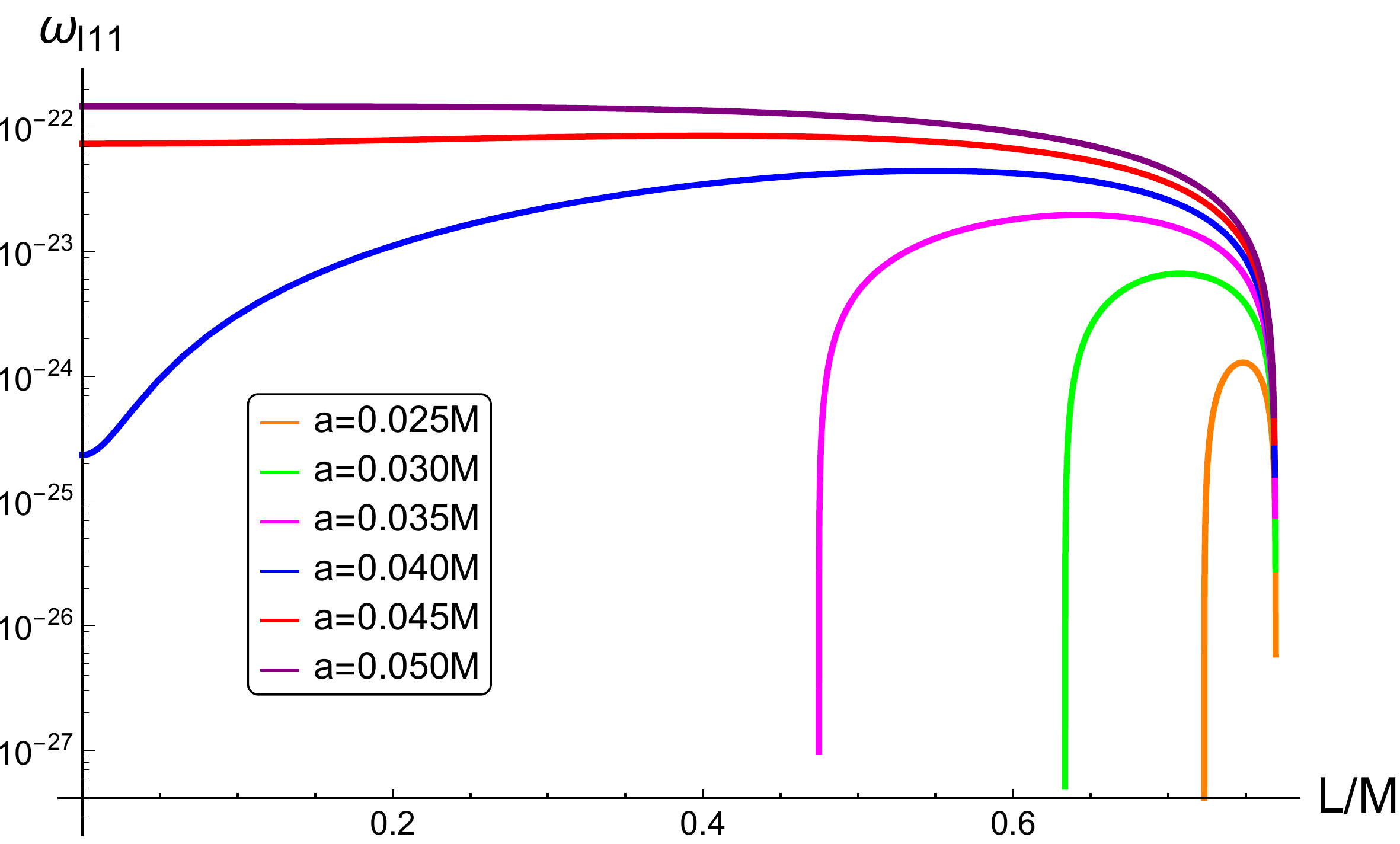}
		\end{minipage}
	}
	\subfigure[]{
		\begin{minipage}[t]{0.4\linewidth}
			\centering
			\includegraphics[width=1\linewidth]{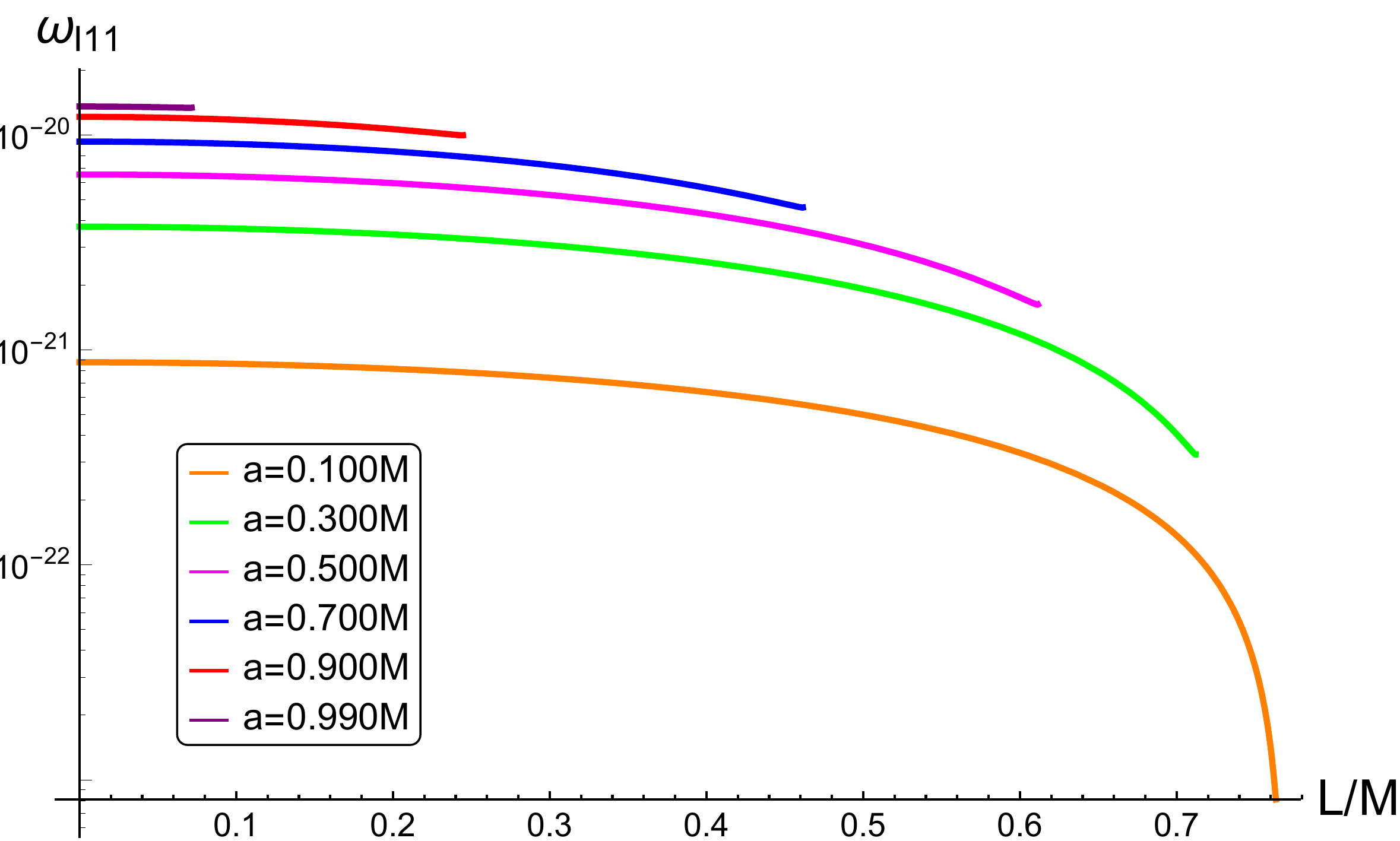}
		\end{minipage}
	}
	\caption{The relationship between the imaginary part of  frequency $\omega_{\rm I}$ and the regularization parameter $L$ for quasi-bound state particles at the leading multipole number ($l=1$, $m=1$, $n=1$) in the rotating Hayward black holes when the rotation parameter $a$ takes various values, where the  diagram (a) is depicted for the case, $0.018M<a\le 0.050M$ and the  diagram (b) for the case, $0.050M<a<1.000M$.}
	\label{fig:Hay_omega_11_a}
\end{figure}

In Fig.~\ref{fig:Hay_omega_11_a} we show the relationship between the imaginary part of  frequency $\omega_{\rm I}$ and the regularization parameter $L$ for quasi-bound state particles at the leading multipole number\footnote{When the fundamental order, $n=1$, is taken, the leading multipole means $l=1$ together with $m=1$ because the superradiance occurs only at $m=1$.} ($l=1$, $m=1$, $n=1$) when the rotation parameter $a$ takes various values. We note that $\omega_{\rm I}$ is positively correlated with the growth rate of amplitude of quasi-bound state particles.
It can be seen that the effect of $L$ on $\omega_{\rm I}$ is related to the value of $a$.
When $a<0.050M$, $\omega_{\rm I}$ increases at first and then decreases with an increase of $L$.
This indicates that the growth rate of superradiance instability increases at first and then decreases when $L$ stays in the low rotation situation,  $a<0.050M$.
Moreover, if $a>0.050M$, $\omega_{\rm I}$ monotonically decreases with an increase of $L$, which indicates that the growth rate of superradiance instability decreases with an increase of $L$.
However, $\omega_{\rm I}$ always increases monotonically with an increase of $a$ for a  fixed $L$.
Thus, an increase of $a$ always makes the superradiance instability grow very fast.
Our summary is that the relationship between the growth rate of superradiance instability and $L$ depends on the rotation parameter $a$, but the relationship between the growth rate of superradiance instability and $a$ is always positively correlated for any fixed $L$.

Next, we focus on how the regularization parameter $L$, together with the rotation parameter $a$, affects the superradiance amplification of massive scalar particles in the scattering process.
Here we fix the particle mass to be $M\mu=0.01$ again, and discuss the behavior of particles at the leading multipole number ($l=1$, $m=1$, $n=1$) in the background of a rotating Hayward black hole.

Under the condition of fixed regularization parameter $L$ and rotation parameter $a$, the superradiance amplification has a peak when the particle frequency $\omega$ changes, which represents the maximum energy extraction efficiency. We note that such an efficiency is mainly affected by $L$ and $a$.
After analyzing the change of peaks with respect to $L$ and $a$, respectively, we find that the change of peaks with respect to $L$ is different in various intervals of $a$. 
Here we call the change in each interval of $a$ ``one mode". We summarize four modes in the parameter region (see Fig.~\ref{fig:Hayward_H}) where the superradiance effect can occur.
The four modes are described in detail as follows.
\begin{enumerate}
\item The mode in the interval of $0.018M<a<0.100M$

In this case, the peak value of superradiance magnification increases at first and then decreases with an increase of $L$, and finally has a very small climb when the rotating Hayward black hole approaches to its extreme configuration.
This shows that when $L$ increases, the energy extraction efficiency of massive scalar particles scattered by the rotating Hayward black hole is initially greater but finally lower than that in the singular Kerr black hole ($L=0$).
In Fig.~\ref{fig:Hay_a=0.05M},  $a=0.050M$ is taken as an example, which is covered by the present interval, $0.018M<a<0.100M$.
When $L$ is in the range of $0<L\le 0.655M$, the peak value of magnification rises if $L$ is increasing, and it reaches the maximum, $9.124\times 10^{-10}$, at $L=0.655M$.
When $L$ is in the range of $0.655M<L\le 0.768176M$,\footnote{Here we use a higher accuracy because the upper limit is so close to the extreme black hole at $L=0.768177M$, and the previous accuracy cannot describe the difference between the near-extreme case and the exetreme case.} the peak value of magnification falls if $L$ is increasing, and it reaches the minimum, $8.836\times 10^{-12}$, at $L=0.768176M$, where the rotating Hayward black hole is very close to its extreme configuration.
When $L$ is in the range of $0.768176M<L<0.768177M$, the rotating Hayward black hole evolves to its final stage --- the extreme configuration, and the peak rises with a slight uptick as the black hole gets closer to its extreme case.

\begin{figure}[htbp]
	\centering
	\subfigure[]{
		\begin{minipage}[t]{0.4\linewidth}
			\centering
			\includegraphics[width=1\linewidth]{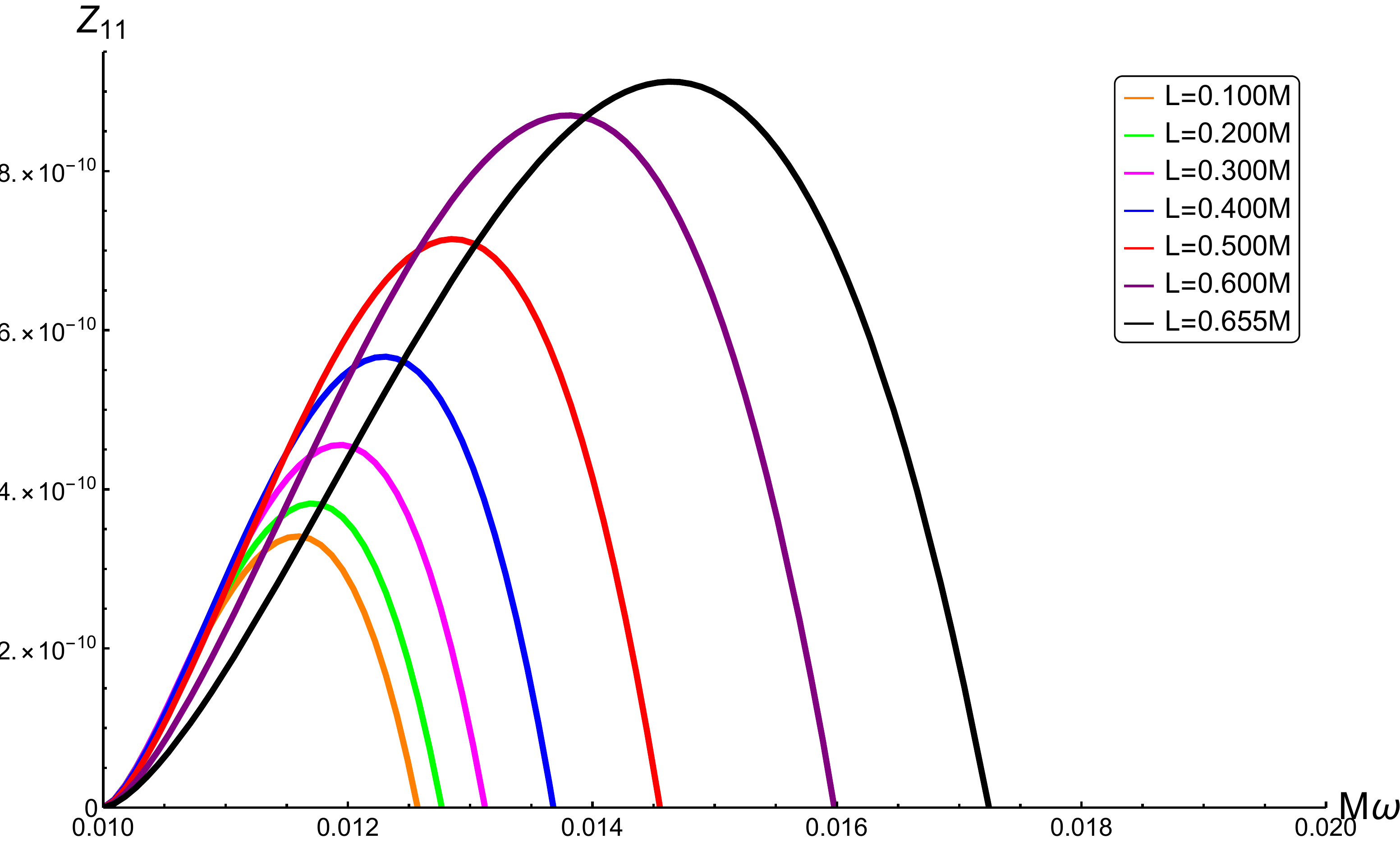}
		\end{minipage}
	}
	\subfigure[]{
		\begin{minipage}[t]{0.4\linewidth}
			\centering
			\includegraphics[width=1\linewidth]{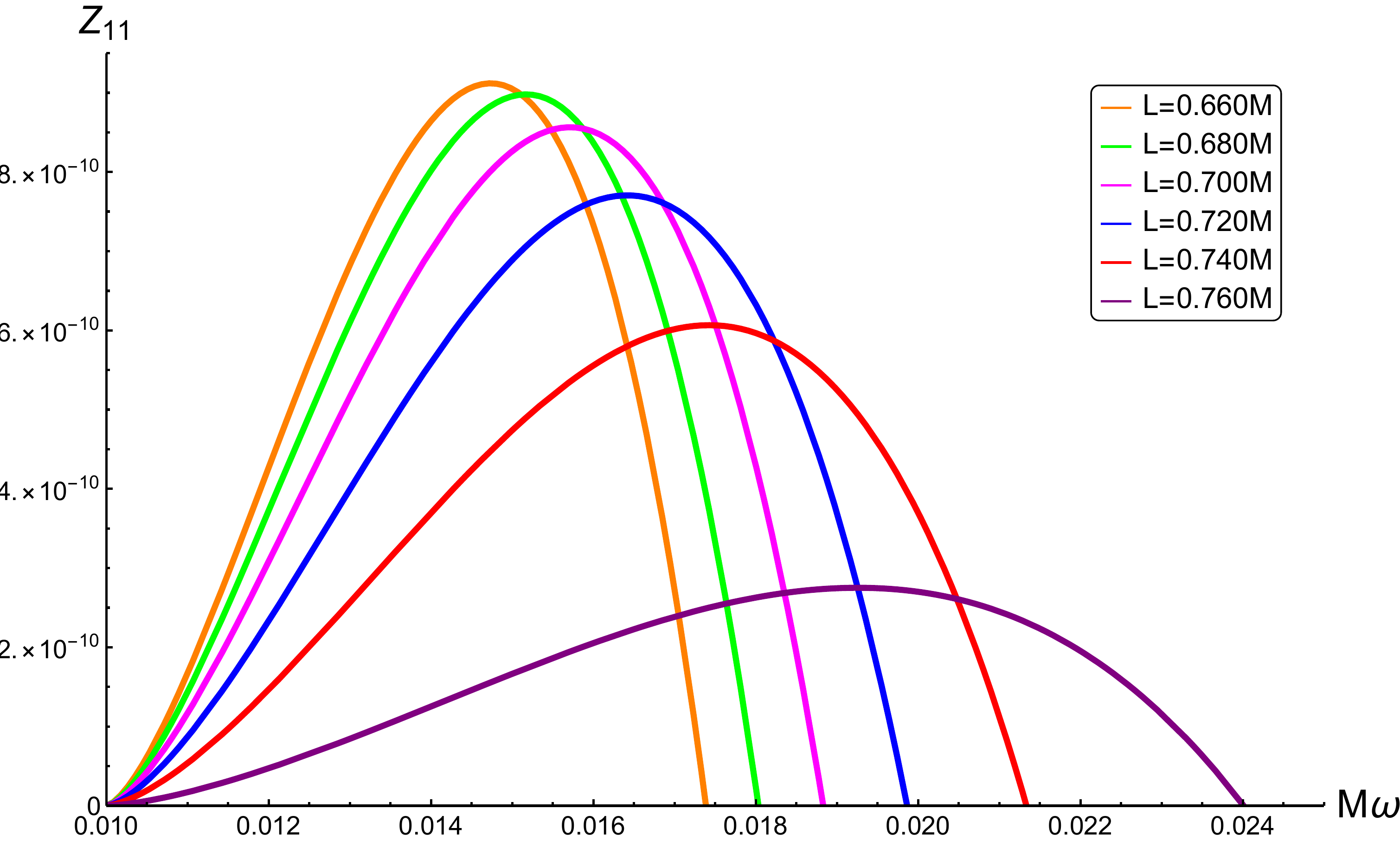}
		\end{minipage}
	}
	\vspace{-3mm}
	\subfigure[]{
		\begin{minipage}[t]{0.4\linewidth}
			\centering
			\includegraphics[width=1\linewidth]{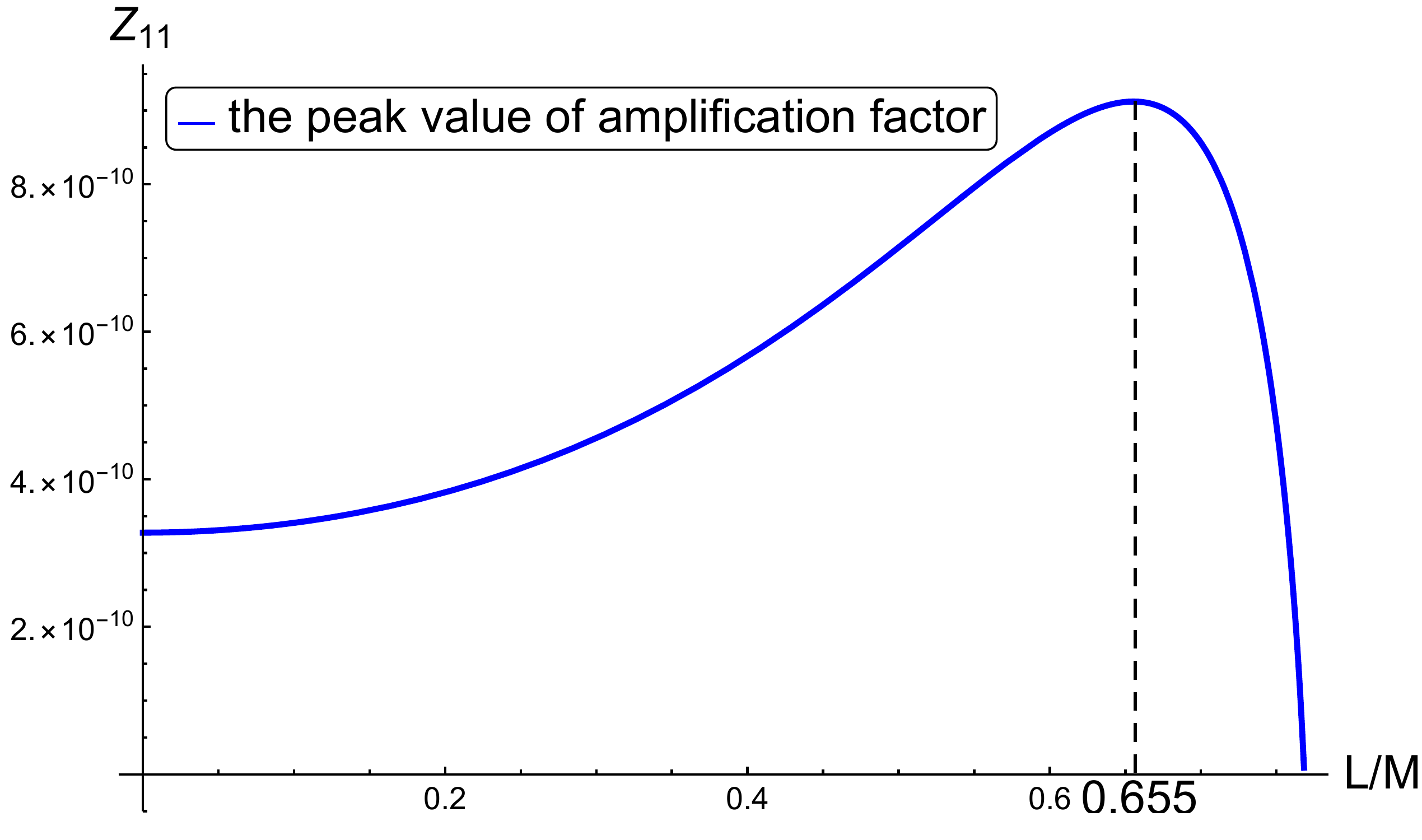}
		\end{minipage}
	}
	\subfigure[]{
		\begin{minipage}[t]{0.4\linewidth}
			\centering
			\includegraphics[width=1\linewidth]{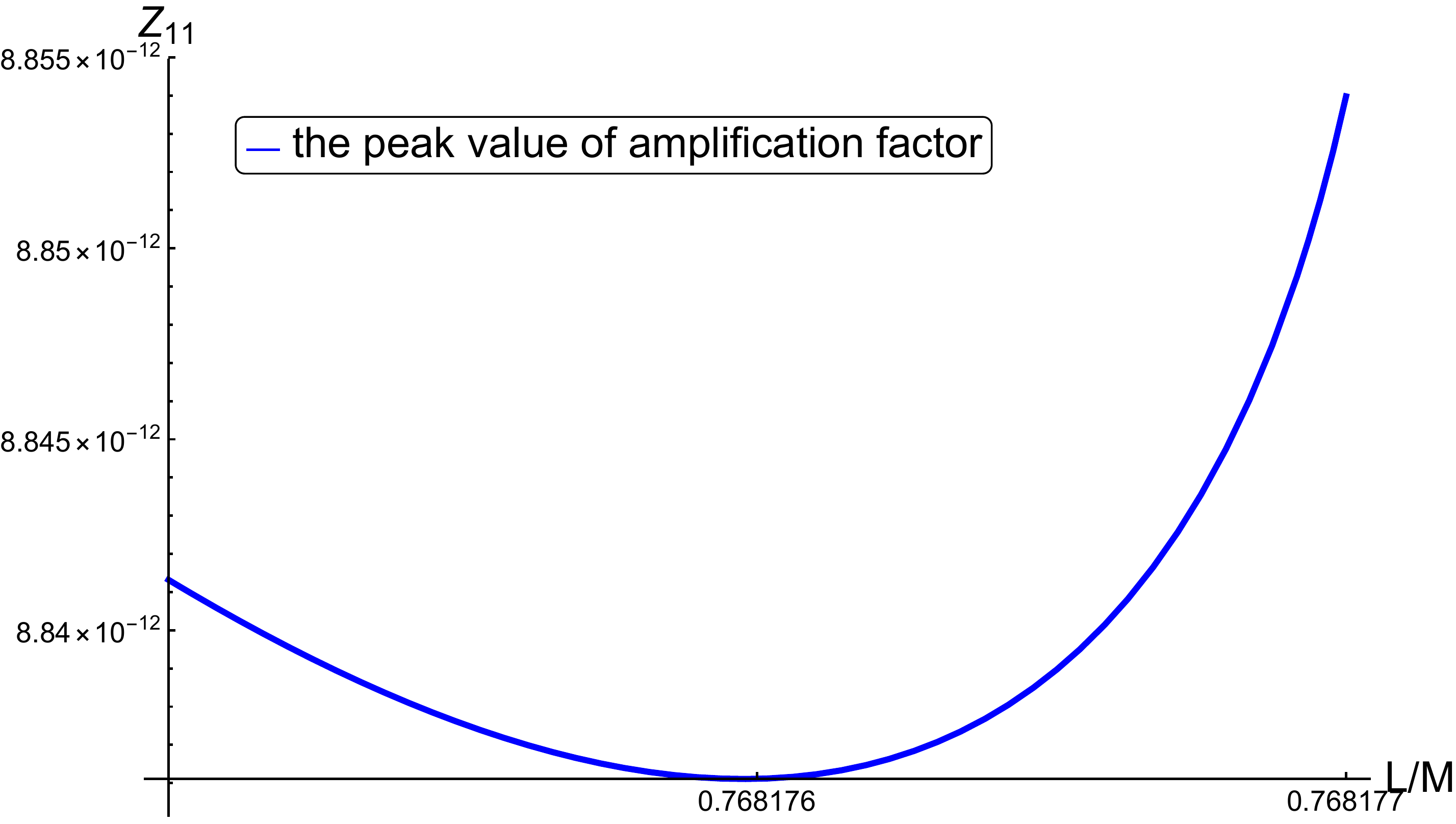}
		\end{minipage}
	}
	\caption{The amplification factor of massive scalar particles scattered by the rotating Hayward black hole is computed by Eq.~\eqref{Zlm}, where the parameters are set to be: $M\mu=0.010$, $a=0.050M$, $(l,m)=(1,1)$,  and a varying $L$. In diagram (a), $0<L\le 0.655M$, and in diagram (b), $0.655M<L\le 0.768176M$. Moreover, diagram (c) depicts the overall variation of the peak of amplification factor with respect to $L$, and diagram (d) describes the behavior of the peak when the rotating Hayward black hole is going to its extreme configuration, $0.768175M\le L < 0.768177M$.}
	\label{fig:Hay_a=0.05M}
\end{figure}

\item The mode in the interval of $0.100M<a<0.859M$

 At first, the peak of superradiance  amplification factor falls with an increase of $L$, and it reaches its lowest value when the rotating Hayward black hole is going to its near-extreme configuration. Then, the peak rises rapidly in the process that the black hole is closer to its extreme configuration, but finally it does not exceed the peak value associated with the Kerr black hole ($L=0$).
It shows that the efficiency is the maximum for the massive scalar particles to extract energy from the Kerr black hole ($L=0$) in this interval of $a$, and that the introduction of a regularization parameter (a non-vanishing $L$) will reduce such an efficiency.
In Fig.~\ref{fig:Hay_a_fixed}, we show the image of the  peak of amplification factor as a function of $L$ for different values of $a$.
As the rotation parameter increases, see the figure from diagram (a) to diagram (d), the peak is continuously rising in the process that the rotating Hayward black hole is approaching to its extreme configuration.
In particular, the peak's order of magnitude grows with an increase of the rotation parameter. 
When the change of the order of magnitude caused by $a$ is compared with that by $L$, we can see that the peak of amplification factor is affected mainly by $a$ rather than by $L$. In Fig.~\ref{fig:Hay_a=0.8M},
$a=0.800M$ is taken as an example, which belongs to the present mode, $0.100M<a<0.859M$.
When $L$ is in the range of $0<L\le 0.345M$, the peak value of magnification falls if $L$ is increasing, and it reaches the minimum, $1.532\times10^{-4}$, at $L=0.345M$.
Moreover, the peak value starts to rise again in the process that the rotating Hayward black hole is approaching to its extreme configuration, but it is still lower than that associated with the Kerr black hole even if the rotating Hayward black hole arrives at its extreme. As a result, the peak value related to the Kerr black hole is the largest in the second mode of $a$-parameter intervals.

\begin{figure}[htbp]
	\centering
	\subfigure[$a=0.400M$]{
		\begin{minipage}[t]{0.4\linewidth}
			\centering
			\includegraphics[width=1\linewidth]{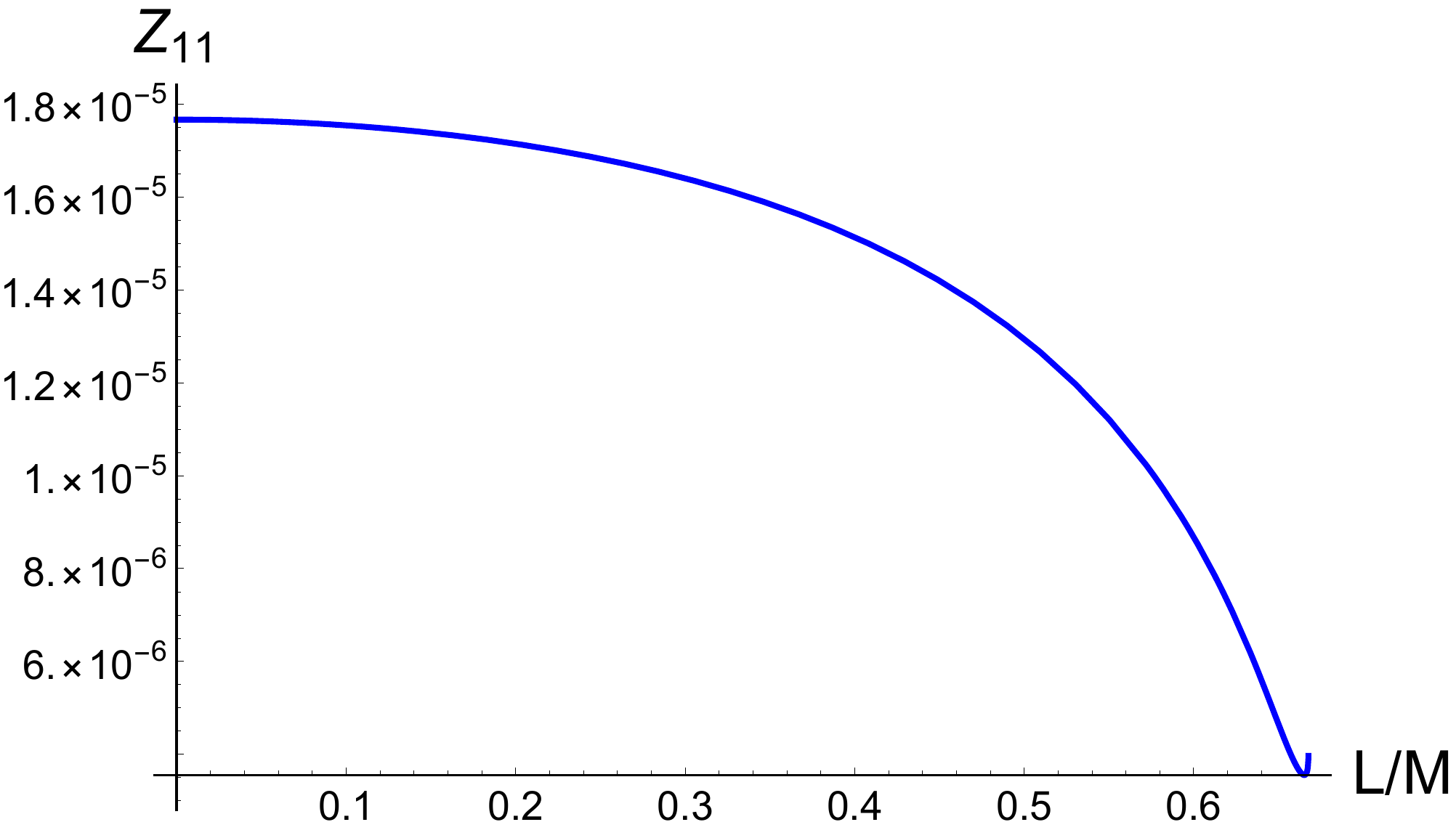}
		\end{minipage}
	}
	\subfigure[$a=0.500M$]{
		\begin{minipage}[t]{0.4\linewidth}
			\centering
			\includegraphics[width=1\linewidth]{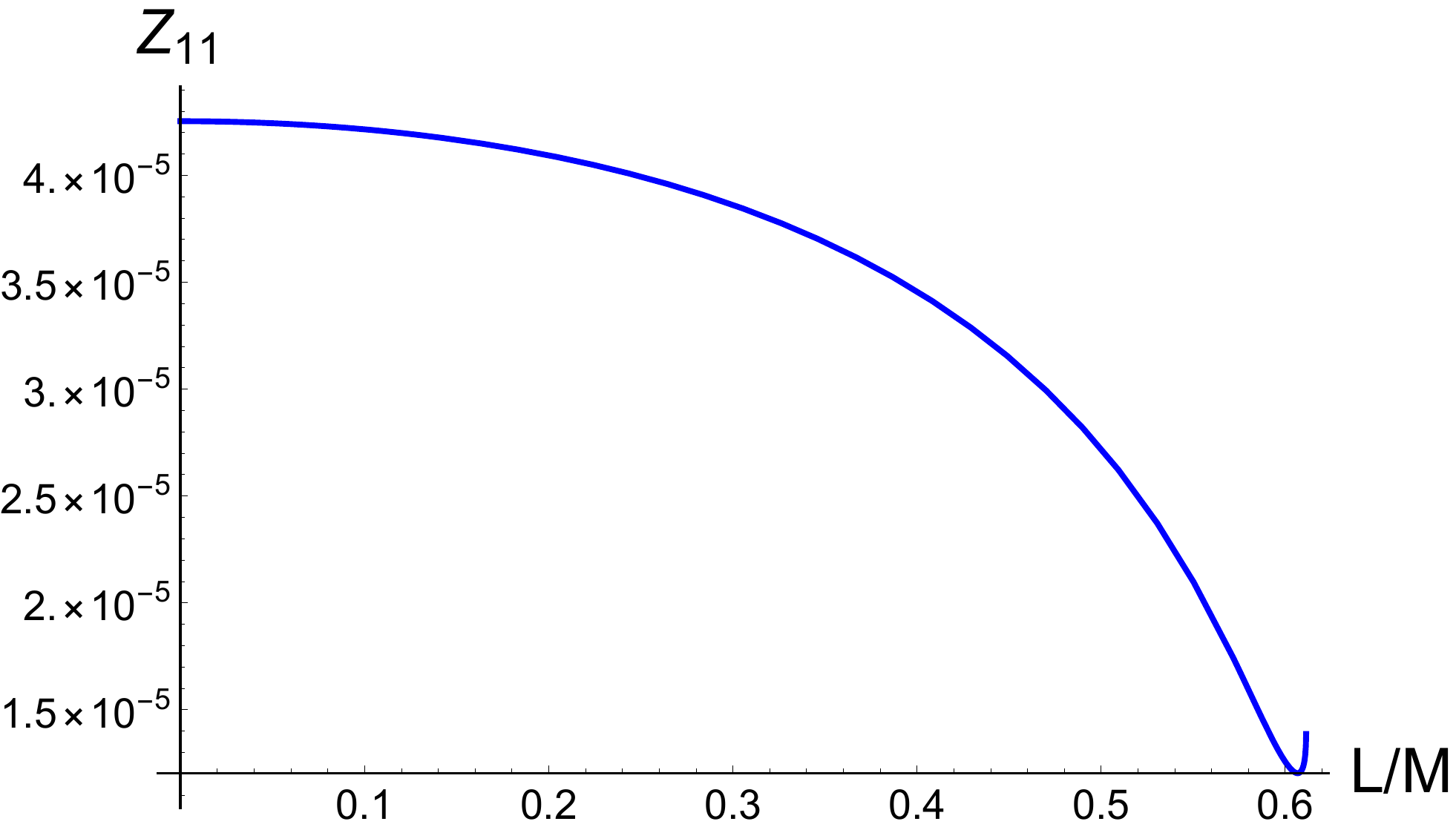}
		\end{minipage}
	}
	\subfigure[$a=0.600M$]{
		\begin{minipage}[t]{0.4\linewidth}
			\centering
			\includegraphics[width=1\linewidth]{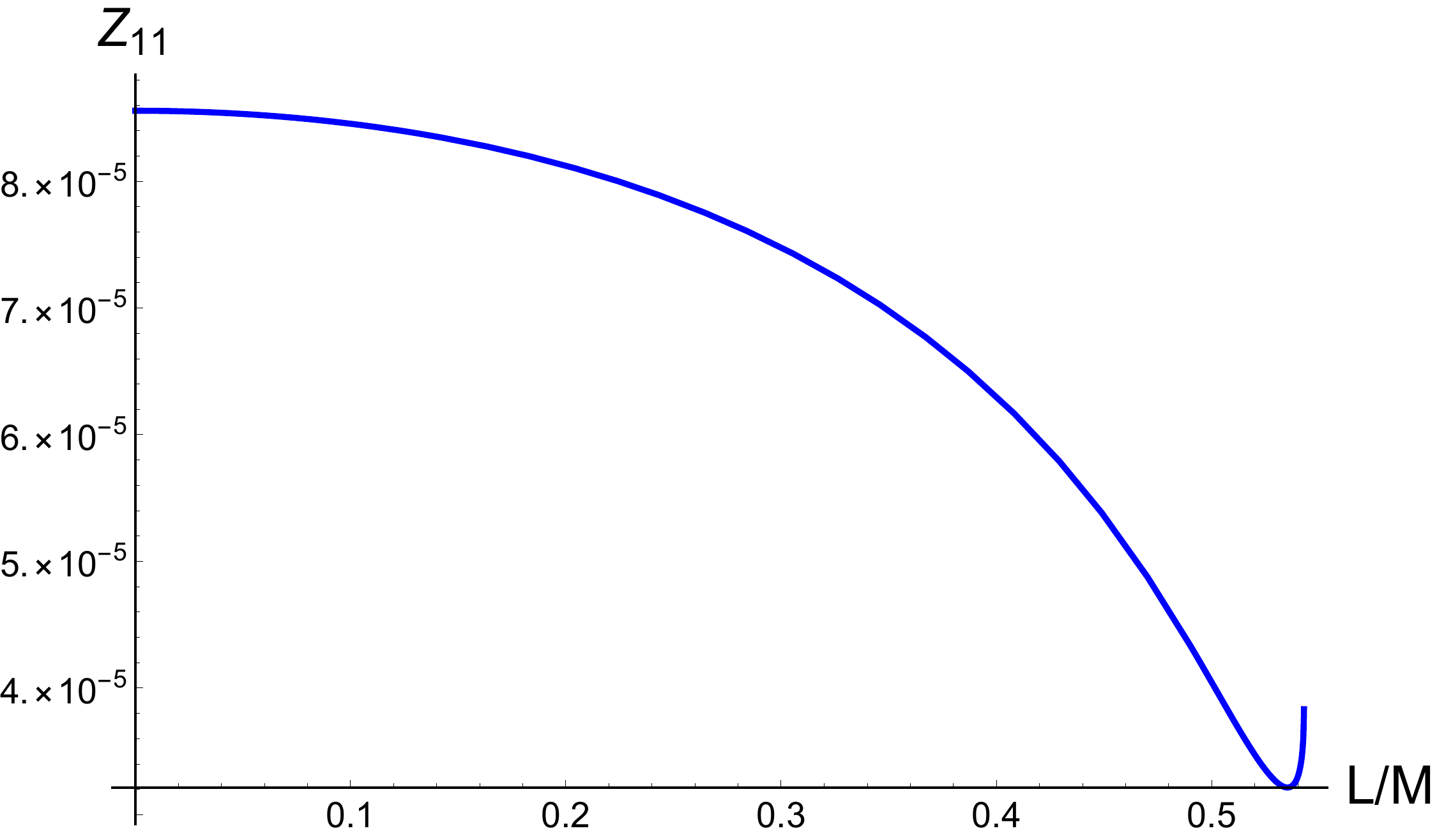}
		\end{minipage}
	}
	\vspace{-3mm}
	\subfigure[$a=0.700M$]{
		\begin{minipage}[t]{0.4\linewidth}
			\centering
			\includegraphics[width=1\linewidth]{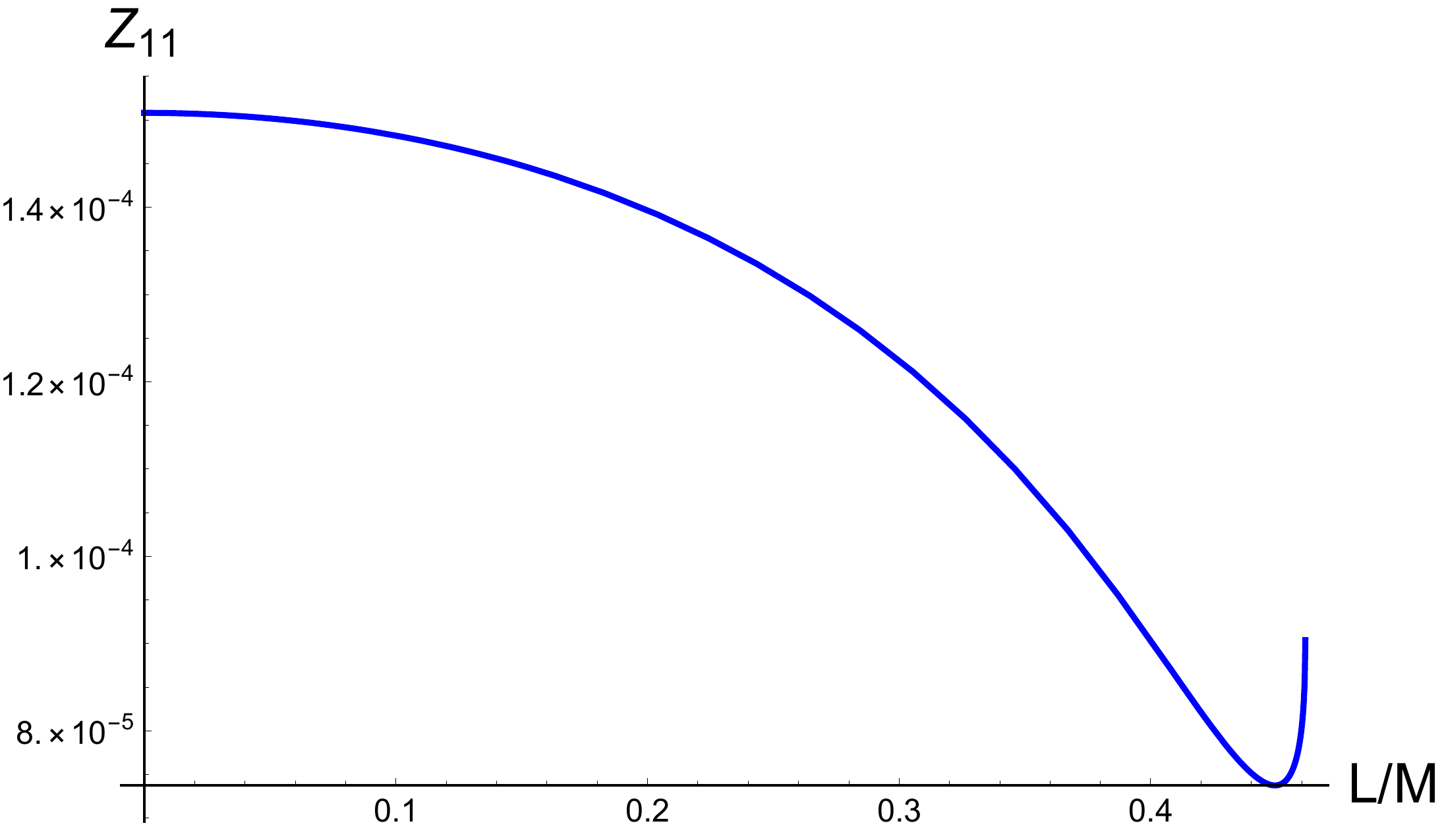}
		\end{minipage}
	}
	\caption{Relationship between the peak of amplification factor and the regularization parameter $L$ for the massive scalar particles scattered by the rotating Hayward black hole, where the parameters are set to be: $M\mu=0.010$, $(l,m)=(1,1)$, and a varying rotation parameter $a$.}
	\label{fig:Hay_a_fixed}
\end{figure}

\begin{figure}[htbp]
\centering
\subfigure[]{
    \begin{minipage}[t]{0.4\linewidth}
    \centering
    \includegraphics[width=1\linewidth]{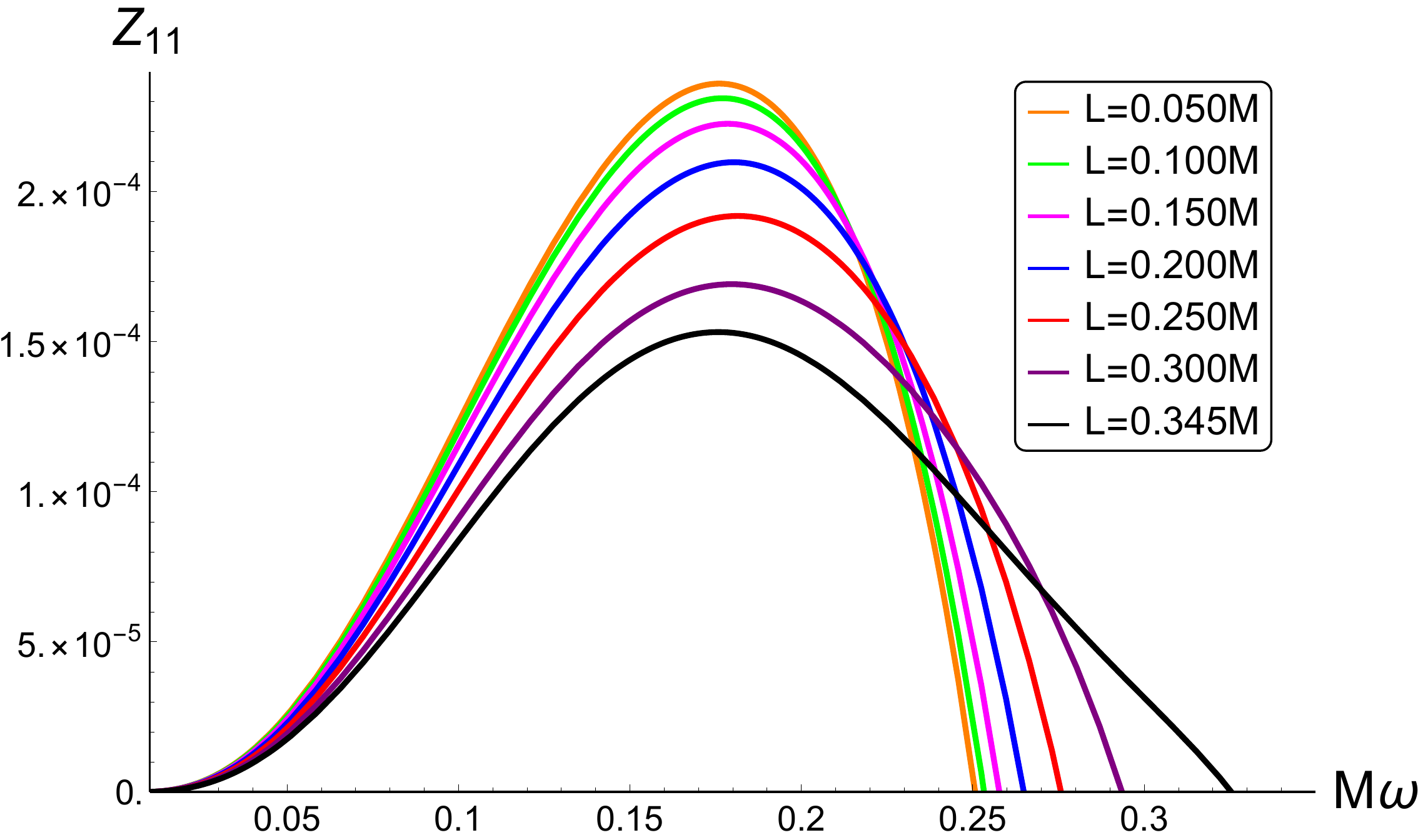}
    \end{minipage}
}
\subfigure[]{
    \begin{minipage}[t]{0.4\linewidth}
    \centering
    \includegraphics[width=1\linewidth]{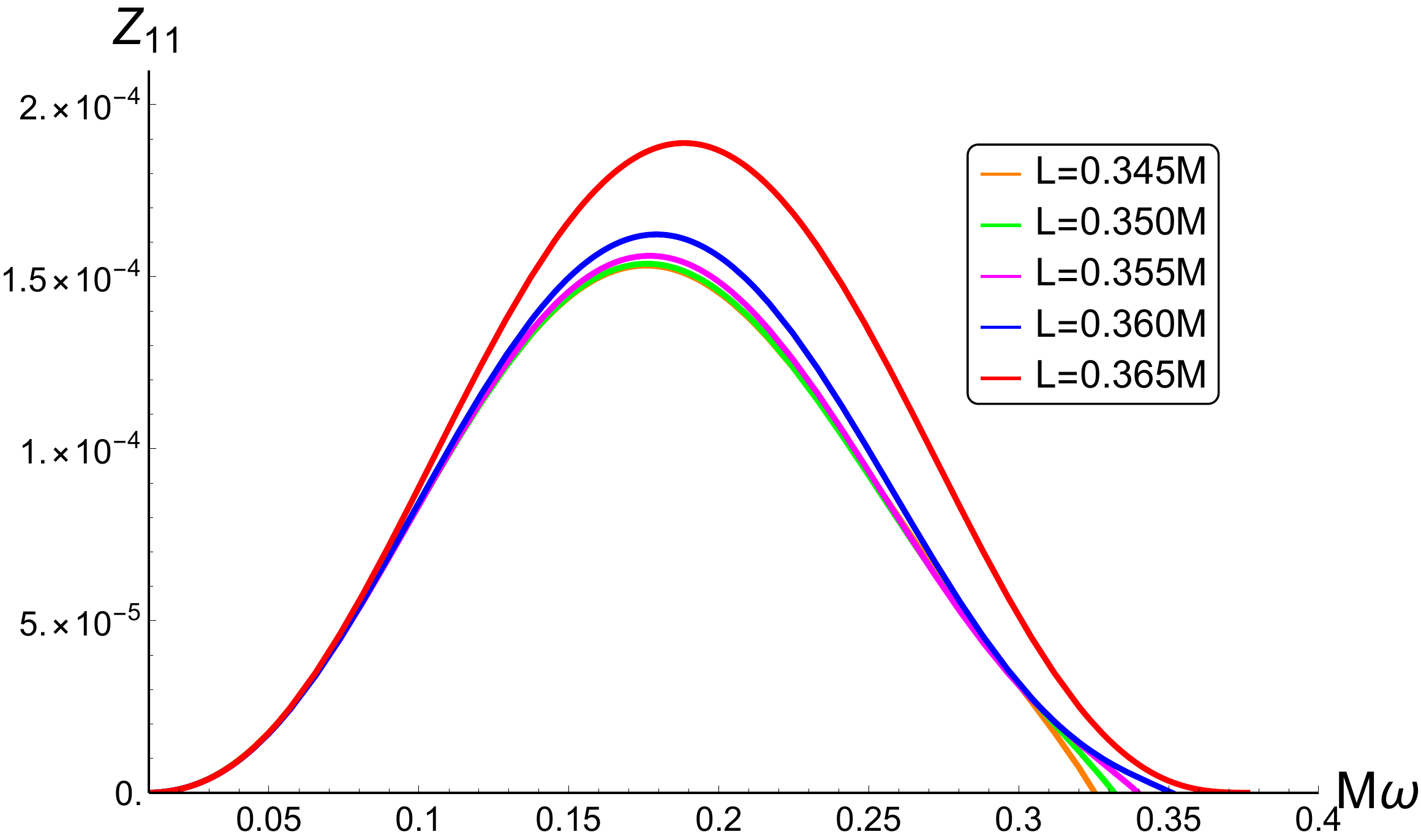}
    \end{minipage}
}
\vspace{-3mm}
\subfigure[]{
    \begin{minipage}[t]{0.4\linewidth}
    \centering
    \includegraphics[width=1\linewidth]{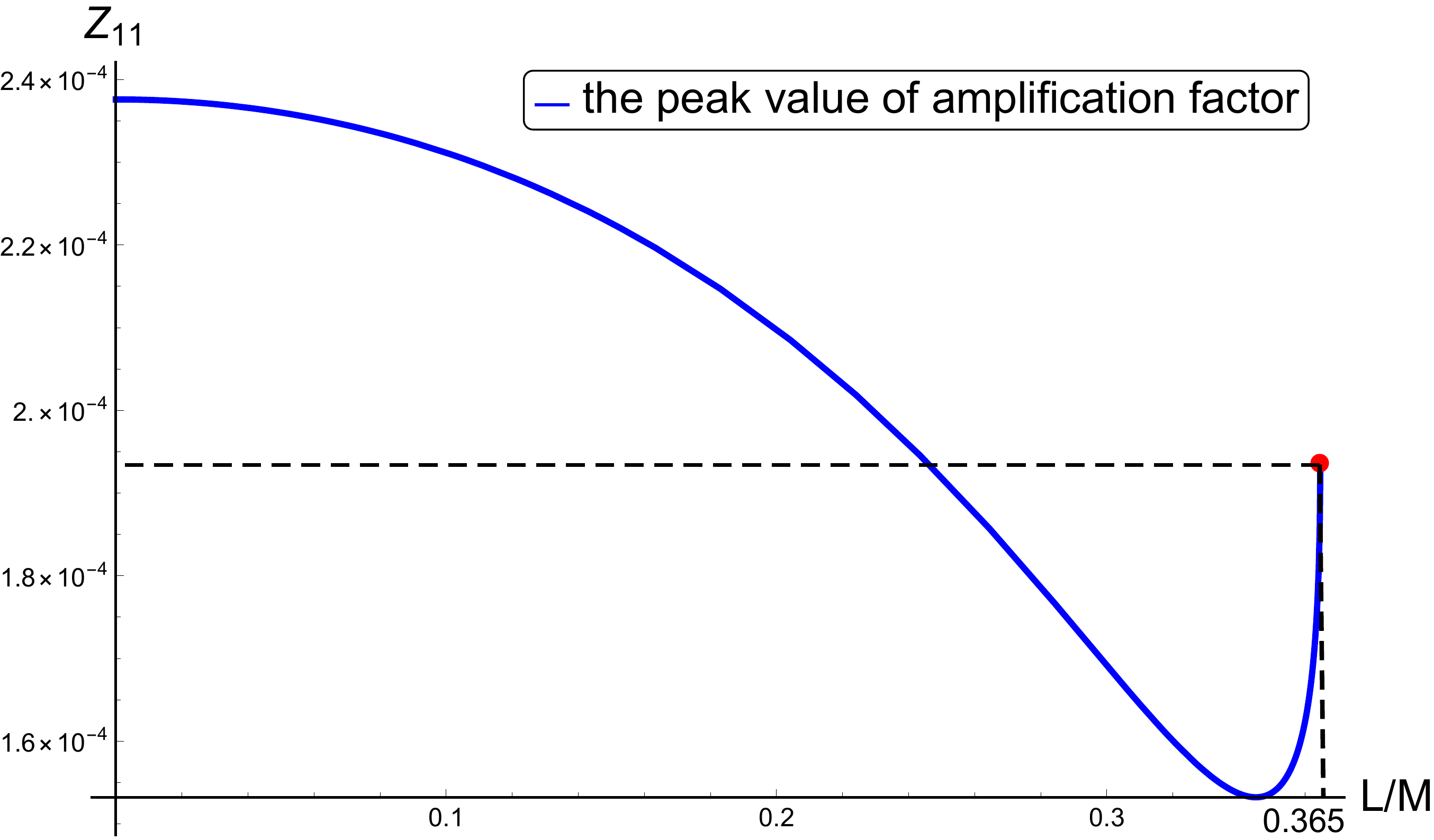}
    \end{minipage}
}
\caption{The amplification factor of massive scalar particles scattered by the rotating Hayward black hole is computed by Eq.~\eqref{Zlm}, where the parameters are set to be: $M\mu=0.010$, $a=0.800M$, $(l,m)=(1,1)$,  and a varying $L$. In diagram (a), $0<L\le 0.345M$, and in diagram (b), $0.345M<L\le 0.365M$. Moreover, diagram (c) depicts the overall variation of the peak of amplification factor with respect to $L$.}
\label{fig:Hay_a=0.8M}
\end{figure}

\item The mode in the interval of $0.859M<a<0.963M$

In this mode, the variation of the peak of superradiance amplification factor with respect to $L$ is the same as that in the second mode. 
The main difference is that the peak is larger than that associated with the Kerr black hole when the rotating Hayward black hole is approaching to its extreme configuration.
This shows that when $L$ increases, the efficiency for massive scalar particles to extract energy from the rotating Hayward black hole is initially lower but finally greater than that from the Kerr black hole ($L=0$).
In Fig.~\ref{fig:Hay_a=0.9M}, $a=0.900M$ is taken as an example, which belongs to the present mode, $0.859M<a<0.963M$.
When $L=0.245M$, the rotating Hayward black hole reaches its extreme configuration, and the corresponding peak of superradiance amplification factor reaches the maximum: $3.967\times 10^{-4}$, which is greater than $3.415\times 10^{-4}$ related to the Kerr black hole.

\begin{figure}[htbp]
\centering
\subfigure[]{
    \begin{minipage}[t]{0.4\linewidth}
    \centering
    \includegraphics[width=1\linewidth]{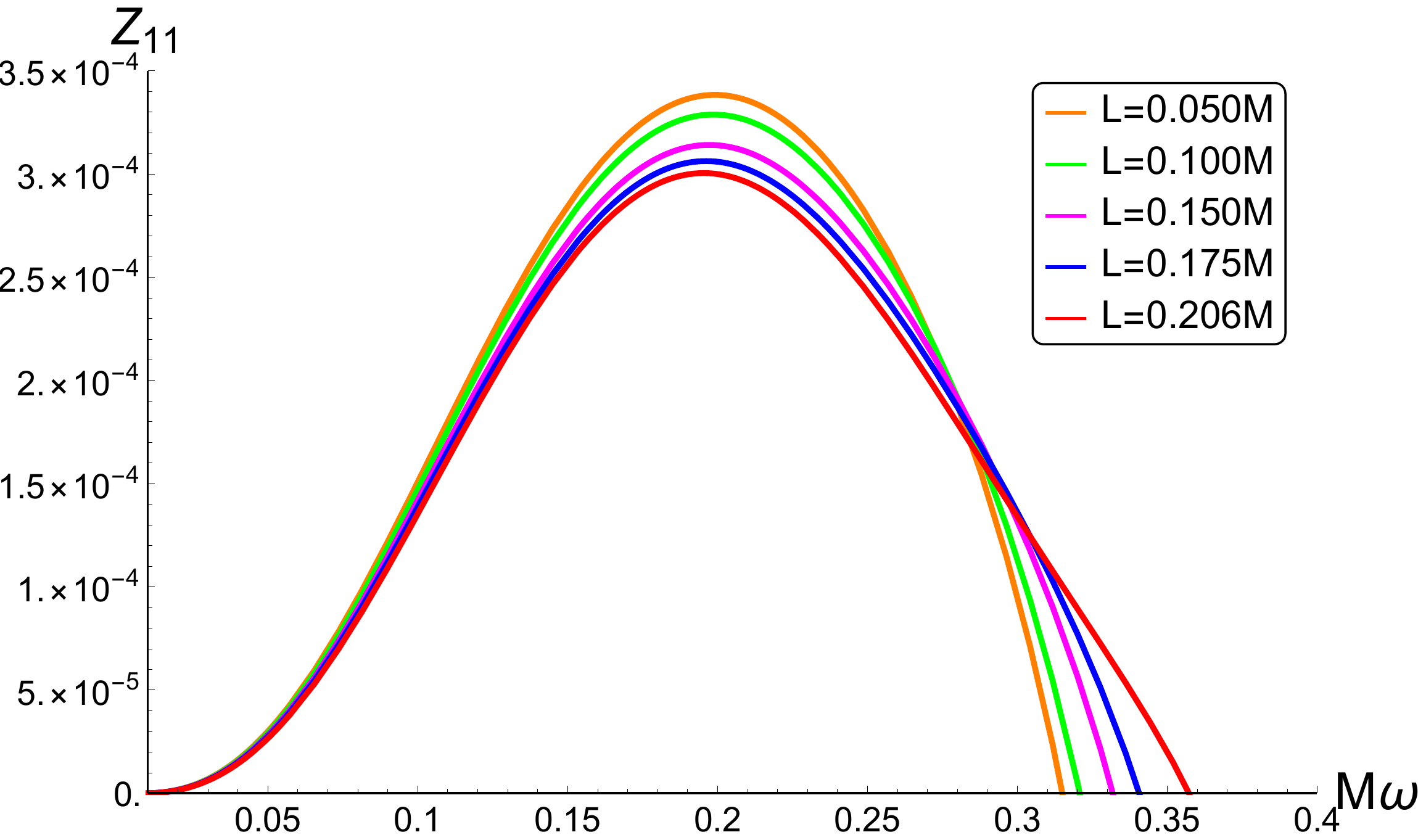}
    \end{minipage}
}
\subfigure[]{
    \begin{minipage}[t]{0.4\linewidth}
    \centering
    \includegraphics[width=1\linewidth]{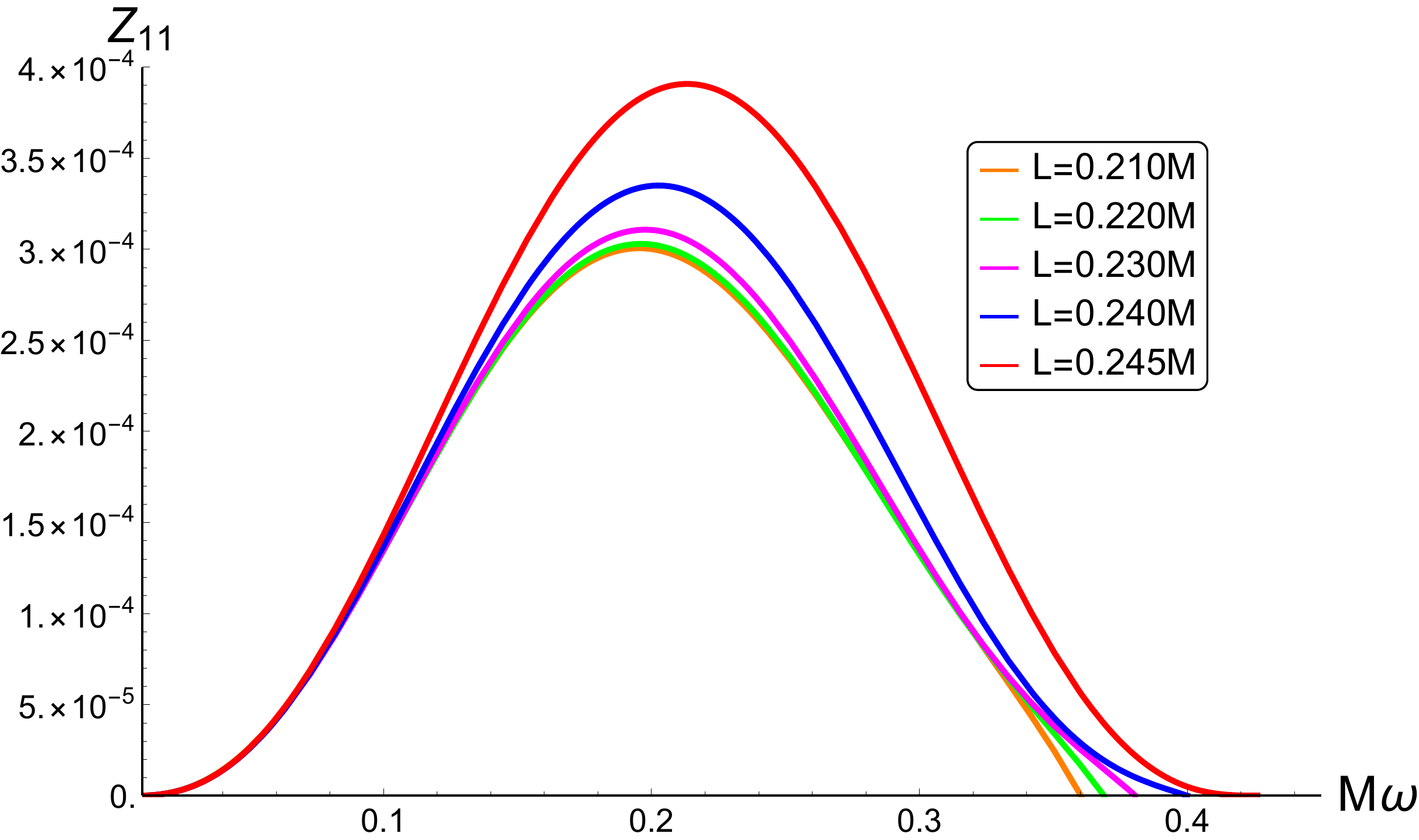}
    \end{minipage}
}
\vspace{-3mm}
\subfigure[]{
    \begin{minipage}[t]{0.4\linewidth}
    \centering
    \includegraphics[width=1\linewidth]{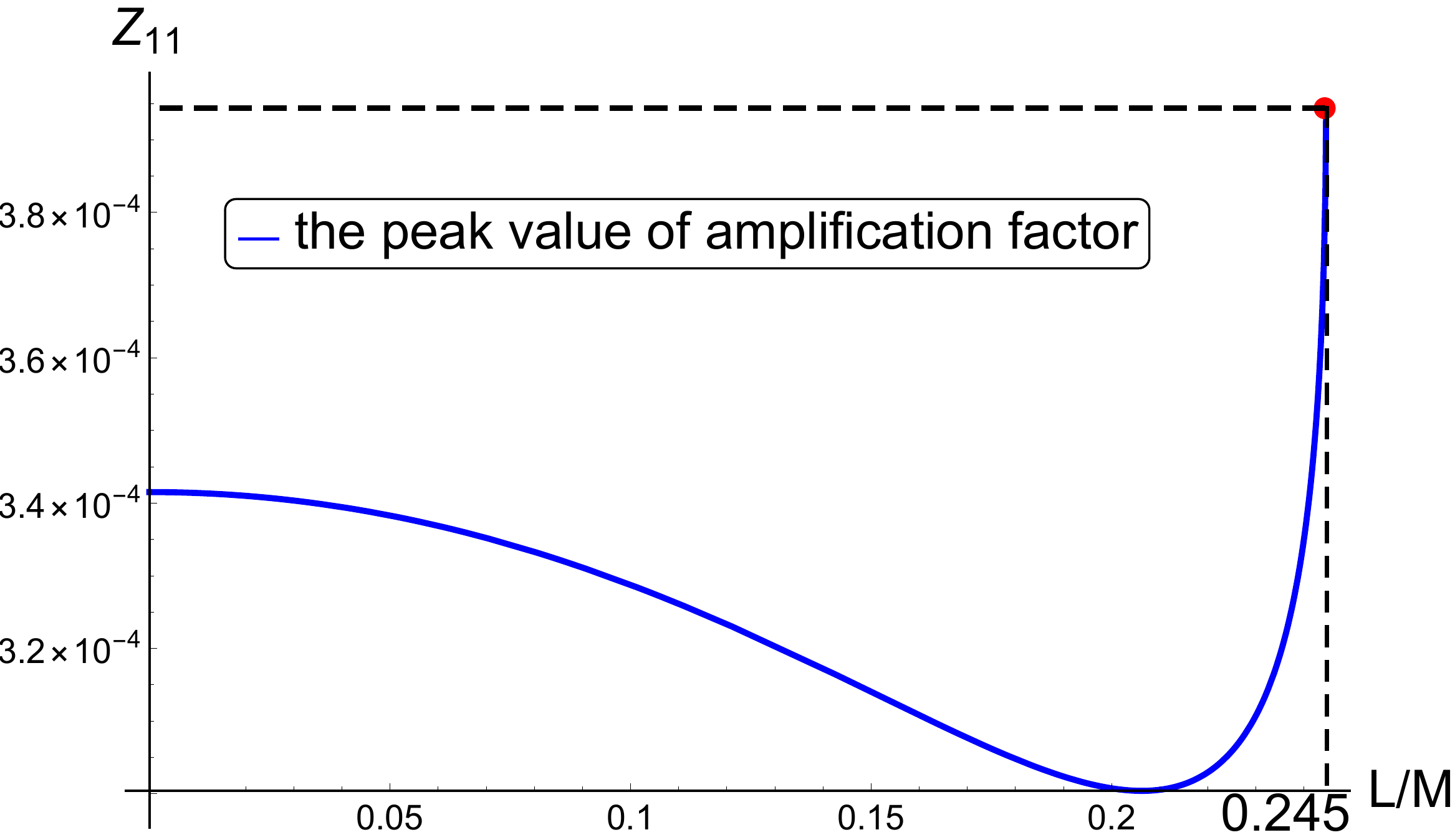}
    \end{minipage}
}
\caption{The amplification factor of massive scalar particles scattered by the rotating Hayward black hole is computed by Eq.~\eqref{Zlm}, where the parameters are set to be: $M\mu=0.010$, $a=0.900M$, $(l,m)=(1,1)$,  and a varying $L$. In diagram (a), $0<L\le 0.206M$, and in diagram (b), $0.206M<L\le 0.245M$. Moreover, diagram (c) depicts the overall variation of the peak of amplification factor with respect to $L$.}
\label{fig:Hay_a=0.9M}
\end{figure}

\item The mode in the interval of $0.963M<a<M$

The peak of superradiance amplification factor rises with an increase of $L$, and it reaches the maximum when the rotating Hayward black hole goes to its extreme configuration.
This means that when  $L$ increases, the efficiency  for massive scalar particles to extract energy from the rotating Hayward black hole is always larger than that from the Kerr black hole ($L=0$).
In Fig.~\ref{fig:Hay_a=0.99M},  $a=0.990M$ is taken as an example, which belongs to the present mode, $0.963M<a<M$.
When $L=0.070M$, the rotating Hayward black hole reaches its extreme configuration, and the  peak of superradiance amplification factor reaches the maximum,  $7.904\times 10^{-4}$, which is larger than $5.942\times 10^{-4}$ related to the Kerr black hole.

\begin{figure}[htbp]
\centering
\subfigure[]{
    \begin{minipage}[t]{0.4\linewidth}
    \centering
    \includegraphics[width=1\linewidth]{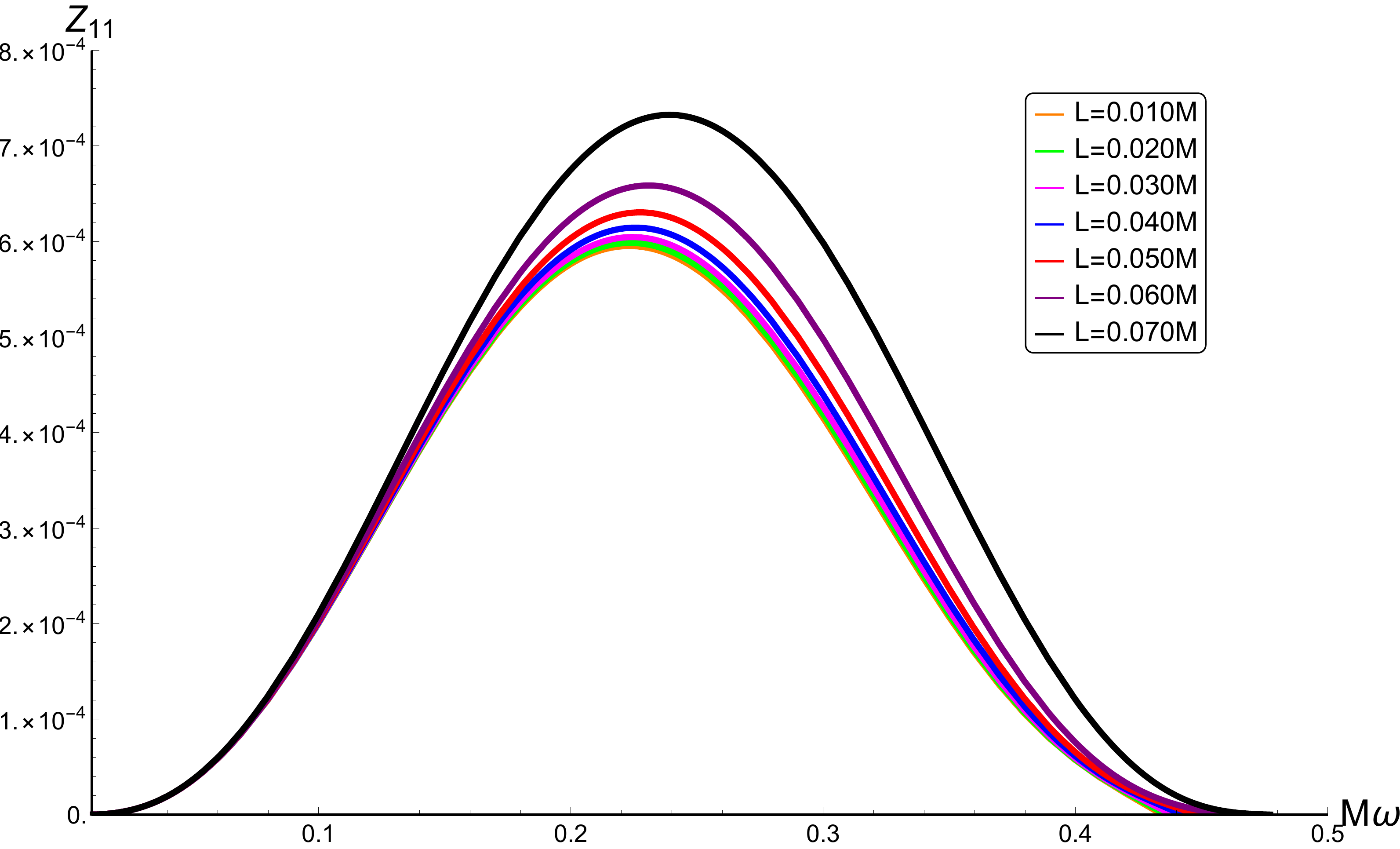}
    \end{minipage}
}
\subfigure[]{
    \begin{minipage}[t]{0.4\linewidth}
    \centering
    \includegraphics[width=1\linewidth]{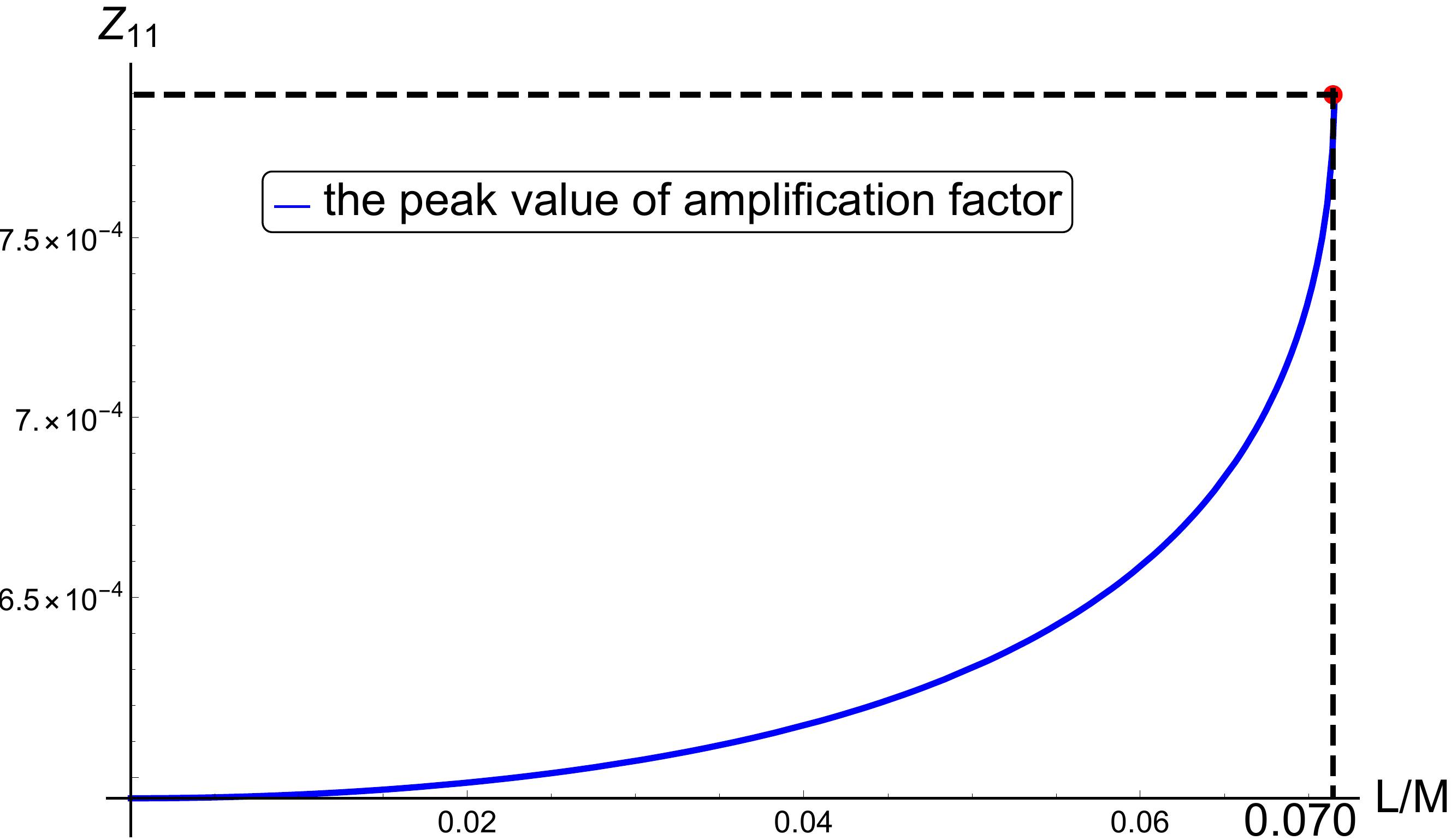}
    \end{minipage}
}

\caption{The amplification factor of massive scalar particles scattered by the rotating Hayward black hole is computed by Eq.~\eqref{Zlm}, where the parameters are set to be: $M\mu=0.010$, $a=0.990M$, $(l,m)=(1,1)$, and $0<L\le 0.070M$. Moreover, diagram (b) depicts the overall variation of the peak of amplification factor with respect to $L$.}
\label{fig:Hay_a=0.99M}
\end{figure}

\end{enumerate}

From the above four modes, we can infer that the introduction of a regularization parameter $L$ has both promotional   and inhibitory effects on the efficiency of energy extraction, and that whether the former or the latter effect is dominant depends on the rotation parameter of the rotating Hayward black hole.
Referring to the above discussions, we divide the values of the rotation parameter into four intervals: the low rotation interval ($0.018M<a<0.100M$), the middle rotation interval ($0.100M<a<0.859M$), the high rotation interval ($0.859M<a<0.963M$), and the near-extreme rotation interval ($0.963M<a<M$).

In the low rotation interval, the introduction of $L$ gives rise to two effects that are opposite to each other, the promotional and inhibitory effects, on the energy extraction efficiency, and it depends on the competition of the two effects whether the peak of amplification factor related to the rotating Hayward black hole ($L\ne 0$) is greater or smaller than that related to the Kerr black hole ($L=0$).
When $L$ begins to increase from zero, the promotional effect grows more quickly than the inhibitory effect, and finally the peak of amplification factor reaches its maximum at a certain $L$.
When $L$ is further increasing, the promotional effect grows more slowly than the inhibitory effect, so that the peak of amplification factor begins to decline.
When $L$ reaches a certain threshold, the promotional  effect is the same as the inhibitory effect, and the peak of amplification factor is the same as that related to the Kerr black hole.
After $L$ exceeds this threshold, the inhibitory effect dominates, and the energy extraction efficiency becomes lower than that associated with the Kerr black hole.
When $L$ goes to the value at which the rotating Hayward black hole is close to its extreme configuration, the promotional effect grows more quickly than the inhibitory effect again, so there is a slight climb in the  peak of amplification factor.
As a summary for the low rotation interval, the dominant role of the promotional effect is gradually replaced by the inhibitory effect when $L$ increases, so that the energy extraction efficiency related to the rotating Hayward black hole is at first higher and then lower than that related to the Kerr black hole.

In the middle rotation interval, the inhibitory effect of $L$ on the energy extraction efficiency is always greater than the promotional effect.
When $L$ increases from zero and reaches the value at which the rotating Hayward black hole is close to its extreme configuration, the promotional effect grows more quickly than the inhibitory effect, so the peak of amplification factor will rise to a certain extent. Although  the peak will rise higher for a larger
rotation parameter $a$, but it is still lower than that related to the Kerr black hole.

In the high rotation interval, the inhibitory effect of $L$ on the energy extraction efficiency is initially greater than the promotional effect.
When $L$ begins to increase from zero, the inhibitory effect grows more quickly than the promotional effect, and finally the peak of amplification factor reaches its minimum at a specific value of $L$.
After $L$ exceeds this specific value, the promotional effect grows more quickly than the inhibitory effect, so the peak of amplification factor begins to rise and reaches its maximum value at which the rotating Hayward black hole is close to its extreme configuration, and such a  maximum value is greater than the peak related to the Kerr black hole.

In the near-extreme rotation interval, the promotional effect of $L$ on the energy extraction efficiency is always greater than the inhibitory effect, so the peak of amplification factor is always greater than that related to the Kerr black hole.
When $L$ begins to increase from zero, the promotional effect grows more quickly than the inhibitory effect, and finally the peak of amplification factor reaches its maximum at which the rotating Hayward black hole is close to its extreme configuration.

From what has been discussed above, we can summarize the following three conclusions:
\begin{itemize}
\item For  massive scalar fields in the background of rotating Hayward black holes, the peak of amplification factor is dominated by the rotation parameter $a$, not by the regularization parameter $L$ which plays only a secondary role.

\item For four rotation parameter intervals, the influence of $L$ on the peak of amplification factor shows various situations. In short, there are four modes in which the energy extraction efficiency varies in accordance with the competition between the two parameters, $a$ and $L$.

\item When a rotating Hayward black hole stays at its near-extreme configuration, the peak of amplification factor always grows with an increase of $L$. Furthermore, the bigger $L$ is, the closer to the extreme configuration the rotating Hayward black hole is.
\end{itemize}

\subsection{Bardeen black holes}\label{sec: model_analysis_Bar}
\begin{figure}[htbp]
\centering
    \begin{minipage}[t]{0.5\linewidth}
    \centering
    \includegraphics[width=1\linewidth]{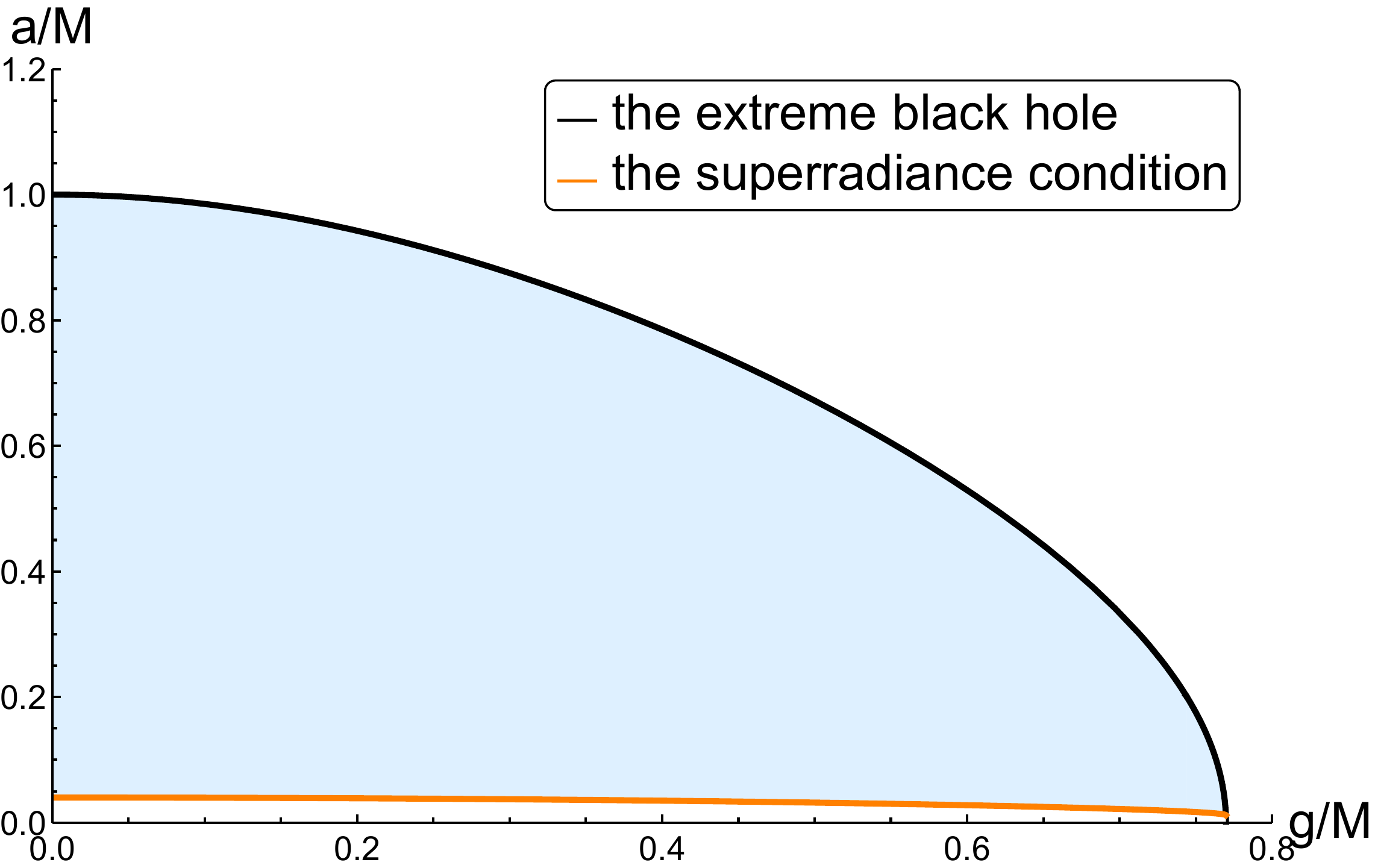}
    \end{minipage}
\caption{The relationship between the rotation parameter $a$ and the magnetic charge $g$ in rotating Bardeen black holes, where the abscissa is $g/M$, the ordinate is $a/M$, and $M$ is the mass of the black hole. The black curve (depicted by Eqs.~(\ref{Bardeen_Horizon}) and (\ref{Bexhorizon})) represents the extreme black hole, the orange curve (depicted by Eqs.~(\ref{case1orglin}) and (\ref{Bardeen_Horizon})) is the lower bound for the occurrence of superradiance, and the blue area bounded by the two curves is the region of black hole parameters in which the superradiance can occur.}
\label{fig:Bardeen_H}
\end{figure}

After extending the static and spherically symmetric Bardeen black hole to the rotating and axially symmetric one that is still regular everywhere~\cite{A16}, we derive the equation that governs the event horizons of the rotating Bardeen black hole,
\begin{equation}
\label{Bardeen_Horizon}
\left(r_{\rm H}^2+a^2\right)^2 \left(r_{\rm H}^2+g^2\right)^3-4 M^2r_{\rm H}^8=0.
\end{equation}
This algebraic equation has two positive and real solutions, $r_{\rm H}^-$ and $r_{\rm H}^+$, which correspond to the inner and outer horizons, respectively.
When the inner and outer horizons are equal, the black hole reaches its extreme configuration. The derivative of Eq.~\eqref{Bardeen_Horizon} with respect to $r_{\rm H}$ gives rise to the equation that governs $r_{\rm H}^{\rm ext}$, the horizon of  the extreme configuration,
\begin{equation}\label{Bexhorizon}
2 \left(\left(r_{\rm H}^{\rm ext}\right)^2+a^2\right) \left(\left(r_{\rm H}^{\rm ext}\right)^2+g^2\right)^3+3 \left(\left(r_{\rm H}^{\rm ext}\right)^2+a^2\right)^2 \left(\left(r_{\rm H}^{\rm ext}\right)^2+g^2\right)^2-16 M^2 \left(r_{\rm H}^{\rm ext}\right)^6=0,
\end{equation}
which, together with Eq.~(\ref{Bardeen_Horizon}), depicts the black curve in Fig.~\ref{fig:Bardeen_H}. 
The area surrounded by the black curve is just the  parameter region where the black hole exists.
As in the rotating Hayward black hole, we still fix the particle mass to be $M\mu=0.01$ for the rotating Bardeen black hole. According to Eq.~\eqref{Super_Gene_con}, the critical condition that
the superradiance phenomenon occurs has the same form as
Eq.~(\ref{case1orglin}), in which $r_{\rm H}^+$ is determined by Eq.~(\ref{Bardeen_Horizon}) instead of Eq.~(\ref{Hayward_Horizon}). That is, Eq.~(\ref{case1orglin}) and Eq.~(\ref{Bardeen_Horizon}) depict the orange curve in Fig.~\ref{fig:Bardeen_H}. 
Therefore, the blue area surrounded by the black curve and the orange curve is the parameter region in which the superradiance can occur.
Note that the intersection of the black and orange curves is $(0.769,0.013)$, which indicates that there will be no superradiance effect when $g>0.769M$ or $a<0.013M$.

\begin{figure}[htbp]
\centering
    \begin{minipage}[t]{0.4\linewidth}
    \centering
    \includegraphics[width=1\linewidth]{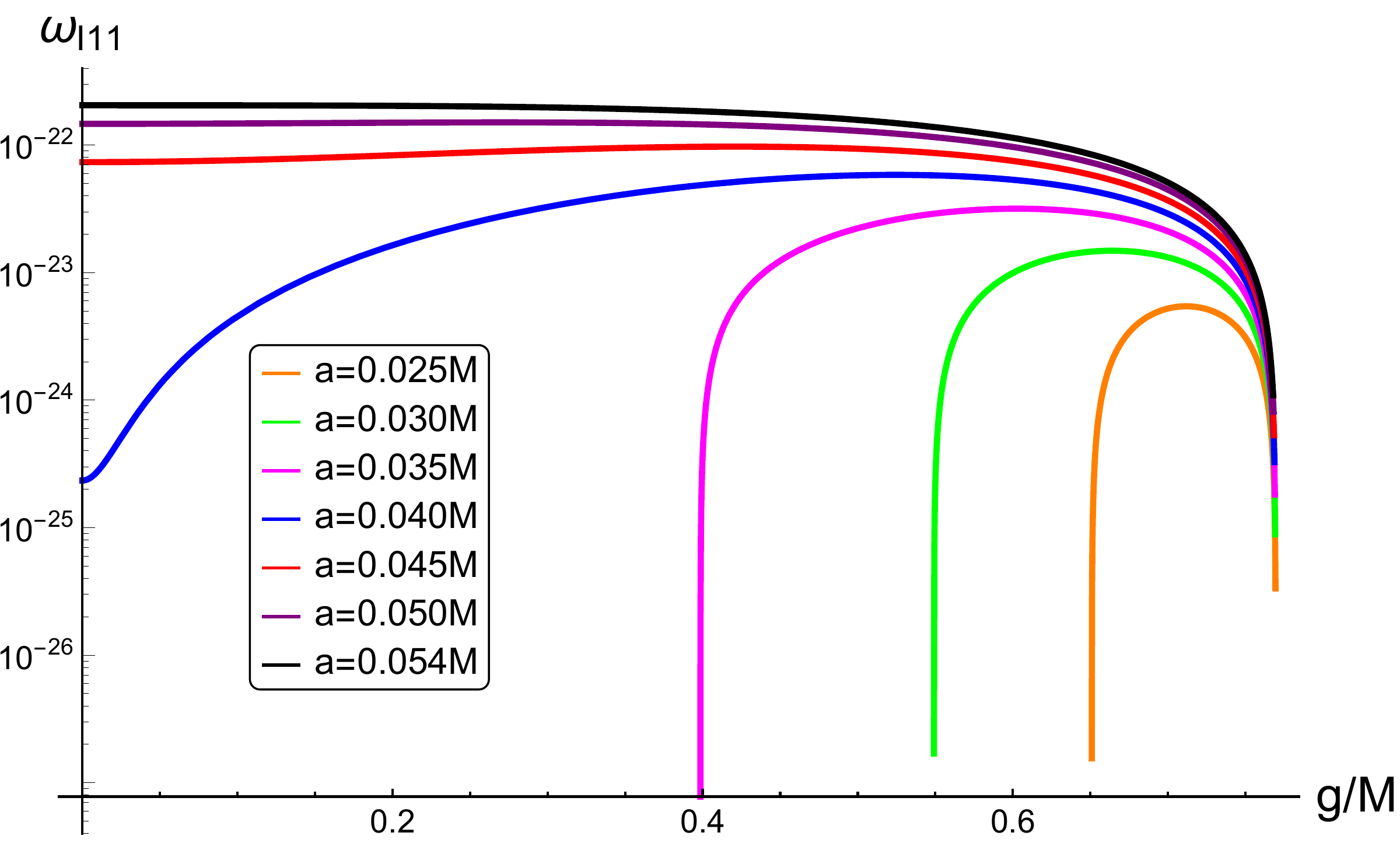}
    \end{minipage}
    \begin{minipage}[t]{0.4\linewidth}
    \centering
    \includegraphics[width=1\linewidth]{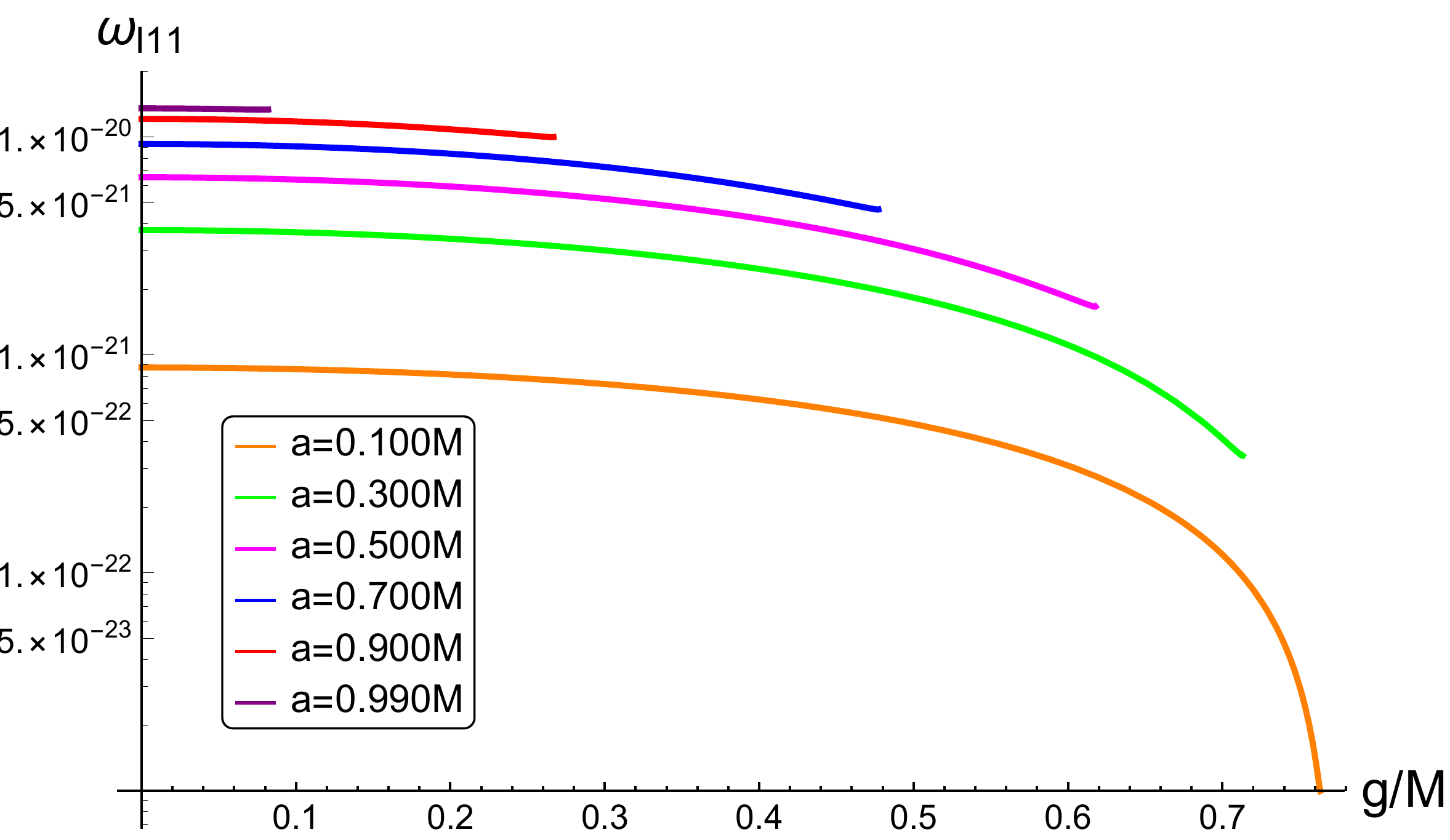}
    \end{minipage}
\caption{The relationship between the imaginary part of frequency $\omega_{\rm I}$ and the magnetic charge $g$ for quasi-bound state particles that are scattered by  rotating Bardeen black holes and are at the leading multipole number $(l=1, m=1, n=1)$ when the rotation parameter $a$ takes various values, where the left diagram is depicted for the case, $0.013M<a<0.055M$ and the right diagram for the case, $0.055M<a<1.000M$.}
\label{fig:Bar_omega_11_a}
\end{figure}

At the moment we discuss how the magnetic charge $g$, together with the rotation parameter $a$, affects the amplitude of a massive scalar particle at its quasi-bound state.
In Fig.~\ref{fig:Bar_omega_11_a}, we show the relationship between the imaginary part of frequency $\omega_{\rm I}$ for quasi-bound state particles at the leading multipole number ($l=1$, $m=1$, $n=1$) when the rotation parameter $a$ takes various values.
We note that $\omega_{\rm I}$ is positively correlated with the growth rate of amplitude of quasi-bound state particles.
It can be seen that the effect of  $g$ on $\omega_{\rm I}$ is related to the value of $a$.
When $a<0.055M$, $\omega_{\rm I}$ increases at first and then decreases with an increase of $g$.
This indicates that the growth rate of superradiance instability increases at first and then decreases when $g$ stays in the low rotation interval $a<0.055M$.
Moreover, if $a>0.055M$, $\omega_{\rm I}$ monotonically decreases with an increase of $g$, which indicates that the growth rate of superradiance instability decreases with an increase of $g$.
However, $\omega_{\rm I}$ always increases monotonically with an increase of $a$ for a fixed $g$.
Thus, an increasing $a$ also makes the superradiance instability  grow very fast in the rotating Bardeen black hole.
Our summary is that the relationship between the growth rate of superradiance instability and $g$ depends on the rotation parameter $a$, but the relationship between the growth rate of superradiance instability and $a$ is always positively correlated for any fixed $g$.

Next, we focus on how the magnetic charge $g$, together with the rotation parameter $a$, affects the superradiance amplification of massive scalar particles.
We discuss the behavior of particles still by setting $M\mu=0.010$ at the leading multipole ($l=1, m=1$) in the background of a rotating Bardeen black hole. 
We also summarize four modes of the influence of $g$ on the peak of superradiance amplification when the rotation parameter $a$ varies in the parameter region where the superradiance effect can occur. 
The four modes are analyzed in detail as follows.

\begin{enumerate}
\item The mode in the interval of $0.013M<a<0.246M$

The peak of amplification factor rises at first and then falls with an increase of the magnetic charge $g$.
This shows that the energy extraction
efficiency of massive scalar particles scattered by the rotating Bardeen black hole is initially greater but finally lower than that scattered by the singular Kerr black hole ($g=0$) when $g$ increases.
In particular, we note that the peak value will rise again when the rotating Bardeen black hole is close to its extreme configuration.
In Fig.~\ref{fig:a=0.1M},  $a=0.100M$ is taken as an example.
When $g$ is in the range of $0<g\le0.545M$, the peak value rises with an increase of $g$, and reaches its maximum value, $5.310\times 10^{-8}$, at $g=0.545M$.
Then, when $g$ is in the range of $0.545M<g\le0.76325M$, the peak value falls with an increase of $g$, and reaches its minimum value, $3.493\times 10^{-9}$, at $g=0.76325M$, where the rotating Bardeen black hole is very close to the extreme configuration.
When $g$ further grows to  the range of $0.76325M<g\le0.76333M$, the peak has a slight uptick as the black hole gets closer to the extreme configuration.
\begin{figure}[htbp]
\centering
\subfigure[]{
    \begin{minipage}[t]{0.4\linewidth}
    \centering
    \includegraphics[width=1\linewidth]{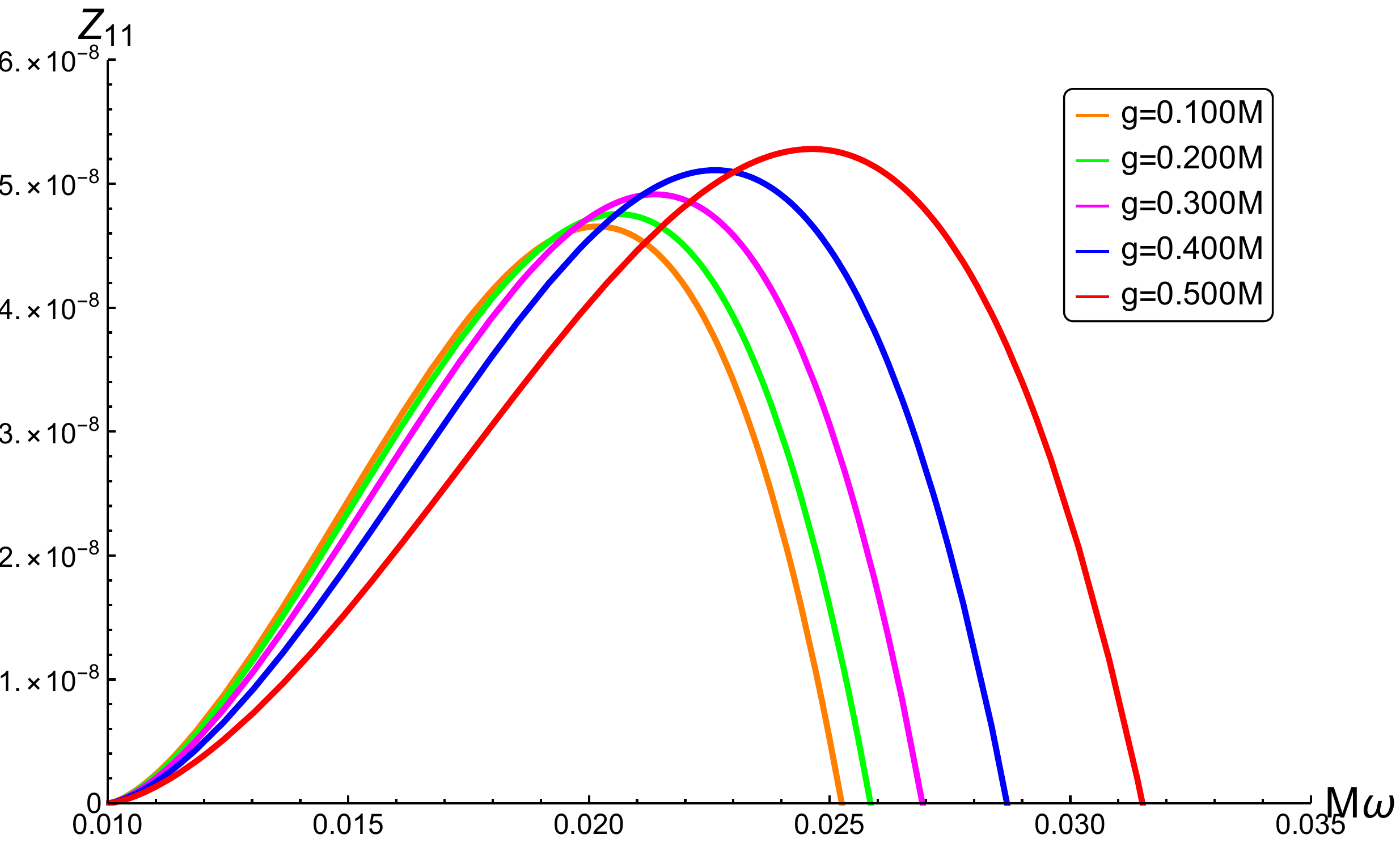}
    \end{minipage}
}
\subfigure[]{
    \begin{minipage}[t]{0.4\linewidth}
    \centering
    \includegraphics[width=1\linewidth]{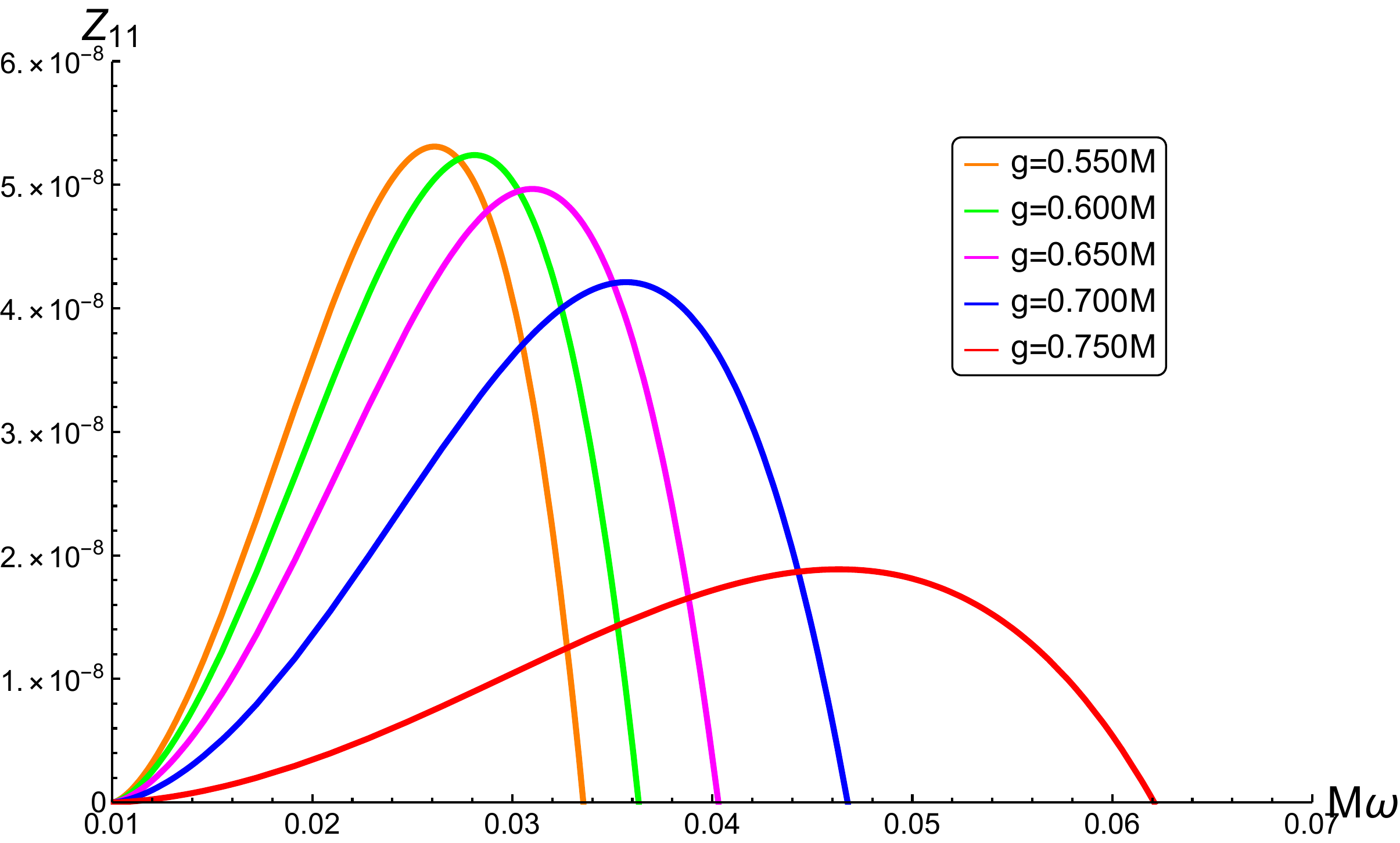}
    \end{minipage}
}
\vspace{-3mm}
\subfigure[]{
    \begin{minipage}[t]{0.4\linewidth}
    \centering
    \includegraphics[width=1\linewidth]{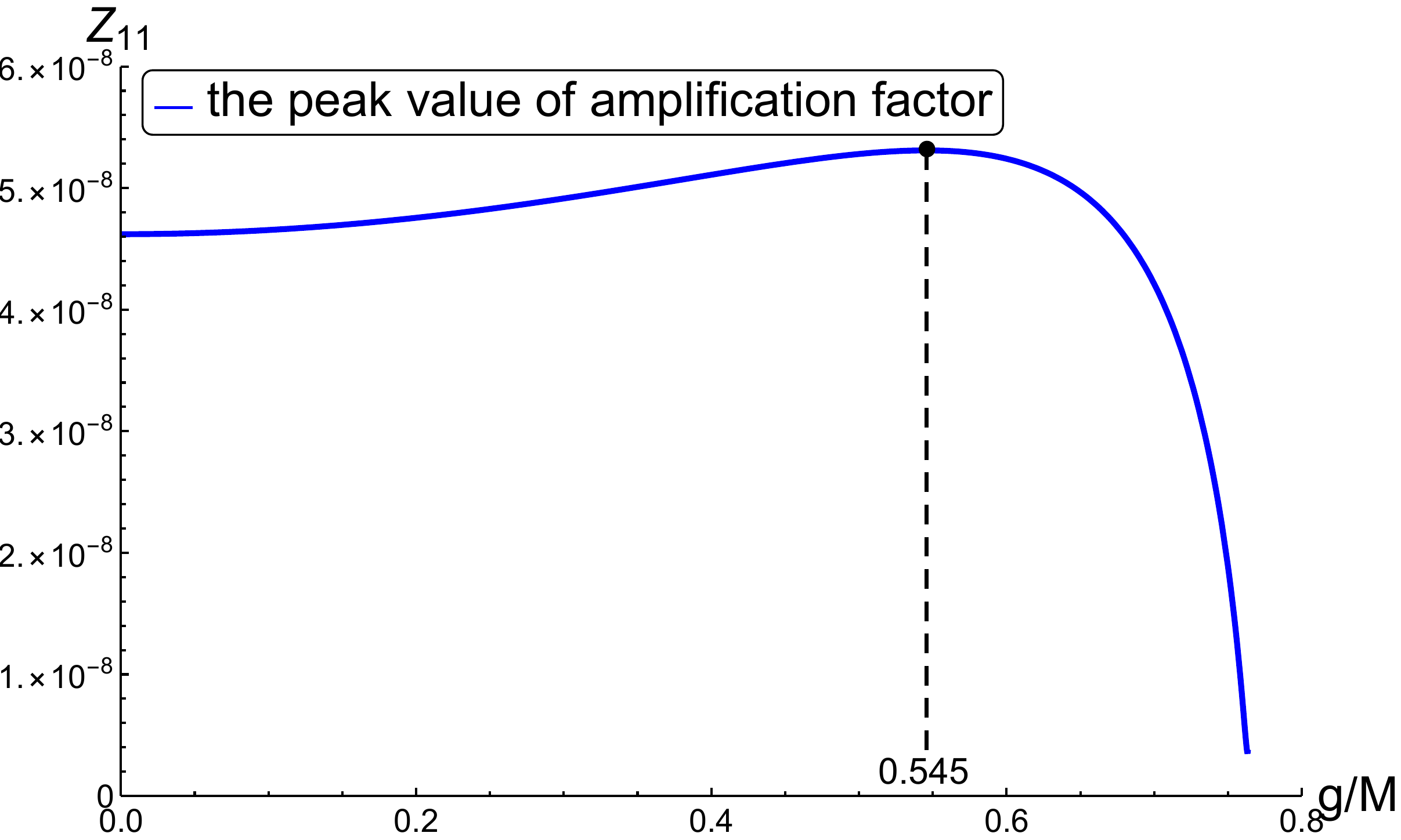}
    \end{minipage}
}
\subfigure[]{
    \begin{minipage}[t]{0.4\linewidth}
    \centering
    \includegraphics[width=1\linewidth]{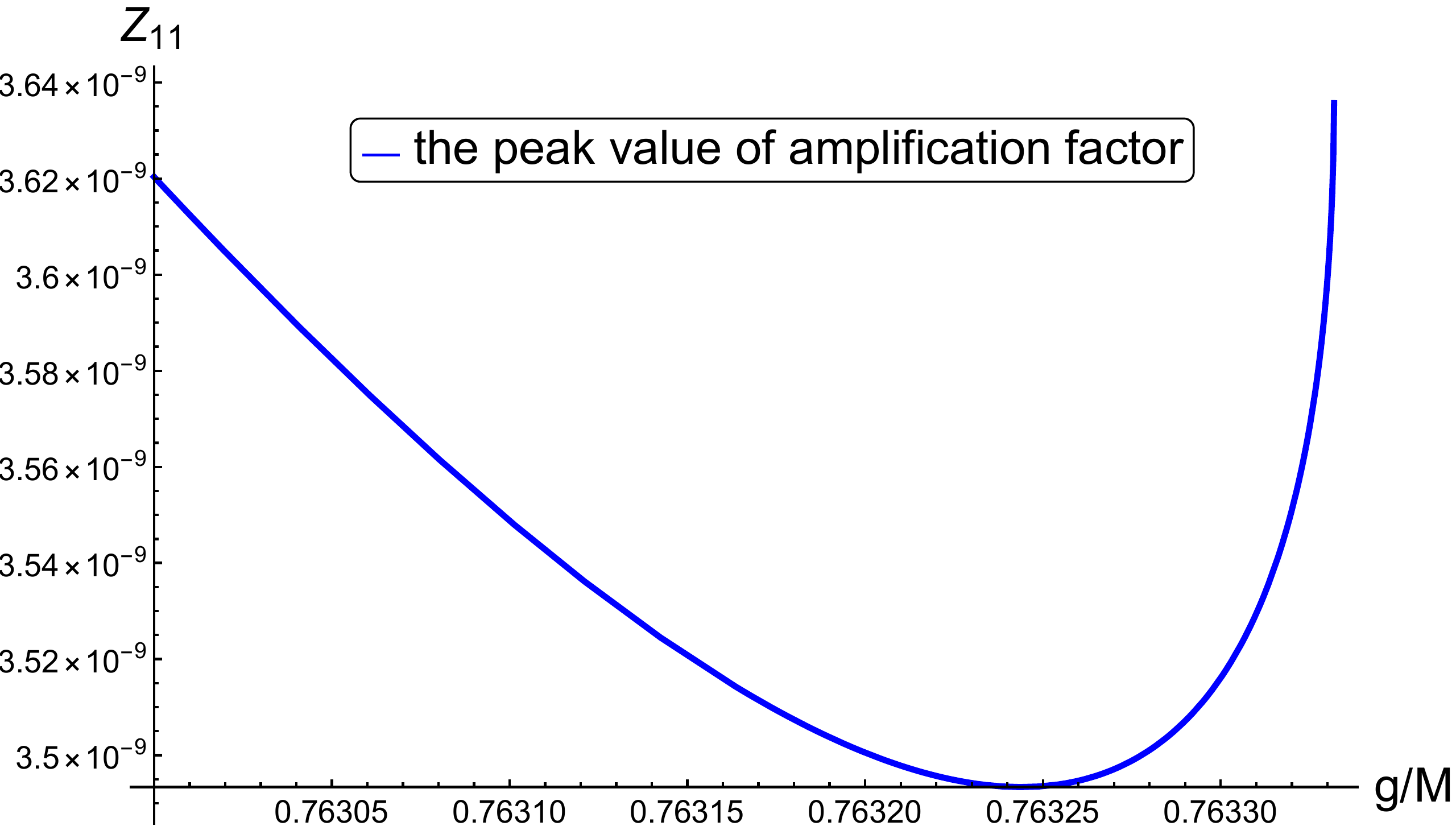}
    \end{minipage}
}
\caption{The amplification factor of massive scalar particles scattered by the rotating Bardeen black hole is computed by Eq.~\eqref{Zlm}, where the parameters are set to be: $M\mu=0.010$, $a=0.100M$, $(l,m)=(1,1)$,  and a varying $g$. In diagram (a), $0<g\le0.545M$, and in diagram (b), $0.545M<g<0.76300M$. Moreover, diagram (c) depicts the overall variation of the peak of amplification factor with respect to $g$, and diagram (d) describes the behavior of the peak when the rotating Bardeen black hole is going to its near-extreme configuration, $0.76300M\le g < 0.76333M$.}
\label{fig:a=0.1M}
\end{figure}

\item The mode in the interval of $0.246M<a<0.737M$

The peak value of superradiance magnification falls monotonically when $g$ increases from zero, and reaches its minimum when $g$ takes a certain range of values in which the rotating Bardeen black hole is going to its near-extreme configuration.
When $g$ increases further, the peak rises very rapidly, and reaches its highest value at which the rotating Bardeen black hole is in its extreme configuration.
However, as shown in Fig.~\ref{fig:a_fixed}, the peak never exceeds the value associated with the Kerr black hole ($g=0$) when $a$ belongs to this mode.
It shows that the efficiency is the maximum for a massive scalar particle to extract energy from the Kerr black hole ($g=0$), and that the introduction of the magnetic charge (a non-vanishing $g$) will reduce such an efficiency.
In particular, the order of magnitude of the peak grows with an increase of $a$. 
When the change of the order of magnitude caused by $a$ is compared with that by $g$, we can see that the peak of amplification factor is affected mainly by $a$ rather than by $g$. In Fig.~\ref{fig:a=0.7M}, $a=0.700M$ is taken as an example.
When $g$ takes the range of $0<g\le 0.453M$, the peak of amplification factor falls with an increase of $g$, and reaches its minimum value, $1.013\times10^{-4}$, at $g=0.453M$.
Moreover, the peak starts to rise again when $g$ increases further from $0.453M$ (This process means that the rotating Bardeen black hole is approaching to its extreme configuration), but it is still lower than that associated with the Kerr black hole even if the rotating Bardeen black hole arrives at the extreme. 
As a result, the peak related to the Kerr black hole is the largest in the second mode of $a$-parameter intervals.

\begin{figure}[htbp]
\centering
\subfigure[$a=0.300M$]{
    \begin{minipage}[t]{0.4\linewidth}
    \centering
    \includegraphics[width=1\linewidth]{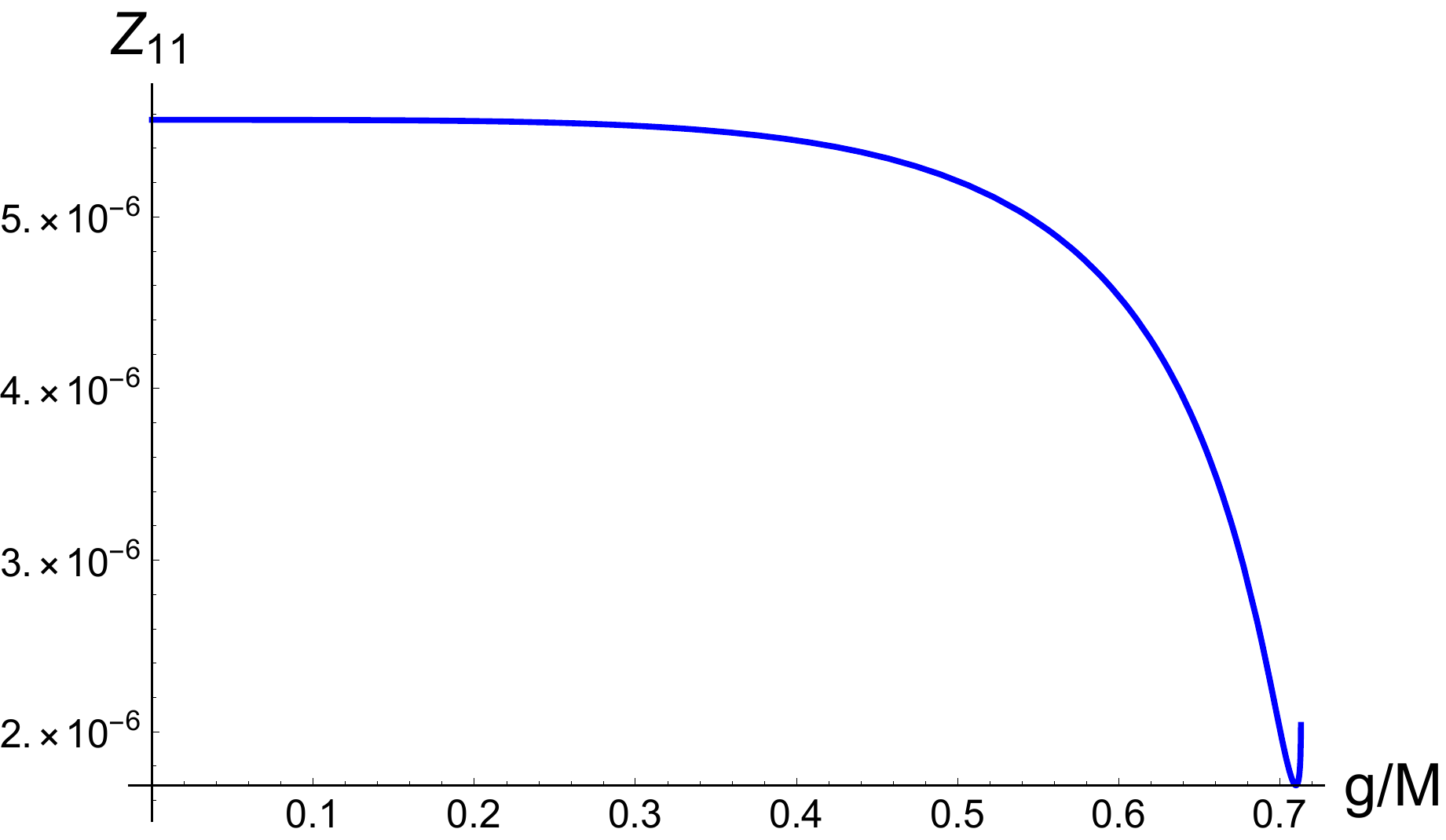}
    \end{minipage}
}
\subfigure[$a=0.400M$]{
    \begin{minipage}[t]{0.4\linewidth}
    \centering
    \includegraphics[width=1\linewidth]{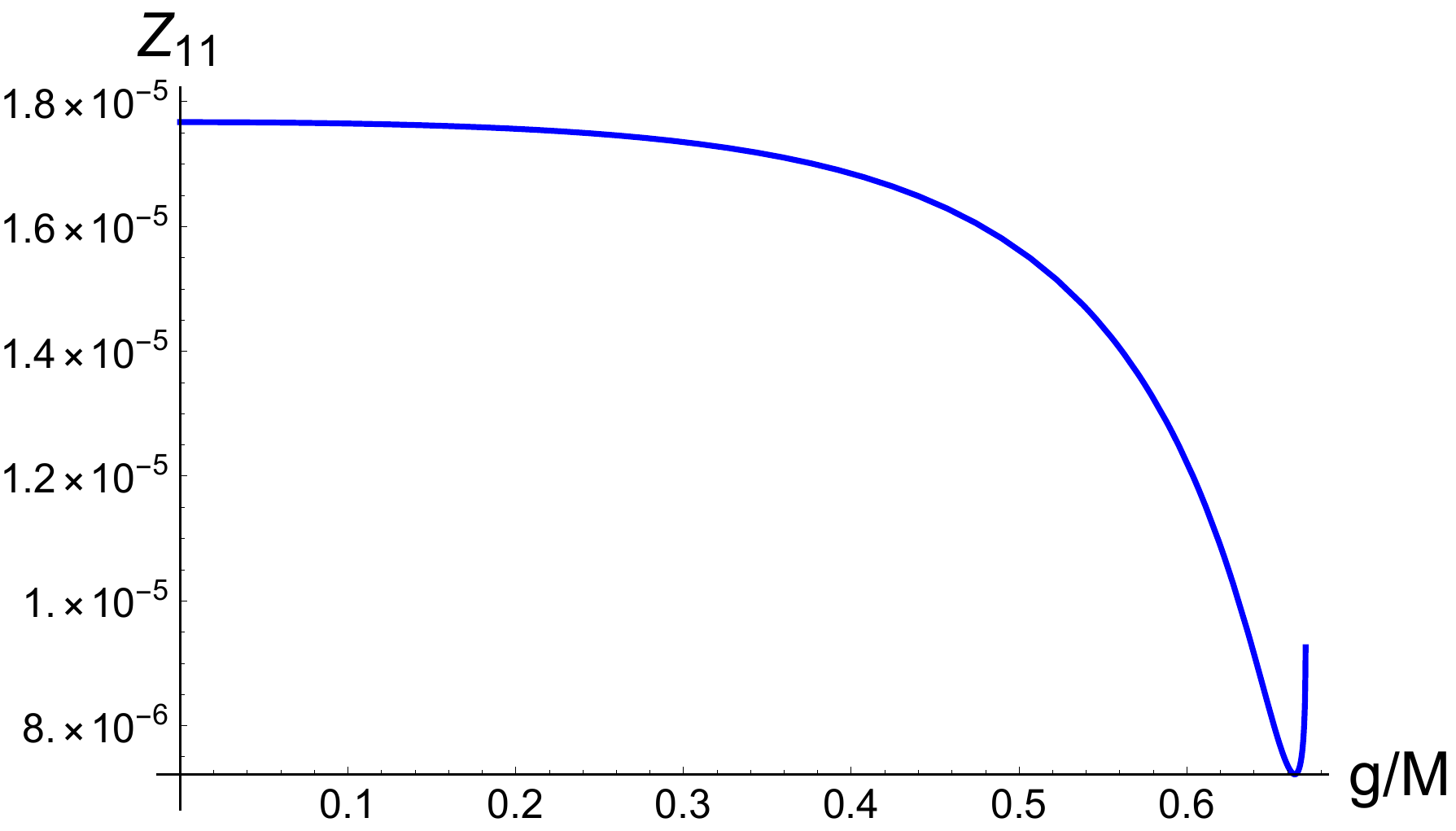}
    \end{minipage}
}
\vspace{-3mm}
\subfigure[$a=0.500M$]{
    \begin{minipage}[t]{0.4\linewidth}
    \centering
    \includegraphics[width=1\linewidth]{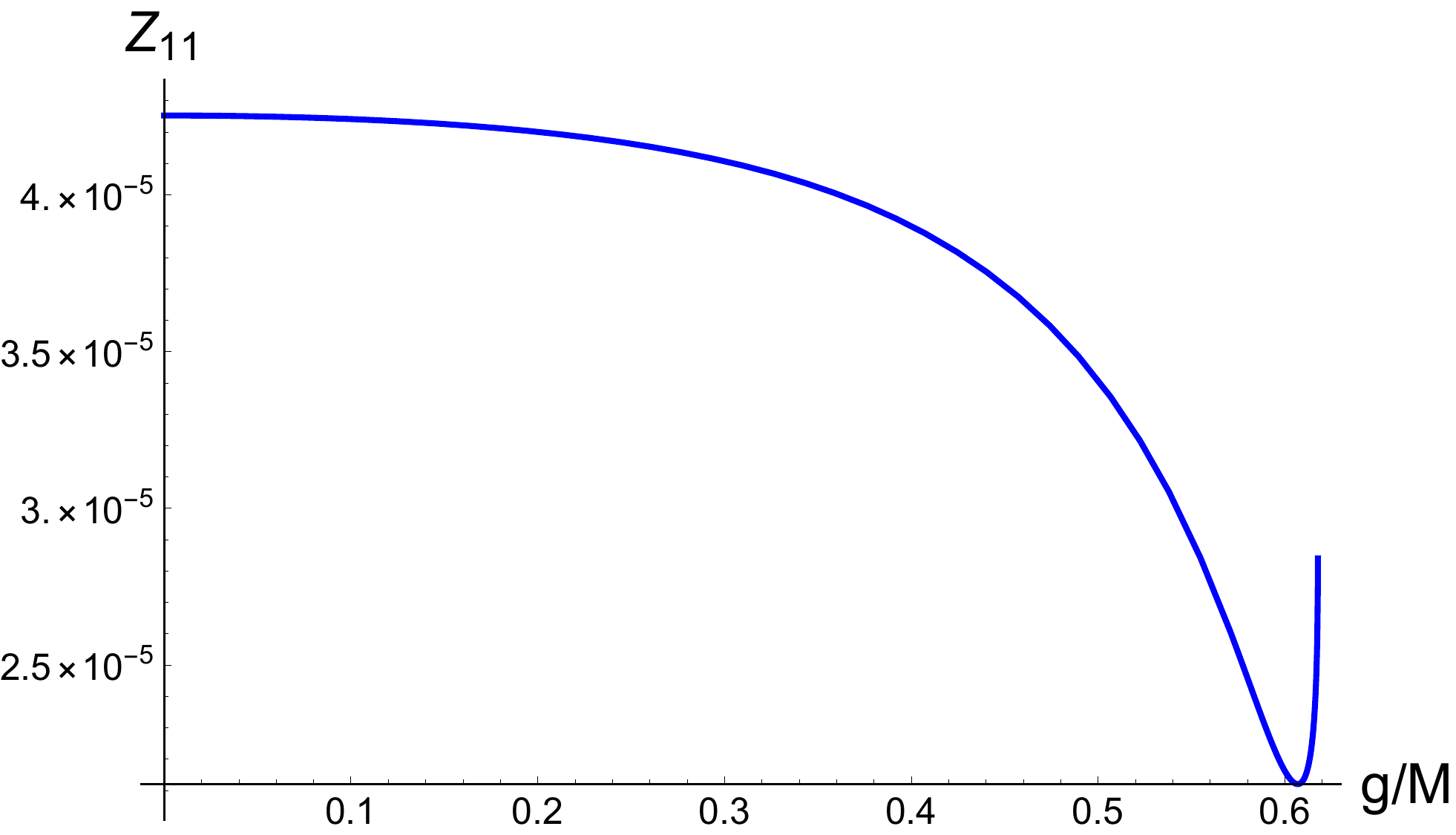}
    \end{minipage}
}
\subfigure[$a=0.600M$]{
    \begin{minipage}[t]{0.4\linewidth}
    \centering
    \includegraphics[width=1\linewidth]{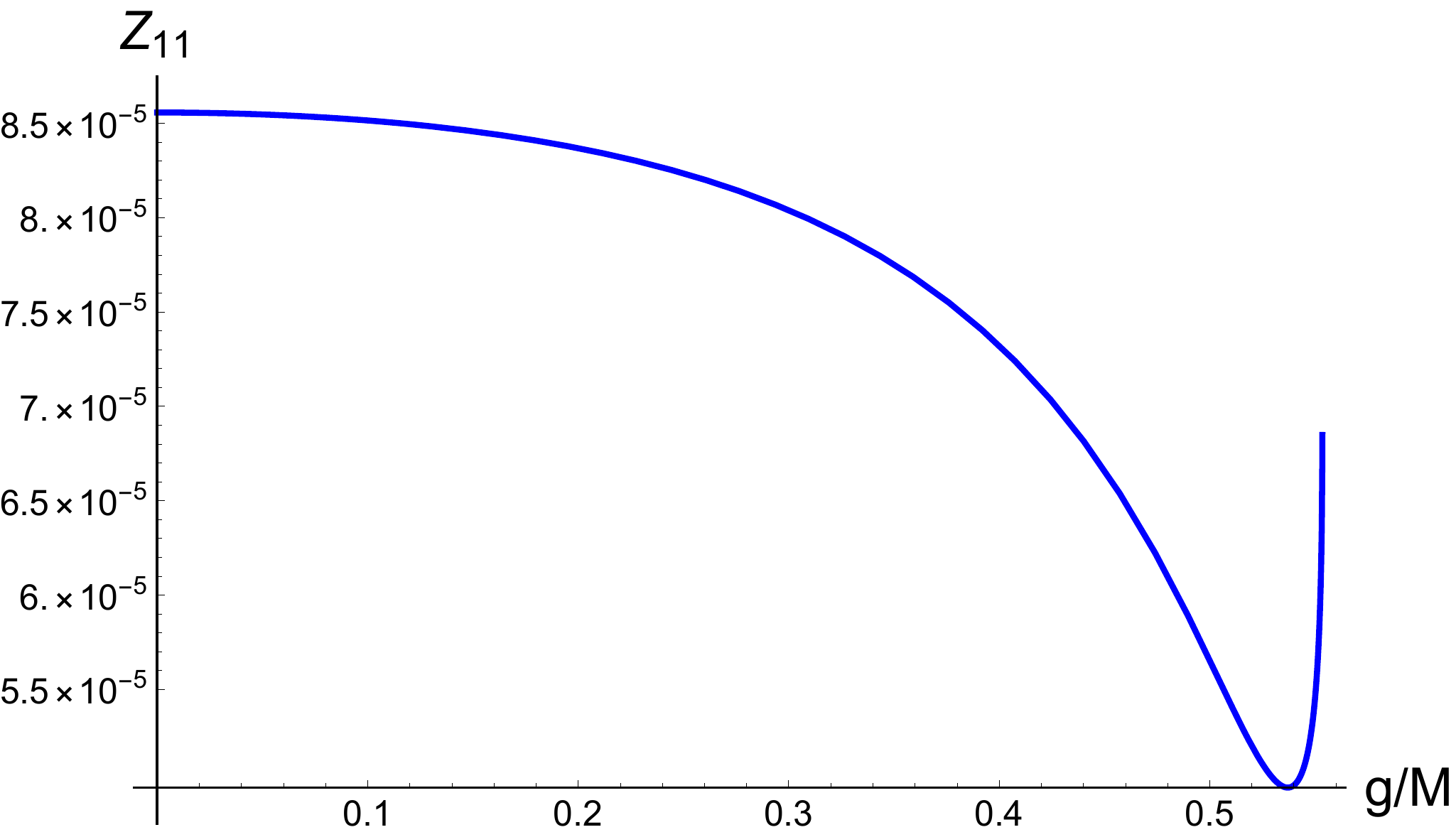}
    \end{minipage}
}
\caption{Relationship between the peak of amplification factor and the magnetic charge $g$ for the massive scalar particles scattered by the rotating Bardeen black hole, where the parameters are set to be: $M\mu=0.010$, $(l,m)=(1,1)$, and a varying rotation parameter $a$.}
\label{fig:a_fixed}
\end{figure}

\begin{figure}[htbp]
\centering
\subfigure[]{
    \begin{minipage}[t]{0.4\linewidth}
    \centering
    \includegraphics[width=1\linewidth]{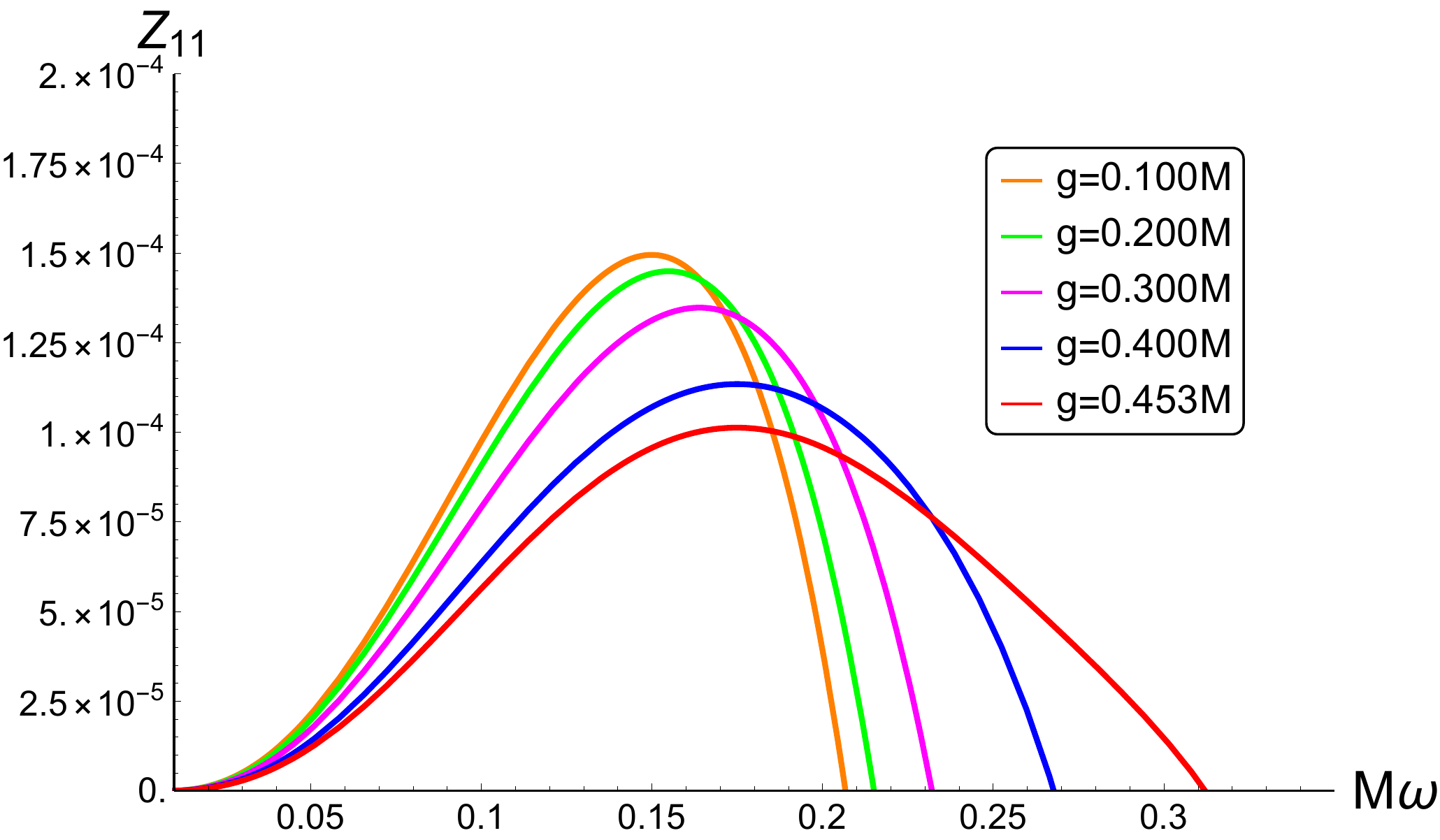}
    \end{minipage}
}
\subfigure[]{
    \begin{minipage}[t]{0.4\linewidth}
    \centering
    \includegraphics[width=1\linewidth]{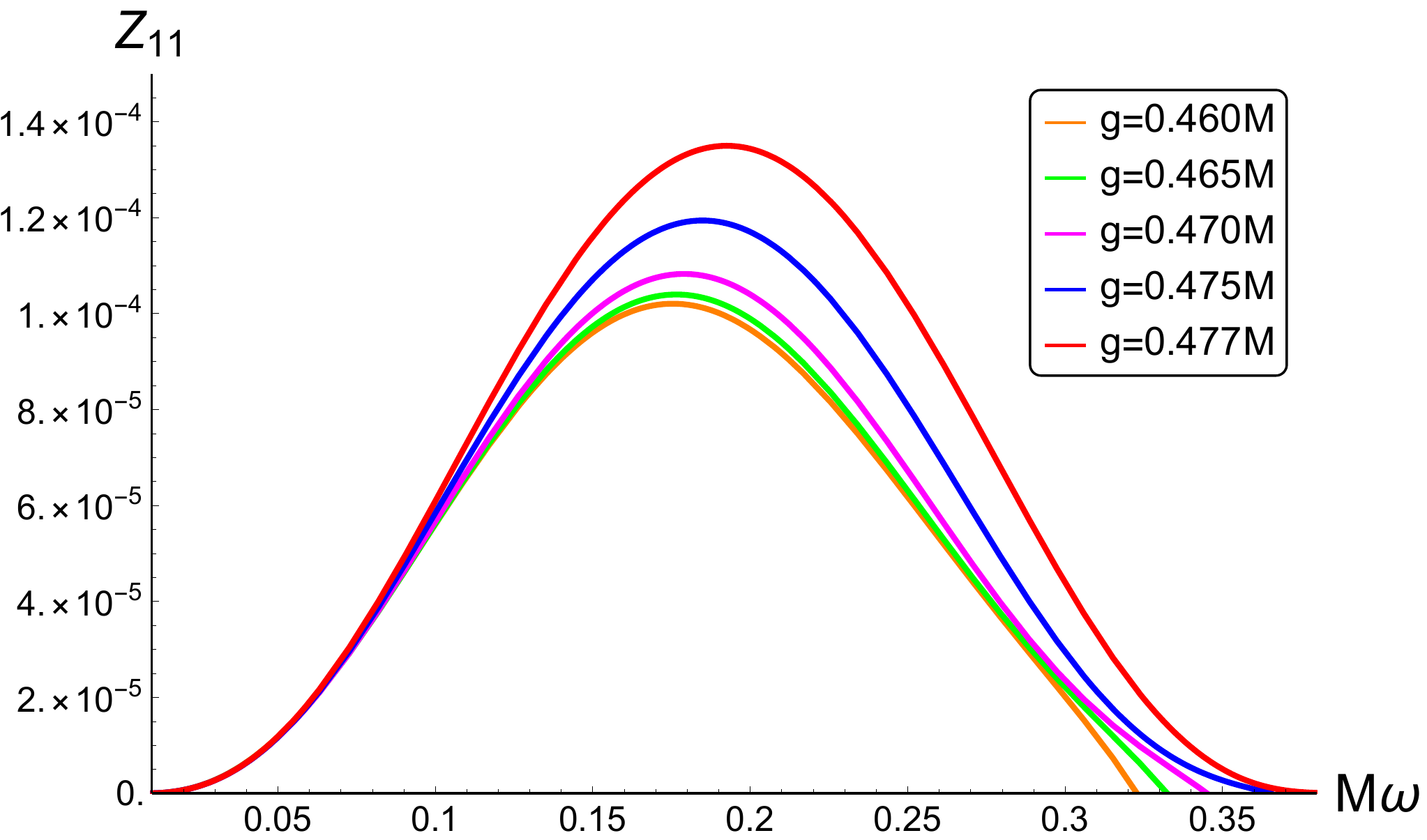}
    \end{minipage}
}
\vspace{-3mm}
\subfigure[]{
    \begin{minipage}[t]{0.4\linewidth}
    \centering
    \includegraphics[width=1\linewidth]{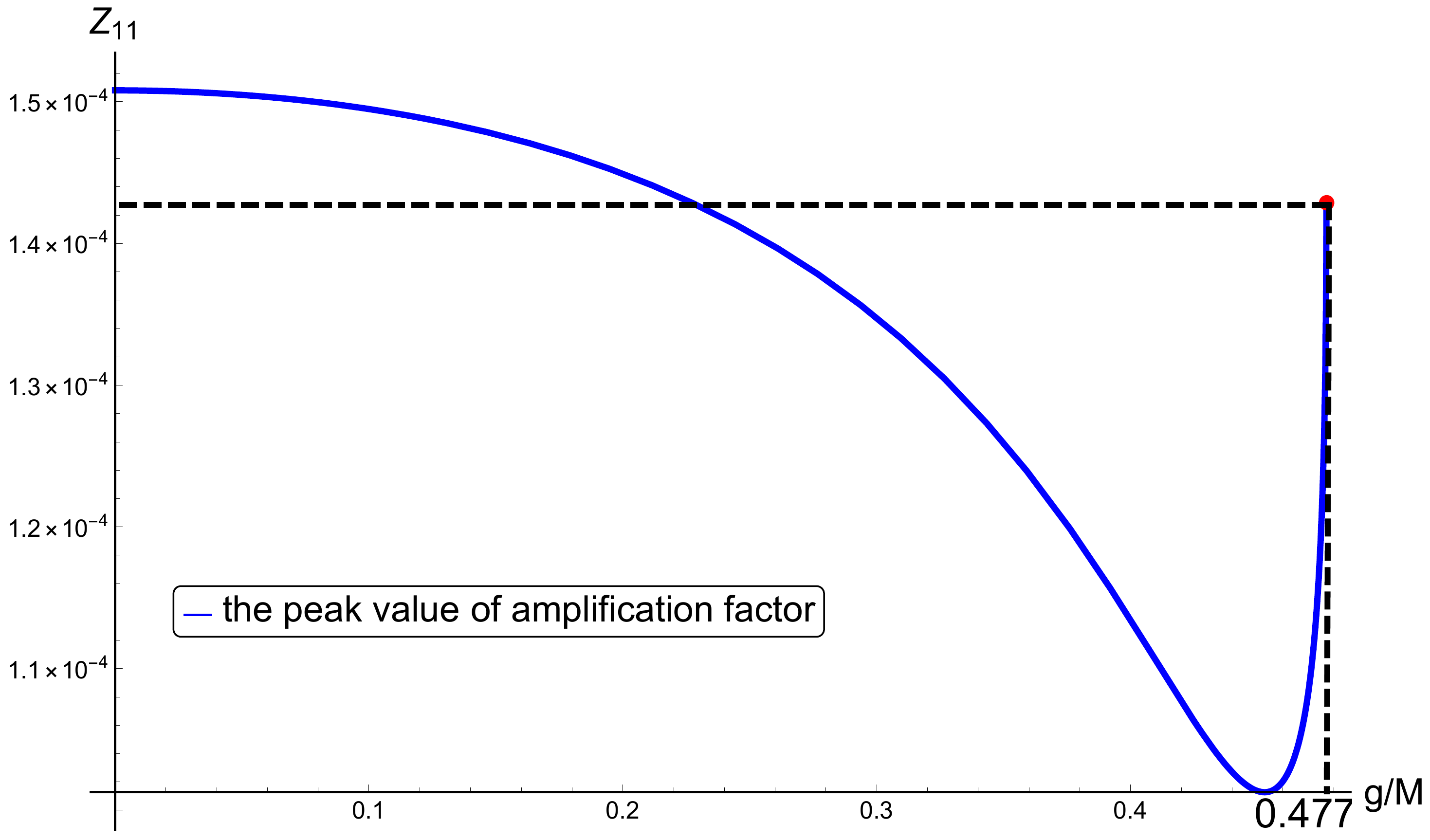}
    \end{minipage}
}
\caption{The amplification factor of massive scalar particles scattered by the rotating Bardeen black hole is computed by Eq.~\eqref{Zlm}, where the parameters are set to be: $M\mu=0.010$, $a=0.700M$, $(l,m)=(1,1)$,  and a varying $g$. In diagram (a), $0<g\le0.453M$, and in diagram (b), $0.453M<g\le 0.477M$. Moreover, diagram (c) depicts the overall variation of the peak of amplification factor with respect to $g$.}
\label{fig:a=0.7M}
\end{figure}

\item The mode in the interval of $0.737M<a<0.947M$

The variation of the peak of superradiance amplification factor with respect to $g$ is the same as that in the second mode.
The main difference is that the peak is larger than that associated with the Kerr black hole when the rotating Bardeen black hole is approaching to its extreme configuration.
This suggests that the effect of magnetic charges makes the efficiency of extracting energy from the rotating Bardeen black hole bigger than that from the Kerr black hole.
In Fig.~\ref{fig:a=0.8M}, $a=0.800M$ is taken as an example.
When $g=0.385M$, the rotating Bardeen black hole reaches its extreme configuration, and the peak goes up to the maximum, $2.721\times 10^{-4}$, which is greater than $2.376\times 10^{-4}$ associated with the Kerr black hole.

\begin{figure}[htbp]
\centering
\subfigure[]{
    \begin{minipage}[t]{0.4\linewidth}
    \centering
    \includegraphics[width=1\linewidth]{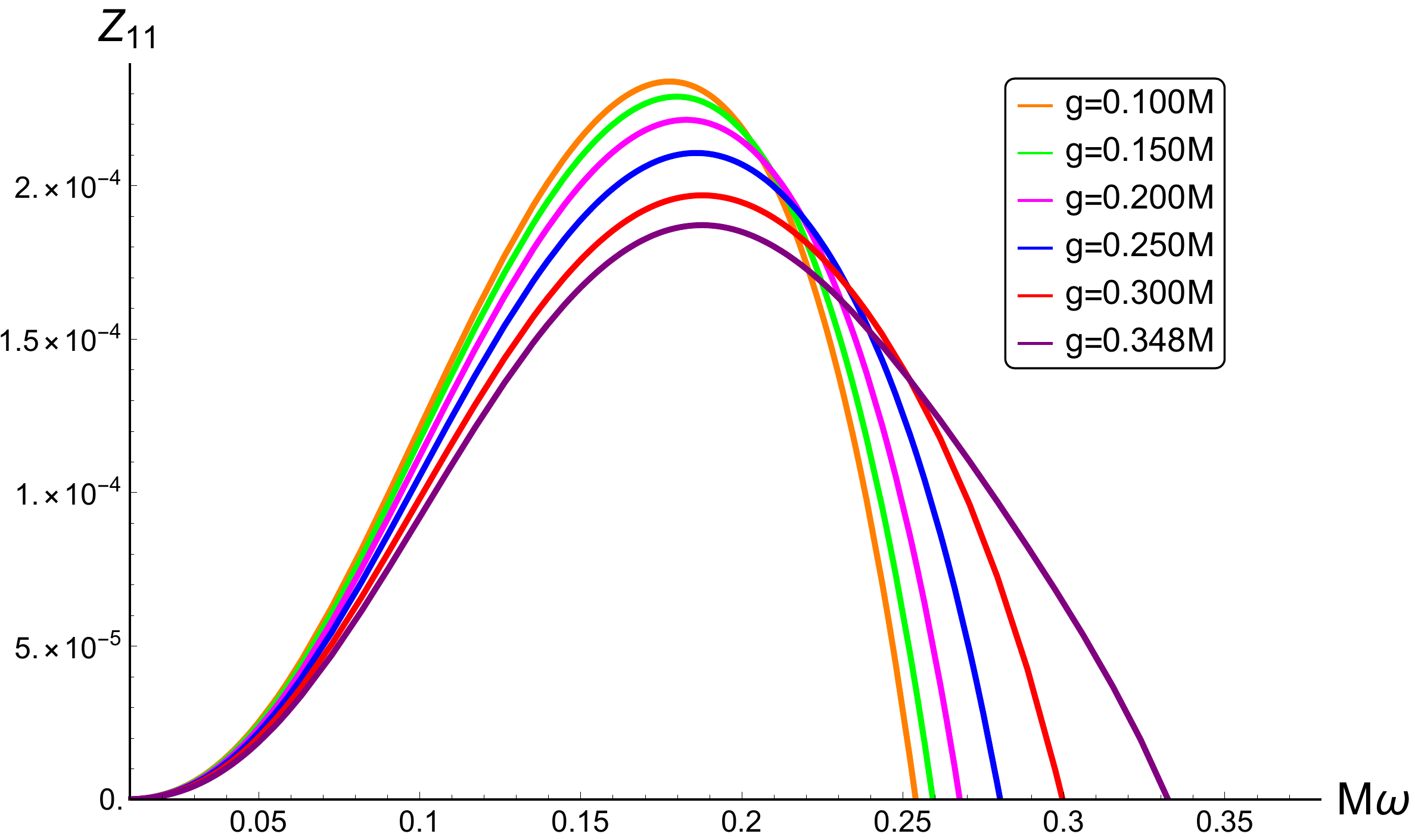}
    \end{minipage}
}
\subfigure[]{
    \begin{minipage}[t]{0.4\linewidth}
    \centering
    \includegraphics[width=1\linewidth]{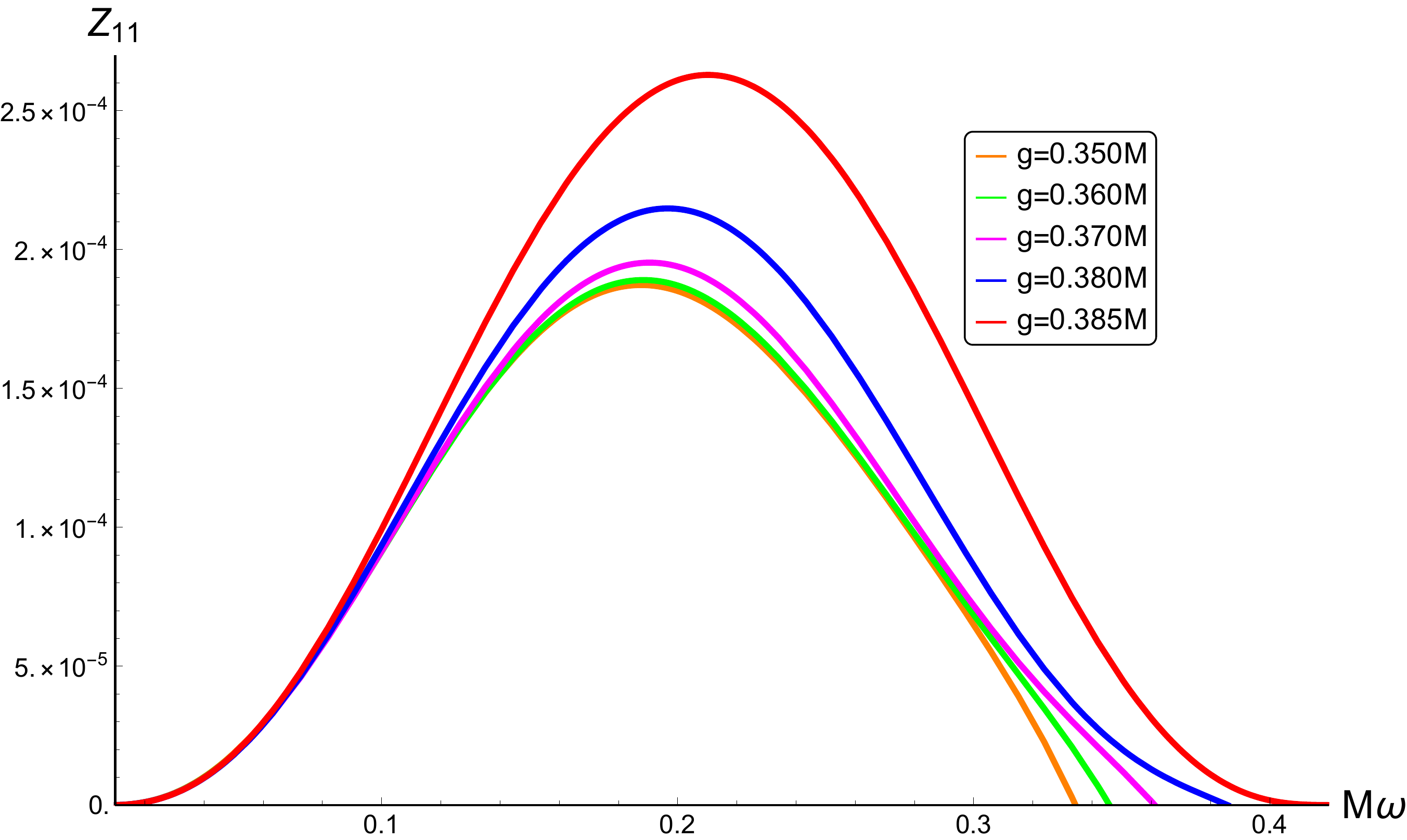}
    \end{minipage}
}
\vspace{-3mm}
\subfigure[]{
    \begin{minipage}[t]{0.4\linewidth}
    \centering
    \includegraphics[width=1\linewidth]{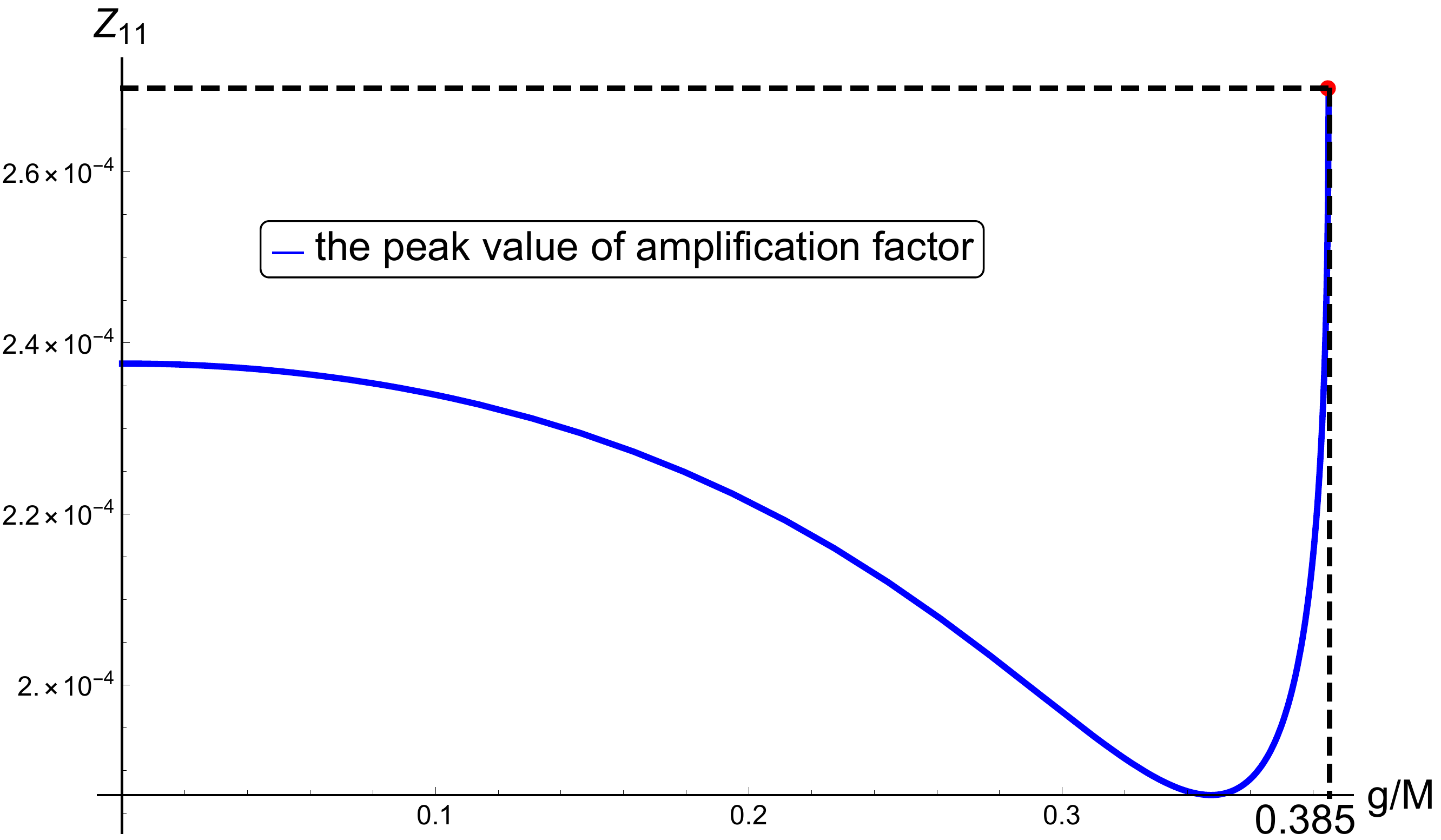}
    \end{minipage}
}
\caption{The amplification factor of massive scalar particles scattered by the rotating Bardeen black hole is computed by Eq.~\eqref{Zlm}, where the parameters are set to be: $M\mu=0.010$, $a=0.800M$, $(l,m)=(1,1)$,  and a varying $g$. In diagram (a), $0<g\le0.348M$, and in diagram (b), $0.348M<g\le 0.385M$. Moreover, diagram (c) depicts the overall variation of the peak of amplification factor with respect to $g$.}
\label{fig:a=0.8M}
\end{figure}

\item The mode in the interval of $0.947M<a<M$ 

The peak rises monotonically with an increase of the magnetic charge, and reaches its maximum when the rotating Bardeen black hole goes to its extreme configuration.
This shows that the introduction of $g$ makes the efficiency of energy extraction continuously grow.
In Fig.~\ref{fig:a=0.99M}, $a=0.990M$ is taken as an example.
When $g=0.082M$, the rotating Bardeen black hole reaches its extreme configuration, and the peak goes up to its maximum, $8.158\times 10^{-4}$, which is larger than $5.942\times 10^{-4}$ related to the Kerr black hole.
\end{enumerate}

\begin{figure}[htbp]
\centering
\subfigure[]{
    \begin{minipage}[t]{0.4\linewidth}
    \centering
    \includegraphics[width=1\linewidth]{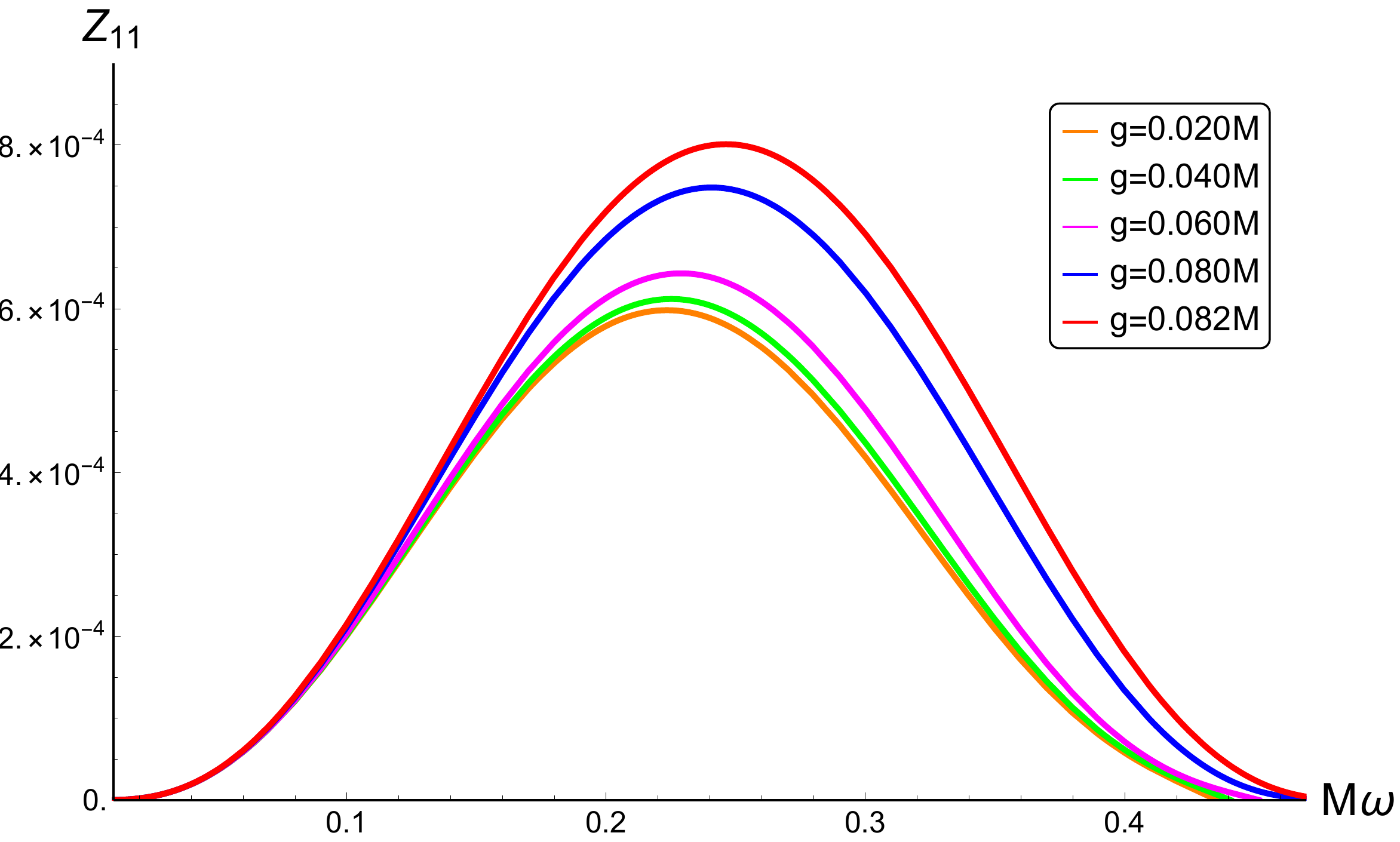}
    \end{minipage}
}
\subfigure[]{
    \begin{minipage}[t]{0.4\linewidth}
    \centering
    \includegraphics[width=1\linewidth]{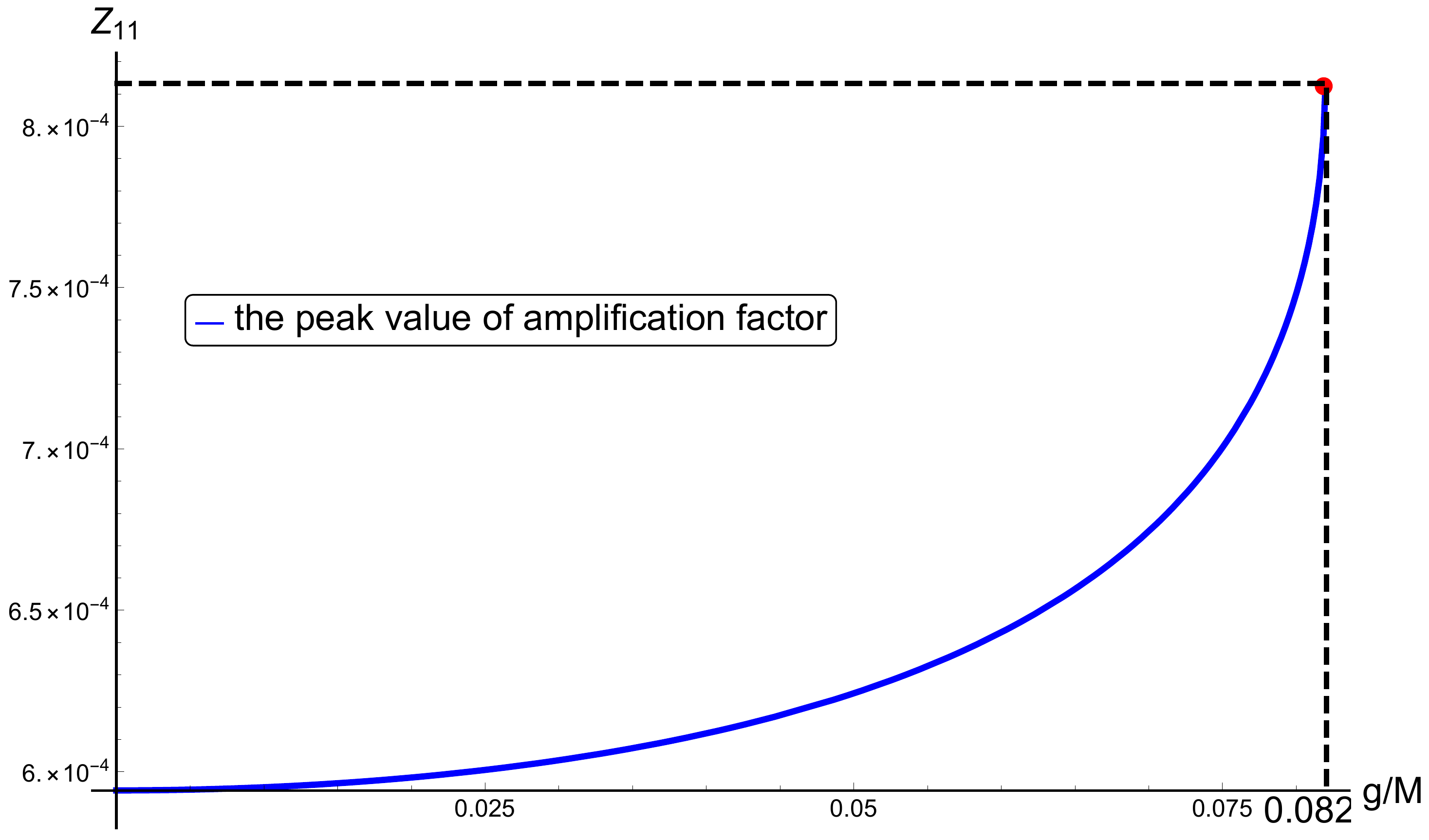}
    \end{minipage}
}

\caption{The amplification factor of massive scalar particles scattered by the rotating Bardeen black hole is computed by Eq.~\eqref{Zlm}, where the parameters are set to be: $M\mu=0.010$, $a=0.990M$, $(l,m)=(1,1)$, and $0<g\le 0.082M$. Moreover, diagram (b) depicts the overall variation of the peak of amplification factor with respect to $g$.}
\label{fig:a=0.99M}
\end{figure}

From the above four modes, we see that the rotation parameter can also be divided into four intervals for rotating Bardeen black holes: The low rotation interval ($0.013M<a<0.246M$), the middle rotation interval ($0.246M<a<0.737M$), the high rotation interval ($0.737M<a<0.947M$), and the near-extreme rotation interval ($0.947M<a<M$).
There also exist the promotional and inhibitory effects of the regularization parameter $g$ on the energy extraction efficiency in these four intervals. The promotional and inhibitory effects are consistent with those in the rotating Hayward black hole.

Thus we can also summarize the following three conclusions:
\begin{itemize}
\item In the background of rotating Bardeen black holes, the peak of amplification factor of  massive scalar fields is dominated by the rotation parameter $a$, not by the magnetic charge $g$ which plays only a secondary role.

\item For four rotation parameter intervals, the influence of $g$ on the peak of amplification factor shows various situations. In short, there are four modes in which the energy extraction efficiency varies in accordance with the competition between the two parameters, $a$ and $g$.

\item When the rotating Bardeen black hole stays in its near-extreme configuration, the peak of amplification factor always grows with an increase of $g$. Furthermore, the bigger $g$ is, the closer to the extreme the rotating Bardeen black hole is.
\end{itemize}

\section{Conclusions}
\label{sec:conclusion}
At first we review the rotating regular black holes constructed by the modified NJA and give the field equations satisfied by massive scalar particles in the background of such black holes. 
Then, for a scalar particle with a small mass, we calculate analytically the eigenfrequency of superradiance instability at its quasi-bound state and the superradiance amplification factor at its free state in terms of the asymptotic matching method. 
For two specific models, the rotating Hayward black hole and the rotating Bardeen black hole, we analyze how the regularization parameter and rotation parameter affect the eigenfrequency and the amplification factor. 
Our main conclusions are summarized as follows.
\begin{itemize}
\item
We provide the condition the superradiance instability can occur for a scalar particle with a small mass when it is at its quasi-bound state, see Eq.~(\ref{superradinscon}). In addition, we give the condition the superradiance amplification can appear for a scalar particle with a small mass when it is at its free state, see  Eq.~(\ref{Super_Gene_con}). Comparing the two conditions, we can see that the two phenomena, the superradiance instability and superradiance amplification, are closely related to each other. In fact, 
they happen or do not happen simultaneously.

\item 
For the two specific models, the rotating Hayward and rotating Bardeen black holes, we find
two modes for the growth of superradiance instability, and four modes for the energy extraction of superradiance amplification. These modes depend on the competition between the regularization parameter and the rotation parameter in each model. In particular, we notice that the behaviors of superradiance instability and superradiance amplification are almost the same in the two models, which is natural because the regularization parameter plays the same role in each model.

\end{itemize}

In general, the static and spherically symmetric regular black holes can be divided~\cite{A42} into four classes, where the Hayward and Bardeen black holes belong to the first class. In this class every regular black hole contains a de Sitter-like core and has two horizons or one overlapped horizon at least, and its causal structure can be depicted by the Carter-Penrose diagram that is similar to that of a Reissner-Nordstr\"om black hole. 
However, the singularity avoidance in the Bardeen black hole can be understood by a mechanism called {\em topology change}~\cite{A43, A44}, that is, the Bardeen black hole has a topology that is different from the Reissner-Nordstr\"om black hole. Inspired by the classification of regular black holes and the mechanism of topology change, we plan to investigate the superradiance relevant effects for the rotating counterparts of the other three classes of regular black holes and explain their superradiance features by analyzing the topology of these black holes in our future work.



\section*{Acknowledgments}
This work was supported in part by the National Natural Science Foundation of China under grant Nos. 11675081 and 12175108.
The authors would like to thank the anonymous referee for the helpful comments that improve this work greatly.
\section*{Appendix}
\begin{appendices}

\section{Derivation of the imaginary part of eigenfrequency  \label{appendix:A}}
\setcounter{equation}{0}
\renewcommand\theequation{A.\arabic{equation}}
At first we solve the eigenfrequency according to Eq.~\eqref{muomere}.
Because $\delta \nu$ is complex with its real part $\delta\nu_{\rm R}$ and imaginary part $\delta\nu_{\rm I}$, we rewrite Eq.~\eqref{muomere} as
\begin{equation}
\begin{split}
\mu^2-\omega^2 &=\frac{A^2\mu^4}{4(l+n+1+\delta\nu_{\rm R}+i\delta\nu_{\rm I})^2}\\
&=\frac{A^2\mu^4(l+n+1+\delta\nu_{\rm R}-i\delta\nu_{\rm I})^2}{4[(l+n+1+\delta\nu_{\rm R})^2+\delta\nu_{\rm I}^2]^2}.\\
\end{split}
\end{equation}
Since the modulus of $\delta\nu$ is much smaller than one, which indicates that both $|\delta\nu_{\rm R}|$ and $|\delta\nu_{\rm I}|$ are much smaller than one, the above equation can be approximated as
\begin{equation}
\begin{split}
\mu^2-\omega^2&\approx \frac{A^2\mu^4(l+n+1+\delta\nu_{\rm R}-i\delta\nu_{\rm I})^2}{4(l+n+1+\delta\nu_{\rm R})^4}\\
&\approx \frac{A^2\mu^4}{4(l+n+1+\delta\nu_{\rm R})^2}-\frac{iA^2\mu^4(l+n+1+\delta\nu_{\rm R})\delta\nu_{\rm I}}{2(l+n+1+\delta\nu_{\rm R})^4}\\
&\approx \frac{A^2\mu^4}{4(l+n+1)^2}-\frac{iA^2\mu^4}{2(l+n+1)^3}\delta\nu_{\rm I}.
\end{split}
\end{equation}
Considering $\mu^2-\omega^2=\mu^2-\omega_{\rm R}^2+\omega_{\rm I}^2-2i\omega_{\rm R}\omega_{\rm I}$ and the slow change approximation, $\omega_{\rm R}\gg\omega_{\rm I}$, we obtain from the above equation,
\begin{equation}
\mu^2-\omega_{\rm R}^2\approx \frac{A^2\mu^4}{4(l+n+1)^2},\label{muomegar}
\end{equation}
and
\begin{equation}
\omega_{\rm R}\omega_{\rm I}\approx\frac{A^2\mu^4}{4(l+n+1)^3}\delta\nu_{\rm I}.\label{frerealfreimag}
\end{equation}
Again considering $\mu A\ll 1$, see Eq.~(\ref{Fexpand}) and the relevant explanation, we derive the real part of eigenfrequency from Eq.~(\ref{muomegar}) approximately, 
\begin{equation}
\omega_{\rm R}
\approx \mu\left[1-\frac{A^2\mu^2}{8(l+n+1)^2}\right],\label{freqreal}
\end{equation}
from which we deduce $\omega_{\rm R}<\mu$ and
the second term is much smaller than the first one, i.e. $\frac{A^2\mu^2}{8(l+n+1)^2}\ll 1$. Note that this inequality ($\omega_{\rm R}<\mu$) means that the massive scalar particle stays at a quasi-bound state.
Substituting Eq.~(\ref{freqreal}) into Eq.~(\ref{frerealfreimag}), we give the imaginary part of eigenfrequency approximately,
\begin{equation}
\omega_{\rm I}
\approx \frac{A^2\mu^3}{4(l+n+1)^3}\left[1+\frac{A^2\mu^2}{8(l+n+1)^2}\right]\delta\nu_{\rm I}.\label{freqimaginary}
\end{equation}

Next, we need to derive $\delta\nu_{\rm I}$ from Eq.~(\ref{imanu}) in which the quantities $k$ and $P$ should be dealt with, for the definitions of $k$ and $P$, see Eq.~(\ref{Def_1}) and Eq.~(\ref{Def_P}), respectively.

According to Eq.~(\ref{Def_1}), $k$ is complex and can be expressed as $k=k_{\rm R}+ik_{\rm I}$, so we obtain the following results up to their respective leading orders by using Eq.~(\ref{freqreal}) and Eq.~(\ref{freqimaginary}), 
\begin{equation}
k_{\rm R}^2-k_{\rm I}^2=\mu^2-\omega_{\rm R}^2+\omega_{\rm I}^2\approx\frac{A^2\mu^4}{4(l+n+1)^2}\equiv C,\label{defquantC}
\end{equation}
\begin{equation}
k_{\rm R}k_{\rm I}=-\omega_{\rm R}\omega_{\rm I}\approx-\frac{A^2\mu^4}{4(l+n+1)^3}\delta\nu_{\rm I}\equiv D.\label{defquantD}
\end{equation}
Note that $C\gg |D|$ due to $|\delta\nu_{\rm I}|\ll 1$.
By solving Eqs.~(\ref{defquantC}) and (\ref{defquantD}), we work out the real and imaginary parts of $k$ approximately,
\begin{eqnarray}
k_{\rm R}^2&\approx& C, \label{squrtkreal}\\
k_{\rm I}^2&\approx&\frac{D^2}{C}.
\end{eqnarray}
Because of  $C\gg |D|$, we deduce 
\begin{equation}
|k_{\rm R}|\gg |k_{\rm I}|,\label{realklargeimgk}
\end{equation}
which will be used to distinguish whether $k_{\rm R}$ or $k_{\rm I}$ gives the main contribution to $\delta\nu_{\rm I}$ in 
Eq.~(\ref{imanu}).

We then turn to $P$. According to Eq.~(\ref{Def_P}), $P$ is complex with its real part  $P_{\rm R}$ and imaginary part $P_{\rm I}$. Let us see the numerator of Eq.~(\ref{Def_P}), the first term  is much larger than the absolute value of the real part of the second term because our approximations are  
$\mu M\ll 1$ and $|\omega| M\ll 1$, which gives rise to
\begin{equation}  
P_{\rm R}\gg |P_{\rm I}|.\label{realPlargeimgP}
\end{equation}
As a result, the main contribution to $\delta\nu_{\rm I}$ comes from the product of $|k_{\rm R}|$ and $P_{\rm R}$  according to Eq.~\eqref{imanu}, Eq.~(\ref{realklargeimgk}), and Eq.~(\ref{realPlargeimgP}).

By using Eqs.~(\ref{defquantC}) and (\ref{squrtkreal}), we obtain
\begin{equation}
	k_{\rm R}\approx\frac{|A|\mu^2}{2(l+n+1)},
\end{equation}
where we have taken $k_{\rm R}>0$ in order to ensure that $x$ defined by Eq.~\eqref{Def_x} does not reverse the direction of $r$.
If the direction of $r$ were reversed, the directions of incoming and outgoing waves at two boundaries would be reversed, which would lead to inconvenience. Moreover, we compute $P_{\rm R}$ in terms of Eq.~(\ref{Def_P}) and Eq.~(\ref{freqreal}), 
\begin{equation}
	P_{\rm R}\approx\frac{ma-2\mu f(r_{\rm H}^+)}{2r_{\rm H}^+-2f'(r_{\rm H}^+)},\label{realPequ}
\end{equation}
where we have ignored the second term in Eq.~(\ref{freqreal}), {\em i.e.}, we have taken $\omega_{\rm R}\approx \mu$. Similarly, we can also ignore the second term in Eq.~(\ref{freqimaginary})
and arrive at Eq.~\eqref{imaomega}.

Substituting Eqs.~(\ref{realklargeimgk})-(\ref{realPequ}) into Eq.~\eqref{imanu}, we compute $\delta\nu_{\rm I}$. Again substituting this $\delta\nu_{\rm I}$ into Eq.~\eqref{imaomega}, we finally derive Eq.~\eqref{eq:Omega_I}, the imaginary part of eigenfrequency.
\end{appendices}


\begin{thebibliography}{99}


\bibitem{A1}
B. P. Abbott {\em et al}. (LIGO Scientific Collaboration and Virgo Collaboration),
    {\em Observation of gravitational waves from a binary black hole merger}, \href{https://doi.org/10.1103/PhysRevLett.116.061102}
    {{Phys. Rev. Lett.} {\bf 116}, 061102 (2016)}.
    \href{https://doi.org/10.48550/arXiv.1602.03837}{[arXiv:1602.03837 [gr-qc]]}

\bibitem{A2}
 K. Akiyama {\em et al}. (Event Horizon Telescope Collaboration),
    {\em First M87 event horizon telescope results. I. The shadow of the supermassive black hole}, \href{https://doi.org/10.3847/2041-8213/ab0ec7}
    {{Astrophys. J.} {\bf 875}, L1 (2019)}.


\bibitem{A3}
K. Akiyama {\em et al}. (Event Horizon Telescope Collaboration),
    {\em First M87 event horizon telescope results. VI. The shadow and mass of the central black hole}, \href{https://doi.org/10.3847/2041-8213/ab1141}
    {{Astrophys. J.} {\bf 875}, L6 (2019)}.

\bibitem{A4}
R. Penrose,
    {\em Gravitational collapse and space-time singularities}, \href{https://doi.org/10.1103/PhysRevLett.14.57}
    {{Phys. Rev. Lett.} {\bf 14}, 57 (1965)}.

\bibitem{A5}
 S. W. Hawking and G. F. R. Ellis,
    {\em The large scale structure of space time},
    {Cambridge University Press, New York, 1973}.

\bibitem{A6}
S. Ansoldi,
    {\em Spherical black holes with regular center: A review of existing models including a recent realization with Gaussian sources}, 
    \href{https://doi.org/10.48550/arXiv.0802.0330}{arXiv:0802.0330 [gr-qc]}.

\bibitem{A7}
 J. M. Bardeen,
    {\em Non-singular general-relativistic gravitational collapse},
    {{ in Proc. Int. Conf. GR5, vol. 174, Tbilisi, 1968}}.
   

\bibitem{A8}
E. Ay$\acute{\rm o}$n-Beato and A. Gar$\acute{\rm c}$a,
    {\em The Bardeen model as a nonlinear magnetic monopole}, \href{https://doi.org/10.1016/S0370-2693%2800%2901125-4}
    {{Phys. Lett.} {\bf B 493}, 149 (2000)}.
    \href{https://doi.org/10.48550/arXiv.gr-qc/0009077}{[arXiv:gr-qc/0009077]}


\bibitem{A9}
S. A. Hayward,
    {\em Formation and evaporation of non-singular black holes}, \href{https://doi.org/10.1103/PhysRevLett.96.031103}
    {{Phys. Rev. Lett.} {\bf 96}, 031103 (2000)}.
    \href{https://doi.org/10.48550/arXiv.gr-qc/0506126}{[arXiv:gr-qc/0506126]}

\bibitem{A10}
P. Nicolini,
    {\em Noncommutative black holes, the final appeal to quantum gravity: A review}, \href{https://doi.org/10.1142/S0217751X09043353}
    {{Int. J. Mod. Phys.} {\bf A 24}, 1229 (2009)}.
    \href{https://doi.org/10.48550/arXiv.0807.1939}{[arXiv:0807.1939 [hep-th]]}   

\bibitem{A11}
N. Bodendorfer, F. M. Mele, and J. M\"unch,
    {\em (b,v)-type variables for black to white hole transitions in effective loop quantum gravity}, \href{https://doi.org/10.1016/j.physletb.2021.136390}
    {{Phys. Lett.} {\bf B 819}, 136390 (2021)}.
    \href{https://doi.org/10.48550/arXiv.1911.12646}{[arXiv:1911.12646 [gr-qc]]}   

\bibitem{A12}
E. T. Newman and A. I. Janis,
    {\em Note on the Kerr spinning-particle metric}, \href{https://doi.org/10.1063/1.1704350}
    {{J. Math. Phys. (N.Y.)} {\bf 6}, 915 (1965)}.

\bibitem{A13}
L. Modesto and P. Nicolini,
    {\em Charged rotating noncommutative black holes}, \href{https://doi.org/10.1103/PhysRevD.82.104035}
    {{Phys. Rev.} {\bf D 82}, 104035 (2010)}.
    \href{https://doi.org/10.48550/arXiv.1005.5605}
    {[arXiv:1005.5605 [gr-qc]]}

\bibitem{A14}
C. Bambi and L. Modesto,
    {\em Rotating regular black holes}, \href{https://doi.org/10.1016/j.physletb.2013.03.025}
    {{Phys. Lett.} {\bf B 721}, 329 (2013)}.
    \href{https://doi.org/10.48550/arXiv.1302.6075}
    {[arXiv:1302.6075 [gr-qc]]}

\bibitem{A15}
B. Toshmatov, B. Ahmedov, A. Abdujabbarov, and Z. Stuchlik,
    {\em Rotating regular black hole solution}, \href{https://doi.org/10.1103/PhysRevD.89.104017}
    {{Phys. Rev. } {\bf D 89}, 104017 (2014)}.
    \href{https://doi.org/10.48550/arXiv.1404.6443}
    {[arXiv:1404.6443 [gr-qc]]}

\bibitem{A16}
M. Azreg-Ainou,
    {\em Generating rotating regular black hole solutions without complexification}, \href{https://doi.org/10.1103/PhysRevD.90.064041}
    {{Phys. Rev.} {\bf D 90}, 064041 (2014)}.
    \href{https://doi.org/10.48550/arXiv.1405.2569}
    {[arXiv:1405.2569 [gr-qc]]} 

\bibitem{A16A}
M. Azreg-Ainou,
    {\em From static to rotating to conformal static solutions: Rotating imperfect fluid wormholes with(out) electric or magnetic field}, \href{https://doi.org/10.1140/epjc/s10052-014-2865-8}
    {{Eur. Phys. J. } {\bf C 74}, 2865 (2014)}.
    \href{https://doi.org/10.48550/arXiv.1401.4292}
    {[arXiv:1401.4292 [gr-qc]]}

\bibitem{A17}
S. G. Ghosh, and S. D. Maharaj,
    {\em Radiating Kerr-like regular black hole}, \href{https://doi.org/10.1140/epjc/s10052-014-3222-7}
    {{Eur. Phys. J.} {\bf C 75}, 7 (2015)}.
    \href{https://doi.org/10.48550/arXiv.1410.4043}
    {[arXiv:1410.4043 [gr-qc]]}   

\bibitem{A18}
I. Dymnikova and E. Galaktionov,
    {\em Regular rotating electrically charged black holes and solitons in nonlinear electrodynamics minimally coupled to gravity}, \href{https://doi.org/10.1088/0264-9381/32/16/165015}
    {{Class. Quant. Grav.} {\bf 32}, 165015 (2015)}.
    \href{https://doi.org/10.48550/arXiv.1510.01353}
    {[arXiv:1510.01353 [gr-qc]]}     

\bibitem{A19}
B. Toshmatov, Z. Stuchlik, and B. Ahmedov,
    {\em Generic rotating regular black holes in general relativity coupled to nonlinear electrodynamics}, \href{https://doi.org/10.1103/PhysRevD.95.084037}
    {{Phys. Rev. }{\bf D 95}, 084037 (2017)}.
    \href{https://doi.org/10.48550/arXiv.1704.07300}
    {[arXiv:1704.07300 [gr-qc]]} 

\bibitem{A20}
S. Brahma, C.-Y. Chen, and D.-h. Yeom,
    {\em Testing loop quantum gravity from observational consequences of non-singular rotating black holes}, \href{https://doi.org/10.1103/PhysRevLett.126.181301}
    {{Phys. Rev. Lett. }{\bf 126}, 181301 (2021)}.
    \href{https://doi.org/10.48550/arXiv.2012.08785}
    {[arXiv:2012.08785 [gr-qc]]} 

\bibitem{A21}
R. Penrose,
    {\em Gravitational collapse: The role of general relativity}, \href{https://doi.org/10.1023/A:1016578408204}
    {{Riv. Nuovo Cim.} {\bf 1}, 252 (1969)};
    {{Gen. Rel. Grav. }{\bf 34}, 1141 (2002)}.

\bibitem{A22}
Ya. B. Zeldovich,
    {\em Generation of waves by a rotating body}, 
    {{Zh. $\acute{\rm E}$ksp. Teor. Fiz. Pisma} {\bf 14}, 270 (1971)}
    [JETP Lett. {\bf 14}, 180 (1971)].

\bibitem{A23}
J. Bekenstein,
    {\em Extraction of energy and charge from a black hole}, \href{http://dx.doi.org/10.1103/PhysRevD.7.949}
    {{Phys. Rev.} {\bf D 7}, 949 (1973)}.

\bibitem{A24}
S. Teukolsky and W. Press,
    {\em Perturbations of a rotating black hole. III - Interaction of the hole with gravitational and electromagnet ic radiation}, \href{http://dx.doi.org/10.1086/153180}
    {{Astrophys. J.} {\bf 193}, 443 (1974)}.

\bibitem{A25}
J. D. Bekenstein and M. Schiffer,
    {\em The many faces of superradiance}, \href{https://doi.org/10.1103/PhysRevD.58.064014}
    {{Phys. Rev. } {\bf D 58}, 064014 (1998)}.
    \href{https://doi.org/10.48550/arXiv.gr-qc/9803033}
    {[arXiv:gr-qc/9803033]}

\bibitem{A26}
R. Brito, V. Cardoso, and P. Pani,
    {\em Superradiance -- the 2020 Edition}, \href{https://doi.org/10.1007/978-3-030-46622-0}
    {{Lecture Notes in Physics} {\bf 971}, (2020)}.
    \href{https://doi.org/10.48550/arXiv.1501.06570}
    {[arXiv:1501.06570 [gr-qc]]} 

\bibitem{A27}
W. H. Press and S. A. Teukolsky,
    {\em Floating orbits, superradiant scattering and the black-hole bomb}, \href{https://doi.org/10.1038/238211a0}
    {{Nature} {\bf 238}, 211 (1972)}.

\bibitem{A28}
V. Cardoso, O. J. C. Dias, J. P. S. Lemos, and S. Yoshida,
    {\em The black hole bomb and superradiant instabilities}, \href{https://doi.org/10.1103/PhysRevD.70.044039}
    {{Phys. Rev.} {\bf D 70}, 044039 (2004)};
    \href{https://doi.org/10.1103/PhysRevD.70.049903}
    {Erratum-ibid. {\bf D 70}, 049903 (2004)}.
    \href{https://doi.org/10.48550/arXiv.hep-th/0404096}
    {[arXiv:hep-th/0404096]} 

\bibitem{A29}
H. R. C. Ferreira and C. A. R. Herdeiro,
    {\em Superradiant instabilities in the Kerr-mirror and Kerr-AdS black holes with Robin boundary conditions}, \href{https://doi.org/10.1103/PhysRevD.97.084003}
    {{Phys. Rev.} {\bf  D 97}, 084003 (2018)}.
    \href{https://doi.org/10.48550/arXiv.1712.03398}
    {[arXiv:1712.03398 [gr-qc]]} 

\bibitem{A30}
T. Damour, N. Deruelle, and R. Ruffini,
    {\em On quantum resonances in stationary geometries}, \href{https://doi.org/10.1007/BF02725534}
    {{Lett. Nuovo Cimento} {\bf 15}, 257 (1976)}.

\bibitem{A31}
T. J. M. Zouros and D. M. Eardley,
    {\em Instabilities of massive scalar perturbations of a rotating black hole}, \href{https://doi.org/10.1016/0003-4916(79)90237-9}
    {{Annals Phys.} {\bf 118}, 139 (1979)}.

\bibitem{A32}
S. Detweiler,
    {\em Klein-Gordon equation and rotating black holes}, \href{https://doi.org/10.1103/PhysRevD.22.2323}
    {{Phys. Rev.} {\bf D 22}, 2323 (1980)}.

\bibitem{A33}
H. Furuhashi and Y. Nambu,
    {\em Instability of massive scalar fields in Kerr-Newman space-time}, \href{https://doi.org/10.1143/PTP.112.983}
    {{Prog. Theor. Phys.} {\bf 112}, 983 (2004)}.
    \href{https://arxiv.org/abs/gr-qc/0402037}
    {[arXiv:gr-qc/0402037]} 


\bibitem{A34}
S. R. Dolan,
    {\em Instability of the massive Klein-Gordon field on the Kerr spacetime}, \href{https://doi.org/10.1103/PhysRevD.76.084001}
    {{Phys. Rev.} {\bf D 76}, 084001 (2007)}.
    \href{https://doi.org/10.48550/arXiv.0705.2880}
    {[arXiv:0705.2880 [gr-qc]]} 

\bibitem{A35}
P. Pani, V. Cardoso, L. Gualtieri, E. Berti, and A. Ishibashi,
    {\em Perturbations of slowly rotating black holes: Massive vector fields in the Kerr metric}, \href{https://doi.org/10.1103/PhysRevD.86.104017}
    {{Phys. Rev.} {\bf D 86}, 104017 (2012)}.
    \href{https://doi.org/10.48550/arXiv.1209.0773}
    {[arXiv:1209.0773 [gr-qc]]} 

\bibitem{A36}
S. Hod,
    {\em No-bomb theorem for charged Reissner-Nordstr\"om black holes}, \href{https://doi.org/10.1016/j.physletb.2012.06.043}
    {{Phys. Lett.} {\bf B 718}, 1489 (2013)}.
    \href{https://doi.org/10.48550/arXiv.1304.6474}
    {[arXiv:1304.6474 [gr-qc]]}

\bibitem{A37}
R. Brito, V. Cardoso, and P. Pani,
    {\em Massive spin-2 fields on black hole spacetimes: Instability of the Schwarzschild and Kerr solutions and bounds on the graviton mass}, \href{https://doi.org/10.1103/PhysRevD.88.023514}
    {{Phys. Rev.} {\bf D 88}, 023514 (2013)}.
    \href{https://doi.org/10.48550/arXiv.1304.6725}
    {[arXiv:1304.6725 [gr-qc]]} 

\bibitem{A38}
W. E. East and F. Pretorius,
    {\em Superradiant instability and backreaction of massive vector fields around Kerr black holes}, \href{https://doi.org/10.1103/PhysRevLett.119.041101}
    {{Phys. Rev. Lett.} {\bf 119}, 041101 (2017)}.
    \href{https://doi.org/10.48550/arXiv.1704.04791}
    {[arXiv:1704.04791 [gr-qc]]}

\bibitem{A39}
M. Khodadi,
    {\em Black hole superradiance in the presence of Lorentz symmetry violation}, \href{https://doi.org/10.1103/PhysRevD.103.064051}
    {{Phys. Rev.} {\bf D 103}, 064051 (2021)}.
    \href{https://doi.org/10.48550/arXiv.2103.03611}
    {[arXiv:2103.03611 [gr-qc]]} 

\bibitem{A40}
Z.-F. Mai, R.-Q. Yang, and H. Lu,
    {\em Superradiant instability of extremal black holes in STU supergravity}, \href{https://doi.org/10.1103/PhysRevD.105.024070}
    {{Phys. Rev.} {\bf D 105}, 024070 (2022)}.
    \href{https://doi.org/10.48550/arXiv.2110.14942}
    {[arXiv:2110.14942 [hep-th]]} 

\bibitem{A41}
F. W. J. Olver, D. W. Lozier, R. F. Boisvert, and C. W. Clark,
    {\em NIST handbook of mathematical functions}, 
    {Cambridge University Press, New York, 2010}.

\bibitem{A42}
K. A. Bronnikov, H. Dehnen, and V. N. Melnikov, {\em Regular black holes and black universes},
\href{	https://doi.org/10.1007/s10714-007-0430-6} 
{Gen. Rel. Grav. {\bf 39}, 973 (2007)}.
\href{	https://doi.org/10.48550/arXiv.gr-qc/0611022}
{[arXiv:gr-qc/0611022]}

\bibitem{A43}
A. Borde,
{\em Open and closed universes, initial singularities and inflation}, \href{https://doi.org/10.1103/PhysRevD.50.3692}
{{Phys. Rev.} {\bf D 50}, 3692 (1994)}.
\href{https://doi.org/10.48550/arXiv.gr-qc/9403049}
{[arXiv:gr-qc/9403049]} 

\bibitem{A44}
A. Borde,
{\em Regular black holes and topology change}, \href{https://doi.org/10.1103/PhysRevD.55.7615}
{{Phys. Rev.} {\bf D 55}, 7615 (1997)}.
\href{https://doi.org/10.48550/arXiv.gr-qc/9612057}
{[arXiv:gr-qc/9612057]} 

\end{thebibliography}
\end{document}